\documentclass[12pt,a4,reqno]{article}
\pdfoutput=1
\usepackage{geometry}                % See geometry.pdf to learn the layout options. There are lots.
\usepackage{bbm}
\usepackage{amsmath}
\usepackage{amsfonts}
\usepackage{amssymb}
\usepackage{latexsym}
\usepackage{graphicx}
\usepackage[latin1]{inputenc}
\usepackage[T1]{fontenc}
\usepackage{mathrsfs}
\usepackage{hyperref}
\usepackage[final]{movie15}
\graphicspath{{./}{figures/}{movies/}}

\newcommand{\clebsch}[6]{\left<{\begin{matrix}#1\\#2\end{matrix}\ \begin{matrix}#3\\#4\end{matrix}}\right\vert\left.{\begin{matrix}#5\\#6\end{matrix}}\right>}
\newcommand{\sixj}[6]{\left\{{\begin{matrix}#1\\#4\end{matrix}\ \begin{matrix}#2\\#5\end{matrix}}\ {\begin{matrix}#3\\#6\end{matrix}}\right\}}

\newcommand{\HS}{\mathcal{H}_{\mathcal{S}}}
\newcommand{\HE}{\mathcal{H}_{\mathcal{E}}}
\newcommand{\HaE}{{H}_{\mathcal{E}}}
\newcommand{\HaS}{{H}_{\mathcal{S}}}
\newcommand{\HaSE}{{H}_{\mathcal{SE}}}
\newcommand{\tr}{\mathrm{tr}}
\renewcommand{\imath}{\mathrm{i}}
\newcommand{\emath}{\mathrm{e}}

\newcommand{\eq}[1]{eq.~(\ref{#1})}
\newcommand{\fig}[1]{fig.~(\ref{#1})}
\newcommand{\Deltaav}{\widehat{\Delta}_{\mathrm{av}}}
\newcommand{\normtwo}[1]{{ | \!  |{#1} | \! |}_2}

\newif\ifincmov
\incmovfalse

\title{A general and solvable random matrix model for spin decoherence}

\author{François David\thanks{francois.david@cea.fr}
\\
Institut de Physique Théorique,\\
CNRS, URA 2306, F-91191 Gif-sur-Yvette, France\\
\& CEA, IPhT, F-91191 Gif-sur-Yvette, France
}
\date{November 14, 2010}       

\begin{document}
\maketitle
\begin{abstract}
We propose and solve a simple but very general quantum model of an SU(2) spin interacting with a large external system with $N$ states. 
The coupling is described by a random hamiltonian in  a new general gaussian SU(2)$\times$U($N$) random matrix ensemble, that we introduce in this paper. 
We solve the model in the large $N$ limit, for any value of the spin $j$ and for any choice of the coupling matrix element distributions in the different possible angular momentum channels $l$ (and provided that the internal dynamics of the spin is slow). 
Besides its mathematical interest as a non-trivial random matrix model, it allows to study and illustrate in a simple framework various phenomena:
the decoherence dynamics, the conditions of emergence  of the classical phase space for the spin, the properties quantum  diffusion in phase space. 
The large time evolution for the spin is shown to be non-Markovian in general, the Markov property emerging in some specific case for the dynamics and the initial conditions. 
\end{abstract}
\newpage
\tableofcontents
\newpage
\section{Introduction}
\label{sIntro}
In this paper we introduce and solve a simple but general model of decoherence for a very simple system: an SU(2) quantum spin coupled in a generic way to a large external system, which plays the role of the environment (external bath, other microscopic degrees of freedom, etc.). 
The purpose of this model is to provide a simple but quite general and exactly solvable model, in order to discuss and illustrate in a pedagogical way some of the basic aspects of decoherence and of the emergence of classical degrees of freedom in  a simple quantum system.
There is of course already a very large scientific literature on these problems and this study shall not bring revolutionary insights on these subjects. 
There are numerous excellent reviews or textbooks on decoherence
\cite{Schlosshauer2007}\cite{Zurek2003} \cite{JoosZehKiefer2003}, 
and more generally on irreversibility and dissipation in open quantum systems
\cite{Weiss2008}\cite{BreuerPetruccione2006}.
Many studies have focused on simple but realistic physical systems like a single free massive particle (Brownian walk) or an harmonic oscillator. 
A quantum spin is in fact simpler than a particle, since its phase space is compact and since the SU(2) symmetry induces many simplifications.
The spin model presented here does not aim at being a realistic one for some specific physical system in relation with experiments.
It incorporates features already introduced and studied by previous authors: quantum spin, coherent states representation, random matrix ensembles.
But it also presents novel features and allows to obtain new and exact results.

Firstly our model describes a general spin, and the results that we obtain are valid for any value of the spin $j$, allowing to study the whole range going from the case $j=1/2$ (the most studied case of the two-level system, i.e. of the q-bit) to $j=\infty$ (the classical limit where the spin becomes a classical top).
The study of dissipation and decoherence in spin systems goes back to \cite{ShibataSaito1975,TakahashiShibata1975}, but since then most studies have focused on the spin 1/2 case, i.e. of the two level system, both for simplicity and for the obvious connection with experiments.
See however Sec. VII of 
\cite{StrunzHaakeBraun2002}
for a discussion of decoherence for a large spin $j$.

Secondly we try to describe the interaction between the spin $\mathcal{S}$ and the external system $\mathcal{E}$ in the most general situation, by considering an interaction Hamiltonian belonging to a random matrix ensemble.
The idea of using random matrix theory (RMT) for such problems is of course not new.
It can be found already in 
\cite{MelloPereyraKumar1988,LutzWeidenmuller1999}.
However, as already stated, most studies involve simpler systems (for instance the two state system corresponding to $j=1/2$), and some simple random matrix ensembles like a single GUE ensemble or au Gaussian ensemble of band matrices. 
In addition in almost all studies, the interaction Hamilonian is taken to be of the form
$H_{\mathrm{int}}=U_{\mathcal{S}}\otimes V_{\mathcal{E}}$, $U_{\mathcal{S}}$ being a well chosen operator for $\mathcal{S}$ (the coupling agent), and $V_{\mathcal{E}}$ the random Hamiltonian for  $\mathcal{E}$.
Here we look for the most general random matrix model which can describe the coupling of an SU(2) spin to an external $N$ states system, irrespective of the value of the spin $j$. 
This leads us to ask the question: what are the most general random matrix ensembles which are invariant under the global unitary group U($N$) for the external system $\mathcal{E}$, and the SU(2) symmetry group for the spin $\mathcal{S}$.
We solve this problem and  define a class of Gaussian SU(2)$\times$U($N$) matrix ensemble, that we denote GU$_{2\times N}$E.
This class of random matrix ensembles is new (as far as we know) and has interesting properties. 
In particular they can be formulated and visualized in terms of random Wigner and Husimi distributions on the Riemann sphere.

 Thirdly, in the limit $N\to\infty$ (i.e. when the external system becomes large)  we are able to write closed equations, and to write in a fully closed form the evolution functional for the density matrix $\rho_\mathcal{S}(t)$ of the spin $\mathcal{S}$ (once the trace over the external system $\mathcal{E}$ has been taken). 
In other word we do not need to make any approximation and to write any master equation for the evolution of the density matrix.
Our calculations rely on the classical methods developed in the physics litterature in 
\cite{BrezinZee1994} \cite{BrezinHikamiZee1995}\cite{Zee1996}
to study sums and products of random matrices, and are closely related to the techniques of free probabilities in mathematics 
(see \cite{Voiculescu1992} and references therein). 
They are also in fact closely related to the methods used by \cite{MelloPereyraKumar1988} and more recently by 
\cite{LebowitzPastur2004} \cite{LebowitzLytovaPastur2007} for studying analytically the dissipation in two level systems. 

One important limitation of the model discussed here is that the internal dynamics of the quantum spin $\mathcal{S}$ is completely neglected.
This means that our results apply in the specific case where the internal dynamics of  $\mathcal{S}$ is much slower than the dynamics of the external system $\mathcal{E}$ and than the dynamics on the whole system $\mathcal{S}{+}\mathcal{E}$ induced by the coupling.
We are thus studying the case of ``interaction dominated decoherence'' already put forward by \cite{StrunzHaakeBraun2002}.
This case is relevant for the study of decoherence and of quantum diffusion effects, but not for dissipation.

\medskip
Rather than discussing the whole literature in this introduction (this would make it much too long), we prefer to present our model now.
Our results will be discussed and compared with previous ones at the end of the paper in section~\ref{ssCompLitt} (the interested reader can go directly to this section).
The paper is organized as follows.

In section \ref{sModel} we introduce our model, define the random matrix ensembles that we shall use, and discuss the connection of these ensembles with the theory of Wigner and Husimi representation for spin operators.

In section \ref{sEvolFun} we study the evolution functional for the spin density matrix, for a general choice of random coupling Hamiltonian, characterized by a set of variances $\Delta(l)$ for the couplings in the different angular momentum channels $l$.
We first write recursion relations for the evolution functional and the associated resolvent operators in the large $N$ limit in section~\ref{sPlanLim}.
We show that these relations takes a simple closed form in each  $l$ channel when expressed in angular momentum components (their Wigner transform double Fourier components).
The general solution is given in section~\ref{sSolEv}. It involves a universal decoherence function $M(t,Z(l))$ which depends on the time $t$ and of a parameter $Z(l)$ which is some linear transform of the variance distribution $\Delta(l')$, involving the SU(2) spin structure through the Racah 6j-coefficients.
The structure of this decoherence function as a function of the parameters $\Delta(l)$ and various scaling limits are discussed at length in section~\ref{sZofDel}, and illustrated in Appendix~\ref{app:fig}.
The relation between the parameters of the decoherence function and the norm of operators  built out from the Hamiltonians and the spin operators is discussed in section~\ref{sDecParNo}.

In section~\ref{sDecCoS} we study explicitely the decoherence of spin states in our model.
We concentrate on the case where the dynamics of the external system $\mathcal{E}$ is also given by a GUE Hamiltonian (for simplicity) and when the initial state for $\mathcal{E}$ is generic.
For large values of the spin $j$ this is of course naturally discussed in term of spin coherent states, whose properties are recalled in section~\ref{sCohSta}.
In section~\ref{ssTS} we compute the various time scales for the system (which corresponds to $\tau_{\mathrm{dyn.}}$ a dynamical timescale of the whole system,  $\tau_{\mathrm{dec.}}$ a decoherence time scale and $\tau_{\mathrm{diss.}}$ a dissipation time scale) as a function of the parameters 
 of the model.
We then compare the evolution of generic non coherent states (random states), coherent states and simple superpositions of coherent states (cat states).
Our solution shows explicitly under which conditions (choices of the $\Delta(l)$) the spin coherent states emerge as robust states under decoherence, and therefore are the pointer states which allow to describe the spin system and its dynamics in semiclassical terms.
Our explicit solutions allow very easily to illustrate the dynamics of decoherence through the dynamics of the Wigner representation of the spin quantum states. This is done in section~\ref{sIllEv} with several pictures and animations.
In section~\ref{ssdiffus} we study the large time evolution of the semiclassical coherent spin states. 
We show that through the coupling with $\mathcal{E}$, they undergo a slow quantum diffusion process on the classical phase space (the sphere $\mathcal{S}_2$).
We show that this diffusion process exhibits universal self similar properties, but is non Markovian at all time scales.
Hence in our case the evolution functional cannot be approximated by a master equation local in time.

In section~\ref{sExtDyn} we extend our calculations to the more general case where the dynamics of the external system $\mathcal{E}$ is given by an arbitrary Hamiltonian with a continuum energy spectrum, and when the initial state for $\mathcal{E}$ is not taken to be a random state, but is an energy eigenstate $|E\rangle$.
The coupling between the spin and $\mathcal{E}$ is still described by a GU$_{2\times N}$E Hamiltonian.
In section~\ref{sInfFEx} we show that we still can compute explicitly the influence functional, but that it depends on more parameters, in particular on the energy $E$ on the initial external state.
The discussion of decoherence in the simple case when the spectrum density $\rho(E)$ for $\mathcal{E}$ is  a semi circle (the explicit results are simpler in this case) is done in section~\ref{sAppWign}.
Finally in section~\ref{ssfast} we concentrate on the case where the internal dynamics of external system $\mathcal{E}$ is much faster than the dynamics induced by the coupling between the spin $\mathcal{S}$ and the $\mathcal{E}$.
In this case we show explicitly that the large time evolution of coherent states is still described by a quantum diffusion process, with depends on the initial state for $\mathcal{E}$ we start from.
The quantum diffusion process has the property that if we start from an energy eigenstate $|E\rangle$ for $\mathcal{E}$ (times a coherent state for $\mathcal{S}$ of course), the quantum diffusion process is now Markovian, and coincides with a diffusion process local in time and space (i.e. the Wiener process).
The effective diffusion constant in phase space depends on $E$, and takes a simple form related to the Fermi golden rule, indicating a fluctuation dissipation relation.
When we start from a general state $|\psi\rangle$  for $\mathcal{E}$, the diffusion process is not Markovian, but  is argued to be described by a randomized Markov process. 
We propose a simple explanation for this randomization, through the decoherence for the energy states of the environment induced by the now large spin.

In section~\ref{sConc} we first summarize the main results on our model, then put it in the context of the existing literature (section~\ref{ssCompLitt}), and finally discuss open problems and possible generalizations of our model (section~\ref{ssGeneral}).

\section{The model}
\label{sModel}
\subsection{A spin coupled to its environment}
\label{sSpin}
We start from the standard paradigmatic system: a small system $\mathcal{S}$  coupled to a large system $\mathcal{E}$ (the environment).
We take for $\mathcal{S}$ a single SU(2) spin with value $j$. This has several advantages, we can use the coherent states representation and the techniques of group representation theory, and we can easlily study the emergence of the classical degrees of freedom in the large spin limit $j\to\infty$. The coupling between $\mathcal{S}$ and $\mathcal{E}$ is formulated and studied with random matrix models. The advantages are that  we can use the mathematical tools of random matrix theory and that the interactions that we consider have are generic.

The Hilbert spaces for the subsystems $\mathcal{S}$ and $\mathcal{E}$, and the whole system  $\mathcal{S}+\mathcal{E}$ are of course
\begin{equation}
\label{HilbjN}
\mathcal{H}_{\mathcal{S}}=\mathbb{C}^{2j+1}
\quad,\qquad
\mathcal{H}_{\mathcal{E}}=\mathbb{C}^{N}
\quad,\qquad
\mathcal{H}\ =\ \mathcal{H}_{\mathcal{S}}\otimes\mathcal{H}_{\mathcal{E}}
\end{equation}
We take as an orthonormal state basis for $\mathcal{H}_{\mathcal{S}}$ the ${S}^3$ eigenstates $\{ |m\rangle,\ m=-j,\cdots,j\}$, the ${S}^\mu,\mu=1,2,3$ are the spin operators (the generators of su(2)).
 An orthonormal basis for $\mathcal{H}_{\mathcal{E}}$ is taken to be $\{ |\alpha\rangle,\ \alpha=1,N\}$.
The Hamiltonian of the whole system is written (altough this separation is partly arbitrary) as a sum of three terms
\begin{equation}
\label{HHSHE}
H=H_{\mathcal{S}}\otimes\mathbf{1}_{\mathcal{E}}+H_{\mathcal{SE}}+\mathbf{1}_{\mathcal{S}}\otimes H_{\mathcal{E}}
\end{equation}
$H_{\mathcal{S}}$ and $H_{\mathcal{E}}$ describe the internal dynamics of ${\mathcal{S}}$ and ${\mathcal{E}}$ respectively. $H_{\mathcal{SE}}$ describes the coupling between ${\mathcal{S}}$ and ${\mathcal{E}}$ and is usually taken to be small (fast internal dynamics).

We are mostly interested in the decoherence of spin and we therefore assume the the dynamics of the spin in slow compared to the other dynamics. 
This  important simplification will be discussed later.
Therefore we simply set 
\begin{equation}
\label{HSzero}
H_{\mathcal{S}}=0
\end{equation}
Moreover we first assume that the internal dynamics of ${\mathcal{E}}$ does not play a special role and is at most as fast as the dynamics of the coupling ${\mathcal{S}}+{\mathcal{E}}$.
We shall therefore recast $H_{\mathcal{S}}$ into $H_{\mathcal{SE}}$, thus setting simply 
\begin{equation}
\label{HEzero}
H_{\mathcal{E}}=0
\end{equation}
We shall go beyond this simplification in section~\ref{sExtDyn}.
In the rest of this section and in the following section~\ref{sDecCoS} we shall simply denote $H_{\mathcal{SE}}$ by $H_\mathrm{int}$ or simply by $H$.

\subsection{Gaussian  SU($2$)$\times$U($N$) random matrix ensembles} 
\label{sGU2NE}

We are interested in the generic situation where $H_\mathrm{int}$ is taken to be a typical (or generic) Hamiltonian allowing the emergence of the spin as a classical degree of freedom. 
Therefore we assume that $H_\mathrm{int}$ is a random Hamiltonian belonging to some $M\times M$ random matrix ensemble, with
 \begin{equation}
\label{M2jplusN}
M=(2j+1)N
\quad.
\end{equation} 
As explained in the introduction, in most models $H_\mathrm{int}$ is taken to be of the form 
\begin{equation}
\label{ }
H_{\mathrm{int}}=U_{\mathcal{S}}\otimes V_{\mathcal{E}}
\end{equation}
$U_{\mathcal{S}}$ being an appropriately chosen operator for $\mathcal{S}$ (the coupling agent), and $V_{\mathcal{E}}$ a random Hamiltonian for  $\mathcal{E}$ (sometimes it is a sum of a few terms like this). 
Here we look for a general random Hamiltonian, simply assuming (mostly for simplicity, but there are some physical motivations too in some specific models)) that it belongs to some Gaussian ensemble.
The question is thus:  can we caracterise the most general Gaussian ensembles (random matrix probability distributions) which are invariant under: (1) the SU(2) group acting on $\mathcal{H}_{\mathcal{S}}$, (2) generic unitary transformations U($N$) acting on $\mathcal{H}_{\mathcal{E}}$?
This does not mean that $H$ is invariant, simply that the distribution of the $H$'s is invariant.

\subsubsection{Spin decomposition of operators}
\label{ssSpinDec}
To answer this question, we simply use standard tools of group representation theory. 
We refer to \cite{VarillyGracia-Bondia89} for a clear reference to the theory of SU(2) representations, coherent states and the theory of Wigner functions, that we shall use heavily
.
Let us first concentrate on the spin part, i.e. on the subspace $\HS$. Any operator $A$ acting on $\HS$, i.e.  $A\in\mathcal{B}(\mathcal{H}_{\mathcal{S}})=M_{2j+1}(\mathbb{C})$ can be considered as an element of the tensor product (or Kronecker product) of two spin $j$ representations of SU(2). Of course such a product is a sum of the irreducible representations
\begin{equation}
\label{jtimesj}
\mathbf{j}\otimes\mathbf{j}=\mathbf{0}\oplus\mathbf{1}\oplus\cdots\oplus\mathbf{2j}=\bigoplus_{l=0}^{2j}\mathbf{l}
\end{equation}
Hence  the operator $A$ can be decomposed into its spin $l$ components $A^{(l)}$
 \begin{equation}
\label{AdecAl}
A=\sum_{l=0}^{2j}A^{(l)}
\end{equation}
The matrix elements of the $A^{(l)}$ are 
\begin{equation}
\label{WInv}
A_{rs}^{(l)}=\langle r|A^{(l)}|s\rangle=\sum_{m=-j}^j(-1)^m \sqrt{\frac{2l+1}{2j+1}}\clebsch{j}{s}{l}{-m}{j}{r} W^{(l,m)}_A
\end{equation}
where the coefficients $W^{(l,m)}_A$ are given in terms of the matrix elements $A_{rs}=\langle r|A|s\rangle$ of $A$ by the inverse transform
\begin{equation}
\label{WTransf}
W^{(l,m)}_A=\sum_{r,s=-j}^j\sqrt{\frac{2l+1}{2j+1}}\clebsch{j}{r}{l}{m}{j}{s} A_{rs}
\end{equation}
The $\clebsch{j}{m_1}{l}{m_2}{j}{m_3}$ are the SU(2) Clebsch-Gordan coefficients.
Of course only the single terms such that $m=s-r$ contribute in the sums.

The spin $l$ component of $A$ satisfies 
\begin{equation}
\label{SSAcom}
[\vec S,[\vec S,A^{(l)}]]=\sum_{\mu=1}^3[S^\mu,[S^\mu,A^{(l)}]]=l(l+1) A^{(l)}
\end{equation}
The coefficients $W_A^{(lm)}$ are complex but satisfy the conjugation constraint
\begin{equation}
\label{WAdagger}
W_{A^\dagger}^{(l,m)}=(-1)^m \overline W_A^{(l,-m)}
\end{equation}

\subsubsection{Relation with the Wigner and the Husimi distributions}
\label{sWiHu}

\paragraph{Wigner representation:} 
The coefficients $W^{(l,m)}_A$ are nothing but the $(l,m)$ coefficients of the Wigner distribution $W_A(\vec n)$ associated to the operator $A$ in the basis of spherical harmonics $Y^m_l(\vec n)$ on the unit sphere $\mathcal{S}_2$.
The Wigner distribution is
\begin{equation}
\label{WWY}
W_A(\vec n)=\sum_{l=0}^{2j}\sum_{m=-j}^j W^{(l,m)}_A Y^m_l(\vec n)
\end{equation}
With the normalisation for the $Y^m_l$ 
\begin{equation}
\label{YYnorm}
\int_{\mathcal{S}_2} d^2\vec n\ Y^m_l(\vec n) \overline{Y_{l'}^{m'}}(\vec n)=\delta_{l,l'}\delta_{m,m'}
\quad,\qquad d^2\vec n = d\theta\,d\phi\,\sin(\theta)
\end{equation}
With these normalisations we have
\begin{equation}
\label{trABnorm}
\tr(AB^\dagger)=\int_{\mathcal{S}_2} d^2\vec n\ W_A(\vec n)\,\overline{W_B(\vec n)}=\sum_{l,m} W_A^{(l,m)}\overline{W_B^{(l,m)}}
\end{equation}
and for the unit operator and the trace
\begin{equation}
\label{WoneTr}
W_\mathbf{1}^{(l,m)}=\sqrt{2j+1}\ \delta_{l,0}\delta_{m,0}\quad,\qquad
\tr(A)=\sqrt{2j+1} \ W_A^{(0,0)}
\end{equation}
NB: In the litterature the normalization for the $W_A$ is often such that 
\begin{equation}
\label{trABbar}
\tr(AB^\dagger)=\frac{2j+1}{4\pi}\int_{\mathcal{S}_2} d^2\vec n\ \mathbf{W}_A(\vec n)\,\overline{\mathbf{W}_B(\vec n)}
\end{equation}
It corresponds to the change of normalisation in the definition of the Wigner distribution
\begin{equation}
\label{Wbar}
\mathbf{W}_A(\vec n)=\sqrt{\frac{4 \pi}{2j+1}}W_A(\vec n)
\end{equation}

\paragraph{Husimi representation:}
Note that we can also construct the Q-symbol, or Husimi representation 
\begin{equation}
\label{Qsymbol}
Q_A(\vec n)=\sum_{l,m}\ \clebsch{j}{j}{l}{0}{j}{j}W^{(l,m)}_A\  Y^m_l(\vec n)
\end{equation}
and the P-symbol
\begin{equation}
\label{Psymbol}
P_A(\vec n)=\sum_{l,m}\ {\clebsch{j}{j}{l}{0}{j}{j}}^{-1}W^{(l,m)}_A\  Y^m_l(\vec n)
\end{equation}
They are such that
\begin{equation}
\label{trABQP}
\tr(AB^\dagger)=\int_{\mathcal{S}_2} d^2\vec n\ Q_A(\vec n)\,\overline{P_B(\vec n)}
\end{equation}
The Husimi function $Q_A(\vec n)$ corresponds to the ``physical'' probability distribution of spin in phase space, since with the proper normalisation
\begin{equation}
\label{Qbar}
\langle\vec n|A|\vec n\rangle=\mathbf{Q}_A(\vec n)\quad,\qquad  \mathbf{Q}_A(\vec n)=\sqrt{\frac{4 \pi}{2j+1}}Q_A(\vec n)
\quad,\qquad
|\vec n\rangle \quad\text{coherent state}
\end{equation}

\subsubsection{Gaussian  SU$_2$ random matrix ensembles:}
\label{ssGSU2}
In order to construct the most general  Gaussian SU(2) invariant ensemble on the self adjoint matrices in $M_{2j+1}(\mathbb{C})$, we simply have to take separately each  matrix component $A^{(l)}$ as Gaussian independent random (non-commutative) variables with zero mean and variance depending on  $l$. 
Namely, we take the  $W^{(l,m)}$ to be independent Gaussian variables, subjected only to the Hermiticity constraint 
\begin{equation}
\label{WWbar}
W^{(l,m)}=(-1)^m \overline W^{(l,-m)}
\end{equation}
which ensures that the $A$ are Hermitian operators, and to the fact that the variance depends on $l$ but not on $m$, which ensures SU(2) invariance of the distribution.
More precisely, we take the $W$'s to be  for $m=0$
\begin{equation}
\label{WAl}
W^{(l,0)}=A^{(l)}
\end{equation}
and for $m>0$
\begin{equation}
\label{WBClm}
W^{(l,m)}=B^{(l,m)}+iC^{(l,m)}\ ,\quad W^{(l,-m)}=(-1)^m \left(B^{(l,m)}-iC^{(l,m)}\right)\ \text{for }\ 0<m\le l
\end{equation}
and to take for the $A^{(l)}$, $B^{(l,m)}$ and $C^{(l,m)}$ random Gaussian independent variables with zero mean and mean square extend $\Delta(l)$ depending only on $l$. The cumulants are
\begin{equation}
\label{EA1}
\mathbb{E}[A^{(l)}]=\mathbb{E}[B^{(l,m)}]=\mathbb{E}[C^{(l,m)}]=0
\end{equation}
\begin{equation}
\label{EA2}
\mathbb{E}[{A^{(l)}}{A^{(l)}}]=\mathbb{E}[B^{(l,m)}B^{(l,m)}]=\mathbb{E}[C^{(l,m)}C^{(l,m)}]=\Delta(l)
\end{equation}
All the others cumulants being zero. The $\Delta(l)$'s are a collection $\bold\Delta$ of $2j+1$ positive numbers 
\begin{equation}
\label{DelColl}
\Delta(l)\ge 0\ ,\quad l=0,\cdots 2j
\end{equation}
which completely characterize the SU(2) gaussian ensemble.
This distribution is given by the Gaussian probability measure on self adjoint $(2j+1)\times(2j+1)$ matrices
\begin{equation}
\label{MesAl}
\mathcal{D}_{\bold\Delta}[A]\propto dA\ \exp\left[-\sum_{l=0}^{2j}\frac{1}{2\Delta(l)} \tr\left[ {A^{(l)}}^2\right]\right]
\end{equation}
where $dA$ is the standard flat measure, so that $\mathbb{E}[F[A]]=\int \mathcal{D}_{\bold\Delta}[A]\,F[A]$.

In this GU$_2$E ensemble, characterised by  $\bold\Delta$, the ``propagator" $ \mathcal{D}_{rs,tu}$ is
\begin{align}
\label{DrstuDef}
 \mathcal{D}_{rs,tu}=\mathbb{E}\left[ A_{rs}\,A_{tu}\right]&=\delta_{s-r,t-u}\sum_{l=0}^{2j} \Delta(l)\frac{2l+1}{2j+1}\clebsch{j}{s}{l}{r-s}{j}{r}\clebsch{j}{t}{l}{u-t}{j}{u}
\end{align}
Of course, if all the $\Delta(l)$ are equals to the same $\Delta$, one recovers the standard GUE ensemble for $(2j+1)\times(2j+1)$ matrices, with
\begin{equation}
\label{EA2GUE}
\mathbb{E}\left[ A_{rs}\,A_{tu}\right]_{\mathrm{GUE}}=\Delta\,\delta_{r,u}\ \delta_{s,t}
\end{equation}

\subsubsection{Gaussian SU$_2\times$U$_N$ random matrix ensembles:}
\label{ssSU2N}
We now take into account the external system $\mathcal{E}$ and want to characterize the most general Gaussian ensemble of self-adjoint matrices $H\in\mathcal{B}(\mathcal{H})=M_{(2j+1)N}(\mathbb{C})$  which is invariant under SU(2) (acting on $\HS$ as the spin $j$ representation) and under U($N$) (acting on $\HE$ as the fundamental representation).  

The solution is simple. We make the same decomposition w.r.t. the spin sector $\HS$ , keeping the sector $\HE$ untouched.
$H$ is decomposed into 
\begin{equation}
\label{HdecHl}
H=\sum_l H^{(l)}
\end{equation}
The matrix elements of $H$ are now denoted as
\begin{equation}
\label{HmatEl}
H_{rs}^{\alpha\beta}=\langle r\alpha|H|s\beta\rangle\quad,\qquad |r\alpha\rangle=|r\rangle\otimes|\alpha\rangle
\end{equation}
and can be written as
\begin{equation}
\label{HDeclm}
H_{rs}^{\alpha\beta}=\sum_{l=0}^{2j}\sum_{m=-j}^j(-1)^m \sqrt{\frac{2l+1}{2j+1}}\clebsch{j}{s}{l}{-m}{j}{r} W^{(l,m)}_{\alpha\beta}
\end{equation}
The hermiticity constraint $H=H^\dagger$ for $H$ reads now
\begin{equation}
\label{HermWCo}
W^{(l,m)}_{\alpha\beta}=(-1)^m\ \overline W^{(l,-m)}_{\beta\alpha}
\end{equation}
The GU$_{2,N}E$ ensemble is obtained by taking the $W^{(l,m}_{\alpha,\beta}$ to be Gaussian independently distributed random variables, subjected to the constraint of \eq{HermWCo} , with zero mean and a variance depending only on $l$, but not on $m$, $\alpha$ and $\beta$.
This ensures the invariance under the group SU($2$)$\times$ U($N$) of the probability distribution on $H$.
The ``propagator'' $\boldsymbol{\mathcal{D}}$ is now the product of the propagator $\mathcal{D}$ for the GU$_{2}$E ensemble (given by \eq{DrstuDef}) times  the propagator for the standard GUE model.
\begin{equation}
\label{PropD}
\boldsymbol{\mathcal{D}}_{rs,tu}^{\alpha\beta,\gamma\delta}=
\mathbb{E}\left[ A_{rs}^{\alpha\beta}\,A_{tu}^{\gamma\delta}\right]=\mathcal{D}_{rs,tu}\ \delta_{\alpha,\delta}\,\delta_{\beta,\gamma}
\end{equation}
This ensemble is characterized by the same ensemble $\bold\Delta=\{\Delta(l),\ l=0,\cdots 2j\}$ of positive parameters.

Let us note that the sector $l=0$ gives a Hamiltonian $H^{(0)}$ independent of the spin, since its matrix elements are of the form 
\begin{equation}
\label{HOdiag}
{H^{(0)}}^{\alpha\beta}_{rs}=\delta_{r,s} H_{(0)}^{\alpha\beta}
\end{equation}
Hence $H^{(0)}$ can be written as
\begin{equation}
\label{HOexp}
H^{(0)}=\mathbf{1}_{\mathcal{S}}\otimes H_{(0)}
\end{equation}
where $H_{(0)}$ is a random Hamiltonian for $\mathcal{E}$, whose distribution is given by the GUE ensemble with (variance)$^2=\Delta(0)$.

Finally we note that the Hamiltonian can of course be rewritten in the general form
\begin{equation}
\label{DsumDW}
H=\sum_{(l,m)} D^{(l,m)}_{\mathcal{S}}\otimes W^{(l,m)}_{\mathcal{E}}
\end{equation}
where the $D^{(l,m)}_{\mathcal{S}}$ are some fixed spin operators (related to the Wigner D-matrices) and the $W^{(l,m)}_{\mathcal{E}}$ some random operators on $\mathcal{E}$.
Previous studies of spin decoherence have dealt with simpler interaction Hamiltonians $\HaSE$ with only  one $D\otimes W$ term or a few   (typically 3 when dealing with a $l=1$ interaction).
Here we keep all the possibles terms in the decomposition \ref{DsumDW}.
But we shall stay with the explicit form \ref{HDeclm} for the decomposition of $H$.

\subsection{Rescaled distributions with $N$ and $j$}
\label{srescDNj}
We thus take for our model a single spin $j$, with for simplicity no internal dynamics $\HaS=0$ and for Hamiltonian $\HaSE$ describing the coupling between the spin and its environment a random Hamiltonian in the SU(2)$\times$U($N$) random Gaussian ensemble characterized 
by the family of coupling amplitudes
$\bold\Delta=\{\Delta(l),\ l=0,\cdots 2j\}$ in the different spin channels $l$. 
In this paper we are interested in the limit $N\to\infty$ (large environment) and $j\to\infty$ (classical spin). 
In these limits it is adequate to rescale these amplitudes (i.e. the time scale for the evolution of the system).
For clarity we define here these rescaled parameters as
\begin{equation}
\label{rescaledD}
\tilde{\Delta}(l)=N \Delta(l)
\quad,\qquad
\bar\Delta(l)={N\over 2j+1}\Delta(l)
\end{equation}
As we shall see later, the large $N$ limit is obtained when keeping the $\tilde{\Delta}(l)$ of order $\mathcal{O}(1)$, while the large $N$ and large $j$ limit is obtained  when keeping the $\bar{\Delta}(l)$ of order $\mathcal{O}(1)$

\section{The evolution functional}
\label{sEvolFun}
\subsection{General framework}
\label{sGenFram}
We start at time $t=0$ from a separable quantum state
\begin{equation}
\label{rhoIn}
\rho(0)=\rho^{\mathcal{S}}(0)\otimes\rho^{\mathcal{E}}(0)
\end{equation}
We do not specify at that stage if the system $\mathcal{S}$ is in a pure or in a mixed quantum state. 
Since the ensemble for the Hamiltonian $H$ is chosen to be invariant under arbitrary unitary transformations $U\in\mathrm{U}(N)$ acting on $\HS$, and since we shall compute only averages w.r.t. the distributions of $H$, the precise initial quantum state $\rho^{\mathcal{E}}(0)$ is not important for what we are interested in.
We may choose it to be the maximal entropy state
\begin{equation}
\label{rhoInOne}
\rho^{\mathcal{E}}(0)={1\over N}\mathbf{1}_\mathcal{E}
\end{equation}
After evolution from time $0$ to $t>0$ of the whole system under the time independent Hamiltonian $H$, the reduced density matrix for the system $\mathcal{S}$ is
\begin{equation}
\label{rho2t}
\rho^{\mathcal{S}}(t)=\frac{1}{N}\tr_{\mathcal{E}}\left( e^{-\imath t H}(\rho^\mathcal{S}(0)\otimes \mathbf{1}_\mathcal{E}) e^{\imath t H}\right)
\end{equation}
We are interested in the evolution of the model for a ``typical'' Hamiltonian $H$ in our GU$_{2\times N}E$ ensemble, and we shall therefore make averages over $H$ in this ensemble. The average reduced density matrix
\begin{equation}
\label{rhoRedDef}
\overline{\rho^\mathcal{S}}(t)=\mathbb{E}\left[\rho^{\mathcal{S}}(t)\right]
\end{equation}
is sufficient to study the observables of the subsystem.
However in order to study decoherence and the evolution of the intrication between the systems $\mathcal{S}$ and $\mathcal{E}$, we must study functions of the reduced density matrix such as the von Neumann entropy $S$ or the purity $P$ (or whichever quantities you prefer)
\begin{equation}
\label{EntPur}
S=-\tr_\mathcal{S}\left[\rho^\mathcal{S}\,\log \rho^\mathcal{S}\right]
\quad,\qquad
P=\tr_\mathcal{S}\left[\rho^\mathcal{S}\rho^\mathcal{S}\right]
\end{equation}

It is well known that simplifications occur for such observables in the limit $N\to\infty$ where the number of degrees of freedom of the external system $\mathcal{E}$ becomes infinite, with a adequate rescaling of time and of the couplings. Indeed, as we shall discuss in the next section, for ``reasonable'' functions $F$ of the density matrix $\rho^{\mathcal{S}}$ (and at least for polynomial functions), in the large $N$ limit we have the factorisation of the expectation of  products of traces with respect to $\mathcal{E}$ due to the fact that in this limit, only planar diagrams contribute in the perturbative expansions.
This means that we have in fact \footnote{In the rest of this paper we shall rather use the physics notation $\overline{F}$ for the average of $F$ rather that the probabilistic notation $\mathbb{E}[F]$ for the expectation of $F$, but this is of course the same quantity.}
\begin{equation}
\label{FrhoAv}
\overline{F\left(\rho^{\mathcal{S}}\right)}=\mathbb{E}\left[F\left(\rho^{\mathcal{S}}\right)\right]
\ = \  F\left(\overline{\rho^\mathcal{S}} \right) \ \left(1+\mathcal{O}(N^{-2})\right)
\end{equation}
This phenomenon is known in physics as ``factorisation'' or emergence of a large $N$ master field. In mathematics it is known as the phenomenon of ``concentration of measures''.  It leads to the formulation of the $N\to\infty$ limit in terms of free probabilities.

It turns out that these kind of Gaussian ensembles are self-averaging (this is discussed at length in the mathematical literature, see for instance \cite{LebowitzLytovaPastur2007}), but we shall not elaborate further this point.

Thus we have to consider the evolution operator $\mathcal{M}(t)$ for the density matrix
\begin{equation}
\label{rhoev}
\overline{\rho^{\mathcal{S}}}_{ru}(t)=\sum_{st}\mathcal{M}_{{ru},{st}}(t)\,\rho^{\mathcal{S}}_{st}(0)
\end{equation}
$\mathcal{M}(t)$ corresponds to the Feynman-Vernon influence functional, it is of course a strictly positive linear trace preserving application on $M_{2j+1}(\mathbb{C})$, thus it is also called a quantum channel, or a POVM.
Its matrix elements are given by
\begin{equation}
\label{MrstuDef}
\mathcal{M}_{{ru},{st}}(t)=\frac{1}{N}\overline{\sum_{\alpha,\beta}\langle r\alpha|\emath^{-\imath t H} | s\beta\rangle\langle t\beta|  \emath^{\imath t H} |u\alpha\rangle }
\end{equation}

\subsection{Perturbation theory and the planar limit}
\label{sPlanLim}
\subsubsection{Resolvents}
\label{ssResolvt}
It is more convenient to consider the double resolvent 
\begin{equation}
\label{GresDef}
\mathcal{G}_{{ru},{st}}(x,y)=
\frac{1}{N}\overline{\sum_{\alpha,\beta}\langle r\alpha| {(x-H)}^{-1}| s\beta\rangle\langle t\beta| {(y-H)}^{-1} |t\alpha\rangle }
\end{equation}
Formally (integration paths to be discussed later)
\begin{equation}
\label{G2Mint}
\mathcal{M}_{{ru},{st}}(t)=\oint\frac{dx}{2\imath\pi}\oint\frac{dy}{2\imath\pi}\,\emath^{-\imath t (x-y)}\,\mathcal{G}_{{ru},{st}}(x,y)
\end{equation}
We consider first the single resolvent
\begin{equation}
\label{HresDef}
\mathcal{H}_{{rs}}(x)=
\frac{1}{N}\overline{\sum_{\alpha}\langle r\alpha| {(x-H)}^{-1}| s\alpha\rangle }
\end{equation}
To compute these generating functions, we use the standard diagrammatic techniques (see for instance
\cite{MelloPereyraKumar1988}\cite{Zee1996} and references therein).
We  represent the propagator $\boldsymbol{\mathcal{D}}_{rs,tu}^{\alpha\beta,\gamma\delta}$ as a double fat line.
\begin{figure}[h]
\begin{center}
$\boldsymbol{\mathcal{D}}_{rs,tu}^{\alpha\beta,\gamma\delta}$\quad =\quad \raisebox{-4ex}{\includegraphics[width=2in]{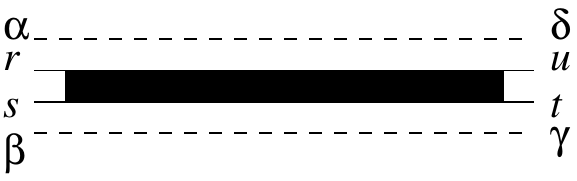}}
\caption{Diagramatic representation of the propagator $\boldsymbol{\mathcal{D}}_{rs,tu}^{\alpha\beta,\gamma\delta}$}
\label{f:prop}
\end{center}
\end{figure}
The dashed lines represent the two external tensors $\delta_{\alpha,\delta}$ and $\delta_{\beta,\gamma}$, and indicates that the $\mathcal{E}$ indices (greek letters) are conserved.
The black ribbon represents the  spin tensor ${\mathcal{D}}_{rs,tu}$. It indicates that the spin indices (roman letters) are mixed. 
But the difference between the left and right indices is conserved.
\begin{equation}
\label{rstucon}
s-t=r-u
\end{equation}

\subsubsection{Recursion equation for the single resolvent}
\label{ssRecSingl}

\begin{figure}[h]
\begin{center}
\includegraphics[scale=.2]{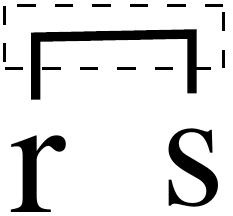}\quad
\includegraphics[scale=.2]{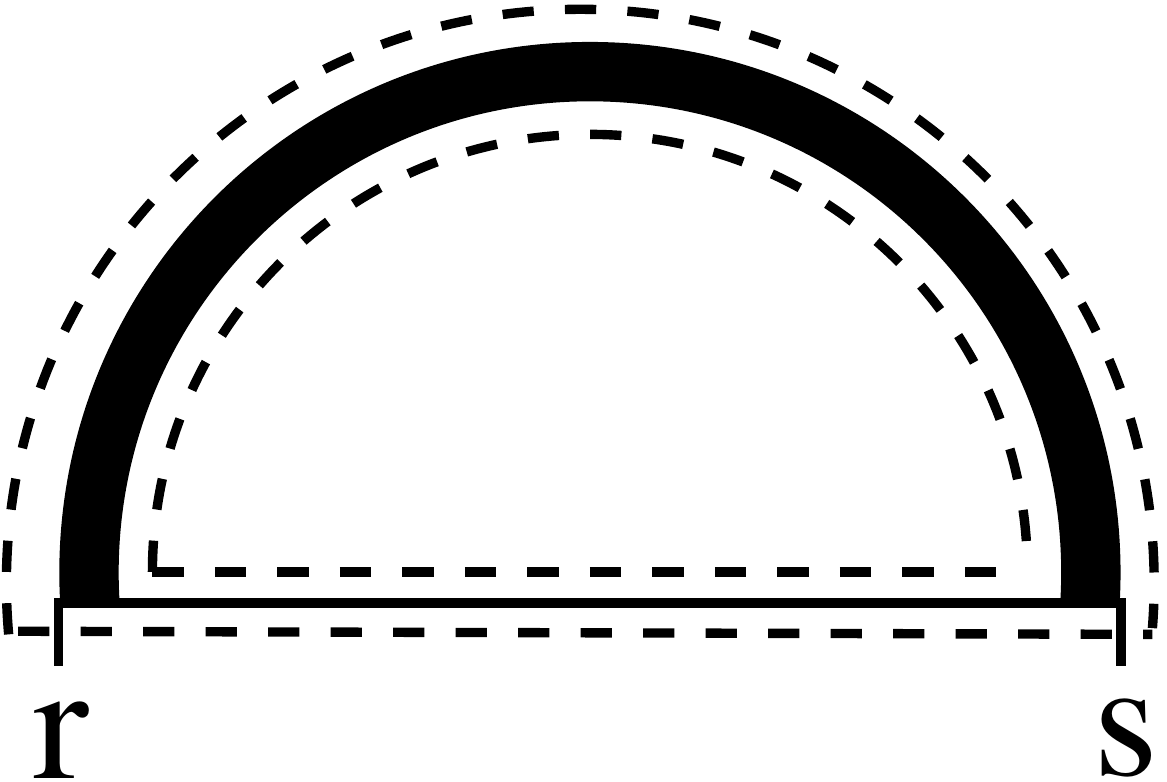}\quad
\includegraphics[scale=.2]{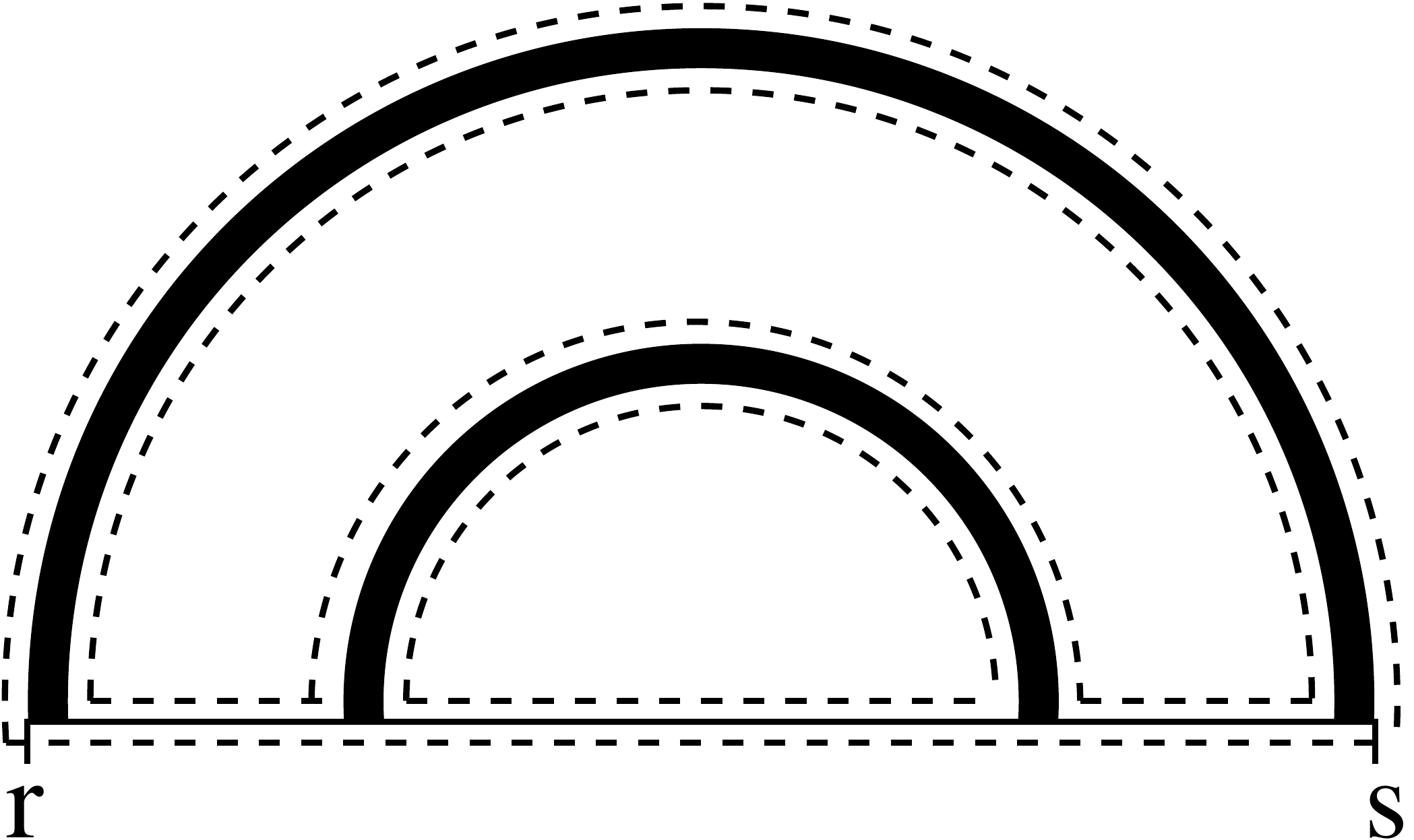}\quad
\includegraphics[scale=.2]{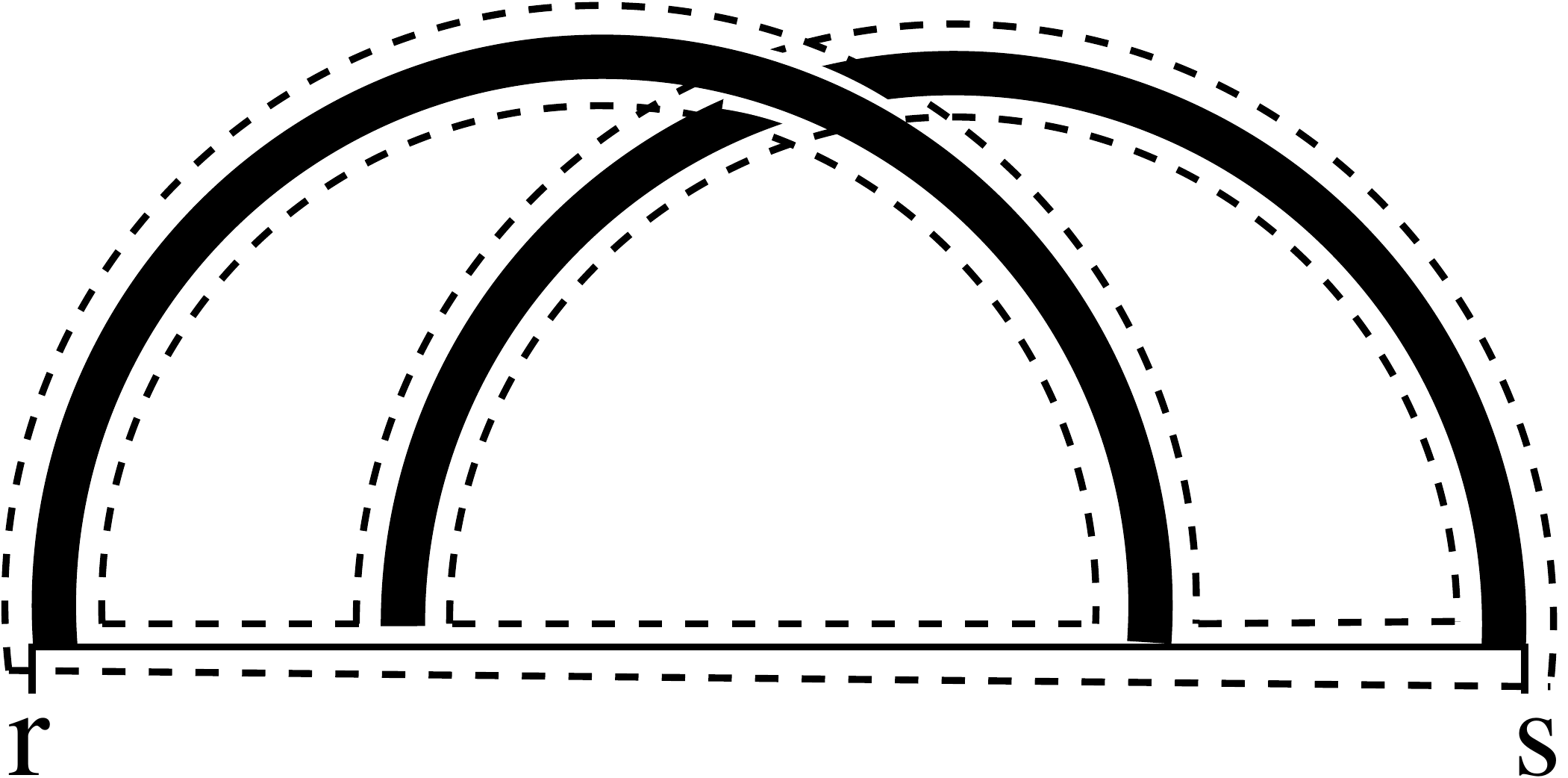}\quad
\caption{The first diagrams for $\mathcal{H}_{{rs}}(x)$}
\label{figHdiag}
\end{center}
\end{figure}

\begin{figure}[h]
\begin{center}
\includegraphics[scale=.2]{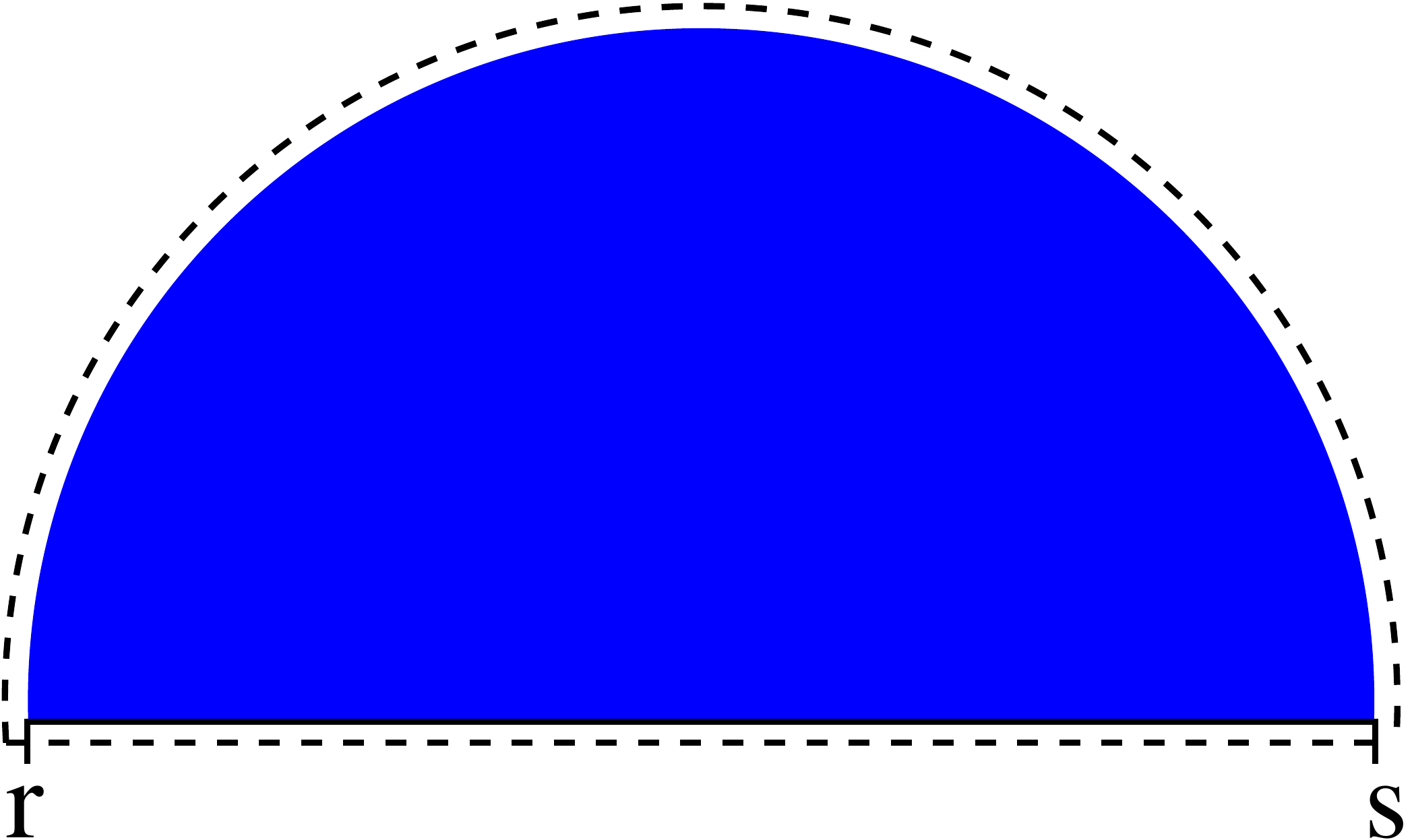}\quad \raisebox{3.ex}{=}\quad
\includegraphics[scale=.2]{arches5.pdf}\quad \raisebox{3.ex}{+}\quad
\includegraphics[scale=.2]{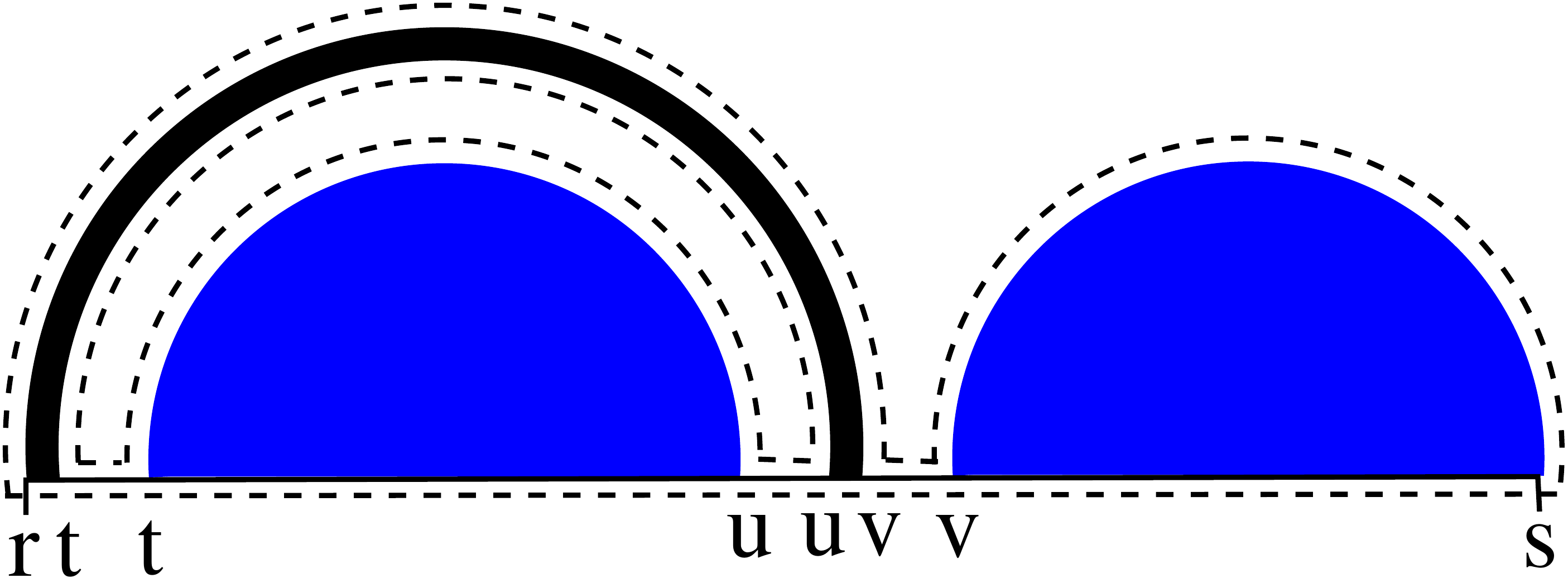}
\caption{Graphical formulation of the recursion relation \eq{RecRelH} for $\mathcal{H}_{{rs}}(x)$}
\label{figHrec}
\end{center}
\end{figure}

To compute $\mathcal{H}_{{rs}}(x)$ we expand in a power series in $x^{-1}$
\begin{equation}
\label{HrsExp}
\mathcal{H}_{{rs}}(x)=\sum_{k=0}^\infty x^{-1-k}\ 
\frac{1}{N}\overline{\sum_{\alpha}\langle r\alpha| {H}^{k}| s\alpha\rangle }
\end{equation}
and use Wick theorem to compute the average $\overline{H^k}$.
We get a sum of contributions associated to diagrams of the form depicted in Fig.\ref{figHdiag}. The propagators form arches above a line going from $r$ to $s$.
Each arch gives a term $\mathcal{D}$ proportional to the $\Delta(l)$, and  each closed dashed loop gives a factor $N$ (sum over the $\mathcal{E}$ indices $\alpha$).
So each diagram is of order
\begin{equation}
\label{EulerRe}
x^{-1-2\#\text{arches}}\, [\Delta]^{\#\text{arches}}\, N^{\# loops-1}=x^{-1}\left(x^{-2} [\Delta]\,N \right)^{\#\text{arches}}\, N^{-\chi}
\end{equation}
$\chi$ being the Euler characteristic of the fat diagram. $[\Delta]$ means any $\Delta(l)$.
Thus in the large $N$ limit, only planar diagrams (rainbow like) survive, provided we rescale the variances by $N$ 
\begin{equation}
\label{tldDel}
%\label{DelTild}
\qquad \tilde\Delta(l)=N \Delta(l)= \mathcal{O}(1)\qquad N\to\infty
\end{equation}
$\mathcal{H}_{{rs}}(x)$ then satisfies the recursion equation
\begin{equation}
\label{RecRelH}
\mathcal{H}_{{rs}}(x)=x^{-1}\delta_{r,s}+ x^{-1} \sum_{t,u,v}\widetilde{\mathcal{D}}_{rt,uv}\mathcal{H}_{{tu}}(x)\mathcal{H}_{{vs}}(x)
\end{equation}
with
\begin{equation}
\label{DrstuDefTilde}
\widetilde{\mathcal{D}}_{rt,uv}=N\,{\mathcal{D}}_{rt,uv}=\delta_{s-r,t-u}\sum_{l=0}^{2j} \widetilde{\Delta}(l)\frac{2l+1}{2j+1}\clebsch{j}{s}{l}{r-s}{j}{r}\clebsch{j}{j}{l}{u-t}{j}{u}
\end{equation}
This recursion equation is depicted graphically in Fig.~\ref{figHrec}.
The solution is of the form, in fact required by SU(2) invariance
\begin{equation}
\label{Hrs}
\mathcal{H}_{{rs}}(x)=\delta_{r,s}\,\mathcal{H}(x)
\end{equation}
Inserting this ansatz \ref{Hrs} into  \ref{RecRelH} and  \ref{DrstuDefTilde} we obtain  the simple recursion equation for $\mathcal{H}(x)$
\begin{equation}
\label{Heq}
\mathcal{H}(x)=x^{-1} + x^{-1} \mathcal{H}(x)^2 \hat\Delta
\quad\text{with}\quad 
\hat\Delta=\sum_{l=0}^{2j} \frac{2l+1}{2j+1}\tilde\Delta(l)
\end{equation}
Hence
\begin{equation}
\label{Hsol}
\mathcal{H}(x)=\frac{1}{2\hat\Delta}\left(x-\sqrt{x^2-4 \hat\Delta}\right)=x^{-1}\mathrm{Cat}(\hat\Delta x^{-2})
\end{equation}
where $\mathrm{Cat}(z)=\sum z^n C_n$ is the generating function of the Catalan numbers $C_n$.

\subsubsection{Recursion equation for the double resolvent}
\label{ssRecDoubl}
\begin{figure}[h]
\begin{center}
\includegraphics[scale=1]{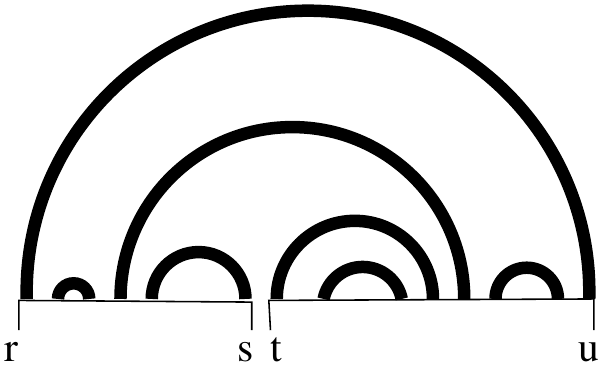}
\caption{The planar diagrams of $\mathcal{G}_{{ru,st}}(x,y)$}
\label{figGdiag}
\end{center}
\end{figure}

Now we can compute the function $\mathcal{G}_{{ru},{st}}(x,y)$.
It is given by the sum of the planar diagrams of the form given in Fig.~\ref{figGdiag}.
It thus obeys the recursion equation
\begin{equation}
\label{RecGenG}
 \mathcal{G}_{{ru},{st}}(x,y)=x^{-1}\delta_{r,s}\mathcal{H}_{tu}(y)
 +x^{-1}\widetilde{\mathcal{D}}_{rv,wx} \mathcal{H}_{vw}(x)\mathcal{G}_{xu,st} (x,y)
 +y^{-1} \widetilde{\mathcal{D}}_{rv,wx} \mathcal{G}_{vw,st}(x,y)\mathcal{H}_{xu} (y)
\end{equation}
\begin{figure}[h]
\begin{center}
\raisebox{-3.ex}{\includegraphics[scale=1]{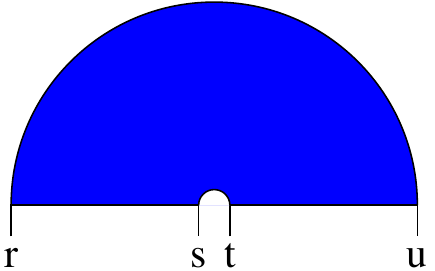}}
\ =\ 
\raisebox{-3.ex}{\includegraphics[scale=1]{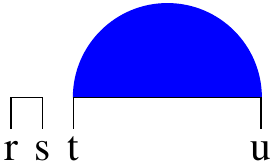}}
\ +\ 
\raisebox{-3.ex}{\includegraphics[scale=1]{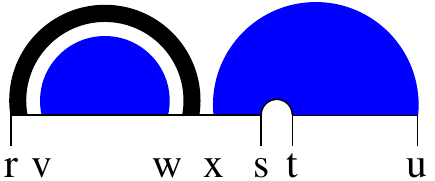}}\\ \vskip 2.em 
\ +\ 
\raisebox{-3.ex}{\includegraphics[scale=1]{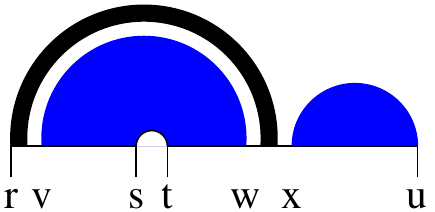}}
\caption{Graphical formulation of the recursion relation \eq{RecGenG} for $\mathcal{G}_{{ru,st}}(x,y)$}
\label{figGRec}
\end{center}
\end{figure}
To solve this equation, it is better to use its SU(2) invariance properties, and to rewrite it for its ``double Wigner transform'' coefficients
\begin{equation}
\label{Mlmlmex}
W_{\mathcal{G}}^{(l_1,m_1),(l_2,m_2)}(x,y)=
\sum_{r,u=-j} ^j \sum_{s,t=-j}^j
\sqrt{\frac{2l_1+1}{2j+1}}\clebsch{j}{u}{l_1}{m_1}{j}{r} 
\sqrt{\frac{2l_2+1}{2j+1}}\clebsch{j}{s}{l_2}{m_2}{j}{t} 
\mathcal{G}_{{ru},{st}}(x,y)
\end{equation} 
Indeed, 
we reexpress the initial propagator, $\widetilde{\mathcal{D}}_{rs,tu}$ (given by \ref{DrstuDefTilde})
in the $(s,t)\to(u,r)$ channel as
\begin{equation}
\label{D2Dhat}
%\widehat{\mathcal{D}}_{ru,st}=
\mathcal{D}_{rs,tu}
=
\widehat{\mathcal{D}}_{ru,st}
\qquad\text{i.e.}\qquad
\raisebox{-6ex}{\includegraphics[width=2.5in]{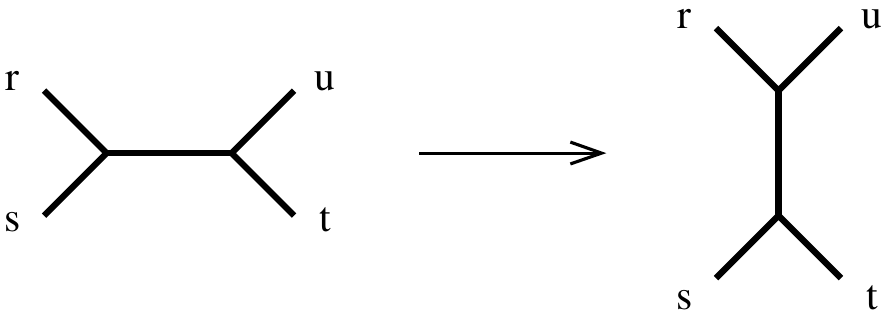}}
\end{equation}
The corresponding double Wigner transform of $\widehat{\mathcal{D}}_{ru,st}$ is
\begin{align}
\label{WDhatI}
W_{\widehat{\mathcal{D}}}^{(l_1,m_1),(l_2,m_2)}&=
\sum_{r,u=-j} ^j \sum_{s,t=-j}^j
\sqrt{\frac{2l_1+1}{2j+1}}\clebsch{j}{u}{l_1}{m_1}{j}{r} 
\sqrt{\frac{2l_2+1}{2j+1}}\clebsch{j}{s}{l_2}{m_2}{j}{t} 
\widehat{\mathcal{D}}_{{ru},{st}}
\end{align}
Using  the original expression \ref{DrstuDef} for $\mathcal{D}_{rs,tu}$, this sum is rewritten as a multiple sum over products of four  Clebsch-Gordan coefficients.
After some SU(2) algebra, it can be reduced to the simple form
\begin{align}
\label{WDhatExpl}
W_{\widehat{\mathcal{D}}}^{(l_1,m_1),(l_2,m_2)}&=
\delta_{l_1,l_2}\,\delta_{m_1+m_2,0}\, (-1)^{m_1} \widehat{\Delta}(l_1)
\end{align}
with $\widehat{\Delta}(l_1)$ given by
\begin{equation}
\label{HatDelOfL}
\widehat{\Delta}(l_1)=\sum_{l'=0}^{2j}\tilde\Delta(l') (2l'+1) (-1)^{2j+l'+l_1}
\sixj{j}{j}{l'}{j}{j}{l_1}
\end{equation}
where $\sixj{j_1}{j_2}{j_3}{j_4}{j_5}{j_6}$ is the Racah 6-j symbol.
In particular, $\widehat{\Delta}(0)$ is nothing but the $\hat\Delta$ of \eq{Heq}
\begin{equation}
\label{HatDelOf0}
\widehat{\Delta}(0)=\hat\Delta=\sum_{l=0}^{2j} \frac{2l+1}{2j+1}\,\tilde\Delta(l)
\end{equation}
Remember that the $\tilde\Delta$'s are just the original $\Delta$'s rescaled by a factor of $N$, $\tilde\Delta(l)=N\Delta(l)$.

The constraints $l_1=l_2$ and $m_1+m_2=0$ are very important! They just express the SU(2) invariance of $\mathcal{D}_{rs,tu}$, i.e. the initial contraint $s-t=r-u$.
But when we take the Wigner transform of the recursion equation \ref{RecGenG} for $ \mathcal{G}_{{ru},{st}}(x,y)$, one see that they are preserved by the equation and that the solution for $W_{\mathcal{G}}^{(l_1,m_1),(l_2,m_2)}(x,y)$ must be of the same form as $W_{\widehat{\mathcal{D}}}$
\begin{equation}
\label{WGgen}
W_{\mathcal{G}}^{(l_1,m_1),(l_2,m_2)}(x,y)=\delta_{l_1,l_2}\,\delta_{m_1+m_2,0}\, (-1)^{m_1}\ \widehat{\mathcal{G}}^{(l_1)}(x,y)
\end{equation}

We can now come back to the recursion equation \ref{RecGenG}.
Using \ref{Mlmlmex} and \ref{WGgen}
it factorizes into independent equations for each $\widehat{\mathcal{G}}^{(l_1)}(x,y)$
\begin{equation}
\label{RecGl}
\widehat{\mathcal{G}}^{(l)}(x,y)=x^{-1} \mathcal{H}(y)+ x^{-1} \widehat{\Delta}(0) \mathcal{H}(x)
\widehat{\mathcal{G}}^{(l)}(x,y)
+y^{-1} \widehat{\Delta}(l) \widehat{\mathcal{G}}^{(l)}(x,y) \mathcal{H}(y)
\end{equation}
The solution is simply (using the explicit form \ref{Hsol} for $\mathcal{H}$)
\begin{equation}
\label{Ghatlexp}
\widehat{\mathcal{G}}^{(l)}(x,y)=\frac{\mathcal{H}(x)\mathcal{H}(y)}{1-\widehat{\Delta}(l)\mathcal{H}(x)\mathcal{H}(y)}
\end{equation}

\subsection{Solution for the evolution functional}
\label{sSolEv}
\subsubsection{General form}
\label{ssGenForEv}
We can now obtain the influence functional $\mathcal{M}(t)$.
As we shall see, the functions $\widehat{\mathcal{G}}^{(l)}(x,y)$ are analytic in $x$ and $y$ around $\infty$, and have a cut in the $x$ and $y$  planes along $[-2\sqrt{\hat\Delta}, -2\sqrt{\hat\Delta}]$.
We can integrate in $x$ and $y$ along a closed anticlockwise curve around the cut to obtain the double Wigner transform of the influence functional $\mathcal{M}_{{ru},{st}}(t)$
\begin{equation}
\label{WlMexp}
W_{\mathcal{M}}^{(l_1,m_1),(l_2,m_2)}(t)=\delta_{l_1,l_2}\,\delta_{m_1+m_2,0}\, (-1)^{m_1}\ \widehat{\mathcal{M}}^{(l_1)}(t)
\end{equation}
where
\begin{equation}
\label{MhlInt}
\widehat{\mathcal{M}}^{(l)}(t)=
\oint\frac{dx}{2\imath\pi}\oint\frac{dy}{2\imath\pi}\,\emath^{-\imath t (x-y)}\,\widehat{\mathcal{G}}^{(l)}(x,y)
\end{equation}
Therefore, the evolution of the reduced density matrix $\rho^{\mathcal{S}}(t)$ becomes a separate simple linear evolution in each $(l,m)$ sector when one considers the components of its Wigner transform.
More precisely, if $W_{\rho^{\mathcal{S}}}^{(l,m)}(t)$ is the $(l,m)$ harmonic, given by \ref{WTransf}, we have simply
\begin{equation}
\label{Wlmt}
W_{\rho^{\mathcal{S}}}^{(l,m)}(t)= \widehat{\mathcal{M}}^{(l)}(t) W_{\rho^{\mathcal{S}}}^{(l,m)}(0)
\end{equation}
Then using \ref{WInv} we can reconstruct $\rho^{\mathcal{S}}(t)$ in the $|r\rangle\langle s|$ basis.

\subsubsection{General decoherence function}
\label{ssGenDecF}
Thanks to the SU(2) invariance, the evolution functional reduces to a single function in each $l$ sector.
This function depends on time and on the distributions of the $\Delta(l)$ which measure the strength of the coupling between the spin and the environment in the different angular momentum sector $l$. This function depends in fact only on two parameters, since it can be rewritten as
\begin{equation}
\label{Mhat2M}
\widehat{\mathcal{M}}^{(l)}(t)= M(t/\tau_0,Z(l)).
\end{equation}
where $\tau_0$ is a time scale
\begin{equation}
\label{tau0Z0}
\tau_0= 1/\sqrt{\widehat{\Delta}(0)}\quad,\qquad \widehat{\Delta}(0)=\sum_{l=0}^{2j} \frac{2l+1}{2j+1}\tilde\Delta(l)
\end{equation}
and $Z(l)$ a parameter depending on the angular momentum $l$
\begin{equation}
\label{Z2l}
Z(l)=\frac{\widehat{\Delta}(l)}{\widehat{\Delta}(0)}=\frac{\sum\limits_{l'=0}^{2j}(2l'+1) (-1)^{2j+l'+l_1}
\sixj{j}{j}{l'}{j}{j}{l_1}  \tilde\Delta(l') }{\sum\limits_{l'=0}^{2j} \frac{2l'+1}{2j+1}\tilde\Delta(l')}
\end{equation}
and the decoherence function $M(t,Z)$ is simply
\begin{equation}
\label{Mint1}
M(t,Z)=\oint\frac{dx}{2\imath\pi}\oint\frac{dy}{2\imath\pi}\,\emath^{-\imath t (x-y)}\,\frac{H(x)H(y)}{1-Z\,H(x)H(y)}
\quad,\qquad H(x)=\frac{1}{2}(x-\sqrt{x^2-4})
\end{equation}
$H(x)$ is nothing but the resolvent of the standard normalized GUE ensemble. It is the Hilbert-Stieltjes transform of the Wigner-Dyson semi-circle density distribution.
It has a cut along the interval $[-2,2]$, behaves as $x^{-1}$ at $\infty$ and its modulus is $|H(x)|<1$ for $x\in\mathbb{C}\backslash [-2,2]$.
Hence the function $M(t,Z)$ is well defined for any real $t$, and analytic in the disc $|Z|\le 1$. 
We shall discuss its properties below.

Thus we have a completely closed and simple formula for the evolution functional of a spin coupled to a large environment via a random coupling Hamiltonian which belongs to an SU(2)$\times$SU(N) invariant ensemble. 
Our formula is valid for any value of the spin $j$, going from $j=1/2$ (the q-bit or two level system) to $j\to\infty$ (the classical spin), and for any distribution $\Delta(l)$ of the strength of the couplings as a function of the total spin $l$ exchanged via the interaction.

Our result separates in two parts: 
(1) the universal decoherence function $M(t,z)$ which comes from the RMT part of the calculation; 
(2)  the parameters $\widehat{\Delta}(l)$ which depend linearly from the initial distribution $\Delta(l)$ of the couplings as a function of the angular momentum $l$, which come from the SU(2) group theory part of the calculation. They give simply the time scale $\tau_0$ and the parameter $Z(l)$
 
\subsection{Properties of the decoherence function $M(t,Z)$:}
\label{sPropMF}
\subsubsection{Analytic representation}
\label{ssAnRepM}
\begin{figure}[h!]
\begin{center}
\includegraphics[width=5in]{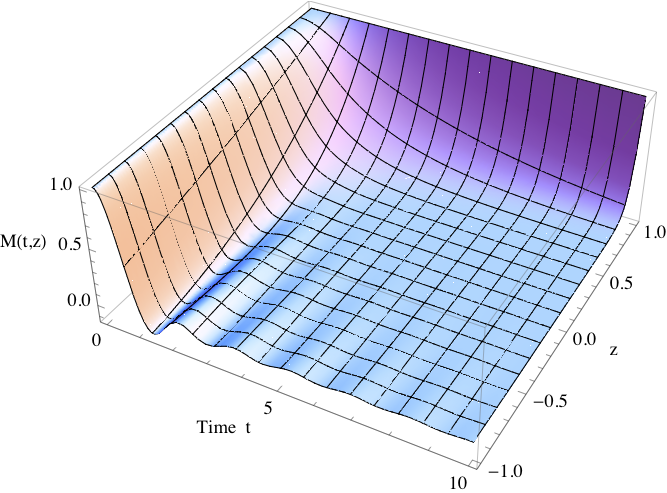}
\caption{The decoherence function $M(t,z)$ as a function of the rescaled time $t$ and the parameter $z\in[-1,1]$. Decoherence is fast when $z<1$ but very slow when $z\simeq 1$. }
\label{f:MtZ}
\end{center}
\end{figure}

Making the standard inversion of variables  $x\to H$  as in \cite{Zee1996} (i.e. going from the Green function $H(x)$ to the so-called ``Blue function'' $B(w)$)
\begin{equation}
\label{H2Bfun}
w=H(x)\quad\iff\quad x=B(w)=w+w^{-1}
\end{equation}
we rewrite $M$ as
\begin{equation}
\label{Mint2}
%w_1=H(x)\ , \ \ w_1=H(y)\quad
 M(t,z)=\oint\frac{dw_1}{2\imath\pi}\oint\frac{dw_2}{2\imath\pi}\,
 %\emath^{-\imath t ((w_1^{-1}+w_1)-(w_2^{-1}+w_2))}
 \emath^{-\imath t B(w_1)}\emath^{\imath t B(w_2)}
  \frac{(w_1-w_1^{-1})(w_2-w_2^{-1})}{1-z\,w_1 w_2}
\end{equation}
where integrating along the cut $[-2,2]$ in \ref{Mint1} amounts to integrate along the unit circle in  \ref{Mint2}.
We can use it to obtain the double $(t,x)$ series expansion of $M(t,z)$ which is found (after a bit of algebra)
\begin{equation}
\label{Mtzdev}
M(t,z)=\sum_{m=0}^\infty \sum_{n=0}^m t^{2m} \,z^n \,(-1)^{m+n}\ \frac{2 (2 m+1) (n+1)^2 (2 m)! }{m! (m+1)! (m-n)! (m+n+2)!}
\end{equation}
Thus $M(t,z)$ is a generalized hypergeometric function of the two variables $t^2$ and $z$.
It is depicted on \fig{f:MtZ}

\subsubsection{Small $t$ limit}
\label{ssSmllT}
The small $t$ behavior of $M(t,z)$ is
\begin{equation}
\label{Mtzt0}
M(t,z)=1+t^2(z-1)+\mathcal{O}(t^4)
\end{equation}

\subsubsection{Large $t$ limit}
The large time behaviour of $M(t,z)$ is most easily calculated from the integral representation \ref{Mint2} by using the steepest descent method at the saddle points $w_1=\pm 1$, $w_2=\pm 1$.
We obtain an algebraic decay as $t^{-3}$, with an oscillatory term negligible when $z\to 1$ and dominant when $z\to -1$.
\begin{equation}
\label{Mtztinf}
M(t,z)=\frac{1}{2\pi}\, t^{-3}\left(\frac{1+z}{(1-z)^3}-\frac{1-z}{(1+z)^3}\sin(4t)\right)\left(1+\mathcal{O}(t^{-1})\right)
\end{equation}

\subsubsection{The $z\to 1$ and $t(1-z)=\mathcal{O}(1)$ scaling}
\label{z1tz1}
\begin{figure}[h]
\begin{center}
\includegraphics[width=3in]{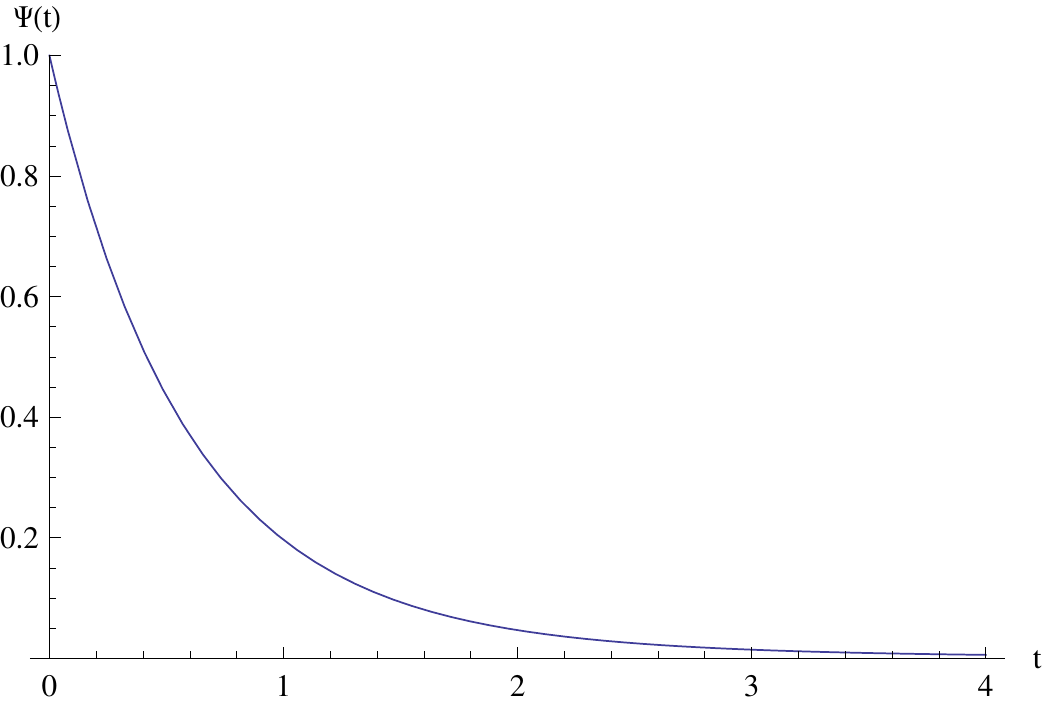}
\caption{The scaling function $\Psi(t)$ for the decoherence function when $z\to 1_-$}
\label{f:psit}
\end{center}
\end{figure}

When $z=1$, we have in fact  
\begin{equation}
\label{Mz1}
\lim_{z\to 1_-} M(t,z)=1
\end{equation}
but the function $M(t,z)$ takes a scaling form when $z\to 1$ while $t$ is large. In fact
\begin{equation}
\label{Mz1tlarge}
M(t,z)=\Psi(t')\quad\text{with}\quad t'=t(1-z)
\quad\text{in the limit}\quad t'=\mathcal{O}(1)\ ,\ \ z\to 1_-
\end{equation}
Indeed, in this limit, the measure in \eq{Mint2} concentrates around $w_2=w_1$ and the integral becomes a single integral representation.
We get a simple hypergeometric function
\begin{align}
\label{Psi1}
\Psi(t')
\ &=\ \oint {dw\over 2\imath \pi}\,\emath^{- t' B(w)} \frac{1}{2w}B(w)^2
\ =\ {2\over\pi}  \int_{-\pi/2}^{\pi/2} d\theta\ \emath^{-2t'\cos(\theta)}\,\cos(\theta)^2
\nonumber\\
&=\ {1\over 2\pi}\int_{-2}^{2} dx\, \sqrt{4-x^2}\  \emath^{-t'\,\sqrt{4-x^2}}
\end{align}
whose series expansion is explicitely
\begin{equation}
\label{Psi2}
\Psi(t')\ =\ \frac{2}{\sqrt{\pi}}\sum_{k=0}^\infty (-2t')^k \frac{\Gamma((3+k)/2)}{k!\,\Gamma(2+k/2)}
\end{equation}
This function is depicted on \fig{f:psit}.
Its asymptotic behavior is 
\begin{equation}
\label{Psi3}
\Psi(t')=1-\frac{16}{3\,\pi} t'+\mathcal{O}({t'}^2)\quad t'\to 0\quad,\qquad \psi(t')=\frac{1}{\pi}\, {t'}^{-3}\ +\ \mathcal{O}({t'}^{-4})\quad t'\to\infty
\end{equation}

\subsubsection{The $z\to 1$ and  $t=\mathcal{O}(1)$ scaling}
\label{ssz21Scal}
Note that $\Psi(t)$ is linear in $t$ at small time, not quadratic in $t$ like $M(t,z)$ for $z<1$.
For $z=1-\epsilon$ close to 1 ($\epsilon\ll 1$) but $t$ of order $1$, the function $M(t,z)$ behaves as
\begin{equation}
\label{Mz1t1}
M(t,z)=1 +\epsilon\, \Phi(t)+\mathcal{O}(\epsilon^2)
\end{equation} 
with $\Phi(t)=1-\, _1F_2\left(-\frac{1}{2};1,2;-4 t^2\right)$ a universal non-linear function which behaves as
\begin{equation}
\label{Phitexp}
\Phi(t)=-t^2+\mathcal{O}(t^4)\quad\text{when}\quad t\to 0\quad,\qquad \Phi(t)=-{1\over\pi} t+\mathcal{O}(1)\quad\text{when}\quad t\to \infty
\end{equation}
\begin{figure}[h!]
\begin{center}
\includegraphics[width=3in]{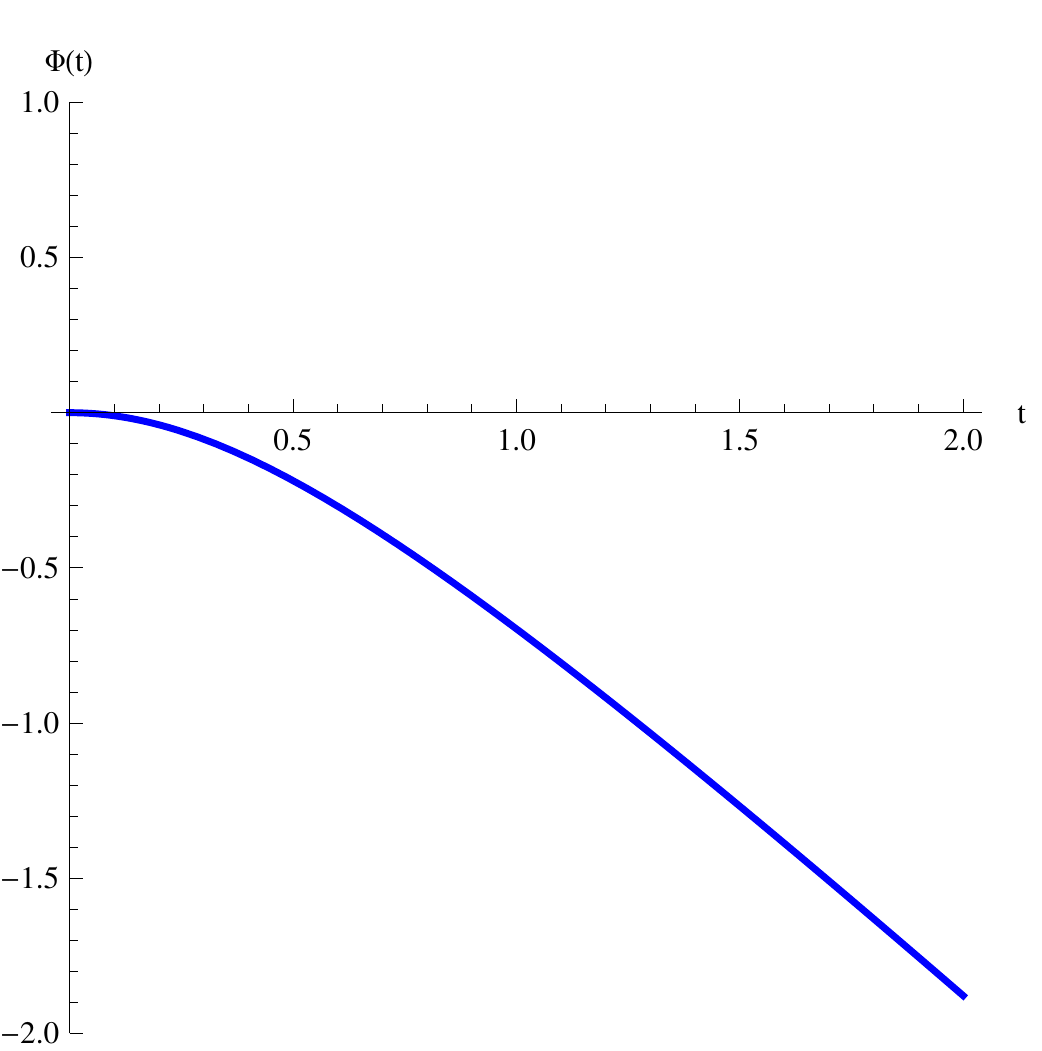}
\caption{The function $\Phi(t)$ of \eq{Mz1t1} and \eq{Phitexp}  that describes the small $t$ and $z\sim 1$ (hence $l\ll j$) behavior of the function $M(t,z)$.}
\label{f:Phi}
\end{center}
\end{figure}
Therefore, the crossover between the non-linear regime \ref{Mz1t1} for small $t$ and the linear regime \ref{Mz1tlarge} for large $t\sim\epsilon^{-1}$ occurs in a domain of  $t$ of size $\mathcal{O}(1)$, hence in a very small interval in $t'$ of size $\mathcal{O}(\epsilon)$.

\subsection{$Z(l)$ dependence on the choice of couplings $\Delta(l)$ and on the total spin $j$.}
\label{sZofDel}
\subsubsection{Various examples}
\label{ssVarExZ}
Now we study how the time scale $\tau_0$ and the parameters $Z(l)$ (which govern the dynamics in each angular momentum sector $l$) depend on the choice of the original distribution of couplings $\Delta(l)$ and of the total spin $j$. 
We remind that
\begin{equation}
\label{tau0Z0b}
\tau_0= 1/\sqrt{\widehat{\Delta}(0)}\quad,\qquad \widehat{\Delta}(0)=\sum_{l=0}^{2j} \frac{2l+1}{2j+1}\tilde\Delta(l)\quad,\qquad\tilde\Delta(l)=N\,\Delta(l)
\end{equation}
and that $Z(l)$
\begin{equation}
\label{Z2lb}
Z(l)=\frac{\widehat{\Delta}(l)}{\widehat{\Delta}(0)}
\quad,\qquad
\widehat{\Delta}(l)=\sum\limits_{l'=0}^{2j}(2l'+1) (-1)^{2j+l'+l}
\sixj{j}{j}{l'}{j}{j}{l}  \tilde\Delta(l') 
\end{equation}

In the trivial case where all the $\Delta$'s are equal we have
\begin{equation}
\label{DelGUE}
\Delta(l)=\Delta\quad\text{for all}\ l \quad\implies\quad \widehat{\Delta}(0)=(2j+1)\, N\,\Delta\quad\text{and}\quad 
\widehat{\Delta}(l)=0\quad\text{if}\ 0<l\le 2j
%Z(l)=1\quad\text{for all}\ l
\end{equation}
Hence $Z(l)=0$ for all $l>0$. We recover the GUE ensemble.

We are interested in the situation where only a finite number $l_0$ of interactions channels are activated, and where this number is much smaller than the total spin
\begin{equation}
\label{Dellb}
\Delta(l)=0\quad\text{if}\quad l>l_0\quad,\qquad l_0\ll j
\end{equation}
We first assume that all the $\Delta(l)$ for $l\le l_0$ are of the same order $\Delta$. Then $\widehat{\Delta}(0)$ is of order $\Delta N l_0^2/j$.
We thus rescale $\Delta\to\bar\Delta$ both with $N$ (as before) and with the spin $j$, having in mind to have both a large environment ($N\to\infty$) and a semiclassical spin ($j\to\infty$).
\begin{equation}
\label{DDbarDtil}
\Delta(l)=(2j+1) N^{-1}\,\bar{\Delta}(l)=(2j+1)\widetilde{\Delta}(l)
\end{equation}
and we are now interested in the limit of a large environment ($N\to\infty$) and a semiclassical spin ($j\to\infty$).
\begin{equation}
\label{ semijlim}
N\to\infty\ ,\quad j\ll 1\quad\text{large but finite or infinite}\ ,\quad\bar{\Delta}(l)\quad\text{of order}\ \mathcal{O}(1)
\end{equation}
We now have $$\widehat{\Delta}(0)=\tau_0^{-2}=\sum_{l=0}^{l_0} (2l+1)\bar\Delta(l)\quad\text{independent of the spin}\ j$$

In the figures presented in the Appendix \ref{app:fig} we plot for several choices of distributions of $\bar\Delta$'s the resulting $Z(l)$, and then discuss the results and prove some of the properties of the function $Z(l)$.

\begin{itemize}
\item In \fig{fZDeEq} we plot $Z(l)$ as a function of $l$ in the case where all the $\Delta(l)$ are equal for $0\le l\le l0$, and zero otherwise.
\item In \fig{fZDelb0} we plot $Z(l)$ as a function of $l$ in the case where $\Delta(0)=0$, all the $\Delta(l)$ are equal for $1\le l\le l0$, and zero otherwise.
\item In \fig{FZoneOdd} only a single and odd $l$ contributes.
\item In \fig{FZallOdd} all odd $l\le l_0$ contribute.
\item In \fig{FZallEven} all even $l\le l_0$ contribute.
\item In \fig{FZlargeZ0} we compare different cases, letting $\Delta(0)$ become large..
\item In \fig{FZranDel} we consider some random distributions of $\Delta(l)$ for $l\le l_0$.
\end{itemize}

First we make the experimental observations: 
\begin{enumerate}
  \item For $0<l\le 2j$, $Z(l)$ is always in the interval $]-1,1[$.
  \item For a fixed coupling distribution $\overline{\bold\Delta}$, when the total spin $j$ is large $Z(l)$ takes a limit scaling form $Y$ (which depends of course of the $\overline{\bold\Delta}$'s)
  \begin{equation}
\label{ZscalY}
Z(l)= Y(l/2j)
%(1+\mathcal{O}(j^{-1})) 
\quad\text{when }\ j\to\infty\quad,\quad l/j\quad \text{fixed}
\end{equation}
  \item By normalisation $Z(0)=1$. Otherwise $Z(l)$ can be close to  $+1$ (respectively to $-1$) only in the limit $j\to\infty$,  when $l\simeq 2j$ and when the $\overline{\Delta}(l)$ are zero for all odd $l$'s (respectively all even $l$'s).
\end{enumerate}

\subsubsection{Limit $j\to\infty$, $l/j$ fixed}
\label{ssj2inf}
The existence of a limit distribution $Y(x)$, $x=l/2j$ when $j\to\infty$ is easily explained.
We use Racah formula of the 6j-symbols to rewrite the formula \ref{Z2lb} for $\widehat{\Delta}(l)$ as
\begin{equation}
\label{RacahDelta}
\widehat{\Delta}(l)=(2j+1) \sum_{l'=0}^{2j} \bar\Delta(l') \ (2l'+1)\sum_{k=0}^{\mathrm{min}(l,l')}\frac{(-1)^k}{(k!)^2}
\frac{(l'+k)!\,(l+k)!\,(2j-k)!}{(l'-k)!\,(l-k)!\,(2j+k+1)!}
\end{equation}
We use Stirling formula to take the limit 
\begin{equation}
\label{jlimitx}
j\to\infty\quad,\quad x=\frac{l}{2j}\ \text{fixed}
\end{equation}
to obtain
\begin{equation}
\label{hatDelSc}
\widehat{\Delta}(l)\to  \sum_{l'=0}^{l_0}\bar\Delta(l')\,(2l'+1)\,F_{l'}(x)
\end{equation}
with $F_{l'}(x)$ the polynomials
\begin{equation}
\label{FlPoly}
F_{l'}(x)=\sum_{k=0}^{l'}\frac{ (-1)^k}{(k!)^2}\frac{(l'+k)!}{(l'-k)!}\,x^{2k}=\ _2F_1(1+l',-l',1,x^2)
\end{equation}
Hence the explicit polynomial form for the limit scaling function $Y$ in \eq{ZscalY}
\begin{equation}
\label{Yscal}
Y(x)=\frac{ \sum\limits_{l'=0}^{l_0}\bar\Delta(l')\,(2l'+1) F_{l'}(x)}{\sum\limits_{l'=0}^{l_0}\bar\Delta(l') (2l'+1)}
\end{equation}

\subsubsection{Limit $j\to\infty$, $l\ll j$}
\label{ssLllJ}
Apart from some very special cases, $Z(l)$ is close to 1 only if $l$ is small. This case is needed for the study of decoherence. 
When $l\ll j$ we need only to keep the terms $k=0$ and $k=1$ in the explicit form \ref{RacahDelta} for $\widehat{\Delta}(l)$.
We obtain
\begin{equation}
\label{HatDexpl}
\widehat{\Delta}(l)=\widehat{\Delta}(0)-\frac{l(l+1)}{4j(j+1)}\sum_{l'=1}^{l_0}\bar{\Delta}(l')\,(2l'+1)\, l'(l'+1)+\mathcal{O}((l/j)^4)
\end{equation}
Hence
\begin{equation}
\label{ZlExp}
Z(l)=1-l(l+1)\ \frac{1}{4}\frac{D_0}{j(j+1)}+\cdots
\quad,\quad D_0=\frac{\sum\limits_{l'=1}^{l_0}\bar{\Delta}(l')\,(2l'+1)\, l'(l'+1)}{\sum\limits_{l'=0}^{l_0}\bar\Delta(l') (2l'+1)}
\end{equation}
This approximation is valid on the top of the curve $Y(x)$ near $x=0$, i.e. provided that
\begin{equation}
\label{ lllj}
l(l+1)\ll j(j+1)
\end{equation}
It is in particular valid when $l\propto\sqrt{j}$, which is the case to consider when studying the coherent states.

Note that if all the $\bar{\Delta}(l)'s$ are of the same order $\bar{\Delta}$ when $l\le l_0$ and zero othervise, the coefficient $D_0$ is of order
\begin{equation}
\label{D0l2}
D_0\sim \frac{1}{2}\,l_0^2
\end{equation}
But note also that the numerator in $D_0$ involves only the $\bar{\Delta}(l)$ for $l>0$.
In particular, if the $l=0$ coupling $\bar{\Delta}(0)$ is much larger than the others $\bar{\Delta}(l)\sim \bar{\Delta}$ for $0<l\le l_0$ 
\begin{equation}
\label{D0ineq}
\bar{\Delta}(0)\gg l_0^2\,\bar{\Delta}\quad\implies\quad  D_0\sim \frac{1}{4}\frac{ l_0^4\,\bar{\Delta}}{\bar{\Delta}(0)}\ \ll\ l_0^2
\end{equation}

\subsubsection{The case of even $l$'s or odd $l$'s}
\label{ssEvOdd}
One sees from \fig{FZallEven} that when there are no couplings for  odd $l$'s, i.e. when
\begin{equation}
\label{levenCo}
l\ \text{odd}\quad\implies\quad \Delta(l)=0
\end{equation}
then for very large spin $j\to\infty$ the condition of slow decoherence $Z(l)\simeq 1$ is satisfied for the $l$'s close to the maximal value $l=2j$.
In fact in this case the scaling function $Y(x)$ defined by \eq{ZscalY} is for $x\simeq 1$
\begin{equation}
\label{Yxscal}
Y(x)\simeq 1-\text{cst.}(1-x)
\end{equation}
The reason why there is a very slow decoherence between opposite states such as $|j\rangle$ and $|-j\rangle$ is of course that the coupling hamiltonian $\HaSE$ has an additionnal $\mathbb{Z}_2$ parity symmetry, the inversion of spin, which protects the states with this symmetry from decoherence. 

On the contrary, one sees on \fig{FZoneOdd} and \fig{FZallOdd} that when there are no couplings for even $l$'s, the condition of maximally fast decoherence $Z(l)=-1$ is satisfied for the $l$'s close to the maximal value $l=2j$.

\subsection{Decoherence parameters and norms of operators}
\label{sDecParNo}
\subsubsection{$\widehat\Delta(0)$ and the norm of $H$}

%$$\widehat{\Delta}(0)=\frac{N}{2j+1}\sum_l (2l+1)\Delta(l)$$
When all the $\Delta(l)$ are equal to the same $\Delta$ (GUE ensemble) all the matrix element of $H$ are of the same order, $\sqrt{\Delta}$, and the whole Hamlitonian $H$ is a random $M\times M$ matrix in a GUE ensemble, with $m=(2j+1) N$. 
The normalization is such that
%$$\mathbb{E}[\tr(H^2)]=(2j+1)^2N^2\Delta$$
\begin{equation}
\label{TrH2D}
\Delta(l) =\Delta
 \quad\implies\quad\overline{\tr(H^2)}=(2j+1)^2N^2\Delta
\end{equation}
In the general case where the $\Delta(l)$ are different, we have
%$$\mathbb{E}[\tr(H^2)]=(2j+1)N\,\widehat{\Delta}(0)$$
\begin{equation}
\label{TrH2Dh}
\overline{\tr(H^2)}=(2j+1)N\,\widehat{\Delta}(0)
\quad\text{with as before}\ 
\widehat{\Delta}(0)=\frac{N}{2j+1}\sum_l (2l+1)\Delta(l)
\end{equation}
so $\sqrt{\widehat{\Delta}(0)}$ is the typical size of an eigenvalue of $H$, that we call its norm ${|\!| H |\!|}_2$,  the norm of an operator $A$  being defined as
\footnote{This norm is the Hilbert-Schmidt norm divided by $\sqrt{\mathrm{dim}(\mathcal{H})}$, and is not the C$^*$ norm
$
{|\!| A |\!|}^2=\sup_{|\psi\rangle} {\langle\psi|A^\dagger A|\psi\rangle\over\langle\psi|\psi\rangle}
$
which corresponds to the modulus of the largest eigenvalue of the operator. }
\begin{equation}
\label{norm2}
{|\!| A |\!|}_2^2={\tr(A^\dagger A)\over\tr(\mathbf{1})}
\end{equation}

In our modified GUE ensembles, if we take for $|m,\alpha\rangle$ a basis of the whole Hilbert space $\HaSE=\mathbb{C}^{(2j+1)N}$, since
\begin{equation}
\label{Hnorm2expl}
{|\!| H |\!|}_2^2={1\over (2j+1)N}\sum_{m,n}\sum_{\alpha,\beta} |\langle m,\alpha|H|n,\beta\rangle|^2
\end{equation}
we see that our norm is 
\begin{equation}
\label{normHav}
{|\!| H |\!|}_2\simeq \sqrt{(2j+1)N}\ \times\ \text{``average value'' of}\  \left| \langle m,\alpha|H|n,\beta\rangle  \right|\ 
\end{equation}

\subsubsection{The spectrum of $H$}
It is also easy to see from the solution \ref{Hrs} and \ref{Hsol} of the single resolvent $\mathcal{H}(x)$ that, although the Hamiltonian $H$ does not belong to a GUE ensemble but to the modified GU$_{2,N}$E ensemble, in the large $N$ limit its density of states (DOS) $\rho(\lambda)$ , i.e. the density distribution of the eigenvalues of the eigenvalues of $H$, is still given by a Wigner semicircle law, on the interval $[-\hat E,\hat E]$ with
\begin{equation}
\label{ }
\rho(\lambda)\propto \sqrt{\hat E^2-\lambda^2}
\quad\text{with}\quad
\hat E=2 \sqrt{\hat\Delta}
\end{equation}

\subsubsection{Mean $\widehat{\Delta}_{\mathrm{av}}$ for the $\widehat{\Delta}(l)$}
It will be convenient to consider the mean value of the  $\widehat{\Delta}(l)$ when averaged over all possible modes $(l,m)$'s, $\widehat{\Delta}_{\mathrm{av}}$.
It is defined as
\begin{equation}
\label{DeltaAv}
\widehat{\Delta}_{\mathrm{av}}=\frac{1}{(2j+1)^2}\sum_{l=0}^{2j}(2l+1) \widehat{\Delta}(l)
%=\frac{N}{2j+1}{\Delta(0)}=\bar{\Delta}(0)
\end{equation}
and in found to be nothing but
\begin{equation}
\label{DeltaAv2}
\widehat{\Delta}_{\mathrm{av}}
%=\frac{1}{(2j+1)^2}\sum_{l=0}^{2j}(2l+1) \widehat{\Delta}(l)
=\frac{N}{2j+1}{\Delta(0)}=\bar{\Delta}(0)
\end{equation}
The average value of $Z(l)$ is
\begin{equation}
\label{ZAv}
Z_{\mathrm{av}}=\frac{1}{(2j+1)^2}\sum_{l=0}^{2j}(2l+1) Z(l)=\frac{\bar{\Delta}(0)}{\widehat{\Delta}(0)}
\end{equation}
and in particular
\begin{equation}
\label{1-Zav}
1-Z_{\mathrm{av}}=\sum_{l=1}^{2j}(2l+1)\frac{\bar\Delta(l)}{\widehat{\Delta}(0)}=\left(\frac{{|\!|H'|\!|}_2}{{|\!|H|\!|}_2}\right)^2
\quad,\qquad H'=H-H^{(0)}
\end{equation}
$H'$ being the purely $\mathcal{S}+\mathcal{E}$ interaction part of the Hamiltonian $H$, and $H^{(0)}$ being the purely external part $\mathcal{E}$
of $H$.

\section{Decoherence and emergence of coherent states}
\label{sDecCoS}
It is easy now to study the dynamics of decoherence and the emergence of the coherent states for spin as the semi-classical states robust against the interaction with the environment.

\subsection{Coherent states}
\label{sCohSta}
\subsubsection{Pure coherent states:}
\label{ssPureCoSt}
Coherent states are the pure states which minimize the uncertaincy relations for spin, i.e. the states with a maximally localized Wigner distribution.
They read explicitely
\begin{equation}
\label{CoStGe}
|\vec n\rangle=\sum_{m=-j}^{j} \sqrt{\frac{(2j)!}{(j+m)!\,(j-m)!}}\cos(\theta/2)^{j+m}\sin(\theta/2)^{j-m}\emath^{-\imath m\phi} |m\rangle
\end{equation}
with $(\theta,\phi)$ the spherical coordinates of the unit vector $\vec n$.
Coherent states are formed by a coherent superposition of modes such that $l\sim\sqrt{j}\ll j$.
Indeed for the single pure state
\begin{equation}
\label{CoSt}
|\vec e_z\rangle=|j\rangle
\end{equation}
the matrix density components are
\begin{equation}
\label{WlmCoh}
W_{|j\rangle\!\langle j|}^{(l,m)}=\delta_{m,0} W^{(l)}_{\mathrm{c.s.}}\quad,\quad W^{(l)}_{\mathrm{c.s.}}=\, \,\sqrt{\frac{\left((2j)!\right)^2\,(2l+1)}{(2j+l+1)!(2j-l)!}}
\end{equation}
and for large $j$ and small $l$ Stirling formula gives
\begin{equation}
\label{WlmCas}
W^{(l)}_{\mathrm{c.s.}}=\, \frac{2l+1}{\sqrt{2j+1}}\exp\left({-\frac{l^2}{2j}}\right)\left(1+\mathcal{O}\left(\frac{l^3}{j^2}\right)\right)
\end{equation}
The Wigner representation of the coherent state $|\vec n\rangle$ is a Gaussian-like positive distribution with width $1/\sqrt{j}$ centered at $\vec n$ on the unit sphere.

\subsubsection{Random pure states:} 
\label{ssRanPuSt}
At variance with coherent states, a random pure spin state $|\psi\rangle$ is such that its density matrix components are independent equally distributed random variables
\begin{equation}
\label{WlmRan}
W_{|\psi\rangle\!\langle \psi|}^{(l,m)}\sim{1\over 2j+1}
\end{equation}
and its Wigner representation is a random function on the sphere (analogous to a random polynomial with zeros obeying Wigner statistics with short distance cut-off $1/\sqrt{j}$).

\subsubsection{Superpositions of coherent states:}
\label{ssSupCoh}
Quantum superpositions of coherent states correspond to more complicated functions. For instance it is well known that the "Schrödinger cat-like" state
\begin{equation}
\label{CatSta}
\left| \uparrow\!{\scriptscriptstyle{+}}\!\downarrow\right>={1\over \sqrt{2}}\left( |+\vec e_z\rangle+|-\vec e_z\rangle\right)={1\over \sqrt{2}}\left(|j\rangle+|-j\rangle\right)
\end{equation}
corresponds to  modes $(l\ \mathrm{even},m=0)$ and $(l=2j,m=\pm 2j)$.
\begin{equation}
\label{WlmCat}
W^{(l,m)}_{\scriptscriptstyle\mathrm{CAT}}={1\over 2} \delta_{m,0}\,\left(1+(-1)^l\right)\sqrt{\frac{\left((2j)!\right)^2\, (2l+1)}{(2j+l+1)!(2j-l)!}}
+{1\over 2}\delta_{l,2j} \left(\delta_{m,2j} (-1)^{2j}  +  \delta_{m,-2j}\right)\sqrt{{1+2j\over 1+4j}}
\end{equation}
In general, the density matrix of a superposition of two coherent states will have large components $W^{(lm)}$ describing the quantum correlations
for 
\begin{equation}
\label{WlmSim}
W^{(lm)}\simeq\mathcal{O}(1)\quad\text{for}\ l\sim j
\end{equation}
In particular, if $|\psi\rangle$ is a superposition of two coherent states $|\vec n_1\rangle$ and $|\vec n_2\rangle$ which are at distance $\theta_{12}$ on the sphere $\mathcal{S}_2$, \begin{equation}
\label{supcoh}
|\psi\rangle=c_1|\vec n_1\rangle+c_2|\vec n_2\rangle\quad\text{with}\ \vec n_1\cdot\vec n_1=\cos \theta_{12}
\end{equation}
the largest elements $W^{(lm)}$ of the ``off diagonal'' part of the density matrix  $\rho_{\mathrm{off}}=|\vec n_2\rangle\langle\vec n_1|+|\vec n_1\rangle\langle\vec n_2| $ occur for
\begin{equation}
\label{lmaxthet}
W^{(lm)}_{\rho_{\mathrm{off}}}\simeq\mathcal{O}(1)\quad\text{for}\ l\simeq 2j\sin(\theta_{12}/2)
\end{equation}

\subsection{The time scales of the system}
\label{ssTS}
From the above discussion, one sees that in our model the evolution of the spin is characterized by at least three time scales, which are such that
%for large spin and 
\begin{equation}
\label{tauideq}
\tau_0<\tau_1\ll\tau_2
\end{equation}

\subsubsection{Dynamical timescale for the whole system ${\tau_0} =\tau_{\mathrm{dyn}}$}
\label{sstau0}
The first characteristic time is
\begin{equation}
\label{t0defa}
\tau_0
%=\left({||H||}_2\right)^{-1}
=\left(\widehat{\Delta}(0)\right)^{-1/2}
%=\left({||H||}_2\right)^{-1}
=\left(\normtwo{H}\right)^{-1}
\end{equation} 
It is the typical evolution time scale for a generic state of the whole system $\mathcal{E}+\mathcal{S}$. 
Thus we denote it also $\tau_{\mathrm{dyn}}$.

\subsubsection{Decoherence timescale for spin states $\tau_1=\tau_{\mathrm{dec}}$} 
\label{sstau1}
The second characteristic time is
\begin{equation}
\label{t1def}
\tau_1=\frac{\tau_0}{1-Z_{\mathrm{av}}}=\frac{{||H||}_2}{{({||H'||}_2)}^2}
\end{equation} 
It is the time scale for the decay of the $W^{(lm)}$ coefficients  of the density matrix for typical $l\sim j$. Hence it corresponds to the decoherence timescale for random pure states $|\psi\rangle$ for the spin, with no particular spin properties. 
Hence we denote it also by $\tau_{\mathrm{dec}}$.

For $t\gg\tau_1$, a pure random state $|\psi\rangle$ has become a complete statistical mixture $\rho\propto\mathbf{1}$.
We have of course $\tau_1>\tau_0$ and in general $\tau_1$ is of the same order than $\tau_0$.
But we have seen that this decoherence time $\tau_1$ can be much larger that $\tau_0$ in the special case where $\Delta(0)\gg \Delta(l)$, $l>0$, i.e; when the internal dynamics of $\mathcal{E}$, given by $\HaE=H^{(0)}$, is much faster than the dynamics of $\mathcal{S}$ induced by the coupling $\mathcal{S}+\mathcal{E}$, and given by $\HaSE=H'$.

\subsubsection{Timescale for evolution of coherent states $\tau_2=\tau_{\mathrm{diff}}$} 
\label{sstau2}
The third characteristic time is the evolution time for a single coherent state. 
It is the decay time scale in sectors such that
\begin{equation}
\label{lsqjj}
l\sim\sqrt{j}\ll j
\end{equation}
In this case the evolution of the density matrix is given by the regime $Z(l)\simeq 1$ studied in \ref{ssLllJ}. 
Using \eq{ZlExp} this time $\tau_2$ is given by
\begin{equation}
\label{t2def}
\tau_2=\tau_0 \frac{j}{D_0}
%=\frac{{||\vec S||}_2 {||H||}_2}{{({||[\vec S,H]||}_2)}^2}
\end{equation}
with $D_0$ given by \eq{ZlExp}.
Note that on general grounds
\begin{equation}
\label{genineqt12}
\frac{j}{l_0(l_0+1)}\tau_1\le\tau_2\le\frac{j}{2}\tau_1
\quad\text{where}\ l_0=\sup\{l:\,\Delta(l)>0\}
\end{equation}
As we shall see in sectionN \ref{ssdiffus}, for $t>\tau_2$ the evolution of coherent states is decribed by a quantum diffusion process. 
Hence we denote it also by $\tau_{\mathrm{diff}}$.

\subsubsection{The conditions for decoherence}
\label{ssCondDec}
The dynamics of coherent states is  much slower that the evolution of non-coherent states when the spin $j$ is large and when $l_0$ is of order $\mathcal{O}(1)$, hence $l_0\ll j$. 
In fact using \eq{ZlExp} for $D_0$, and \eq{SSAcom} the ration of the two caracteristic times $\tau_1$ and  $\tau_2$ can easily be rewriiten in term of the ratio of operator norms
\begin{equation}
\label{t1overt2}
{\tau_1\over \tau_2}=    \frac {\left(\normtwo{ [{\vec S},H'] }\right)^2}{ \left(\normtwo{\vec S}\right)^2 \left(\normtwo{\vphantom{\vec S}H'}\right)^2}
\end{equation}
$H'$ is the interaction Hamiltonian $H'=\sum\limits_{l>0}H^{(l)}$.
Of course the norm of the spin operator $\vec S$ is simply $\normtwo{\vec S}=\sqrt{j(j+1)}$.
The numerator is the squared norm of the commutator of the spin operators with $H'$
\begin{equation}
\label{normSH2}
\left(\normtwo{ [{\vec S},H'] }\right)^2=\sum_{\mu=1}^{3}\left(\normtwo{[S^\mu,H']}\right)^2
={1\over (2j+1)N}\sum_{\mu=1}^{3} \tr\left[  \left([S^\mu,H']\right)^2  \right]
\end{equation}
One sees explicitly that it is when the commutator of the spin with the interaction Hamiltonian is small compared to the product of these two operators 
\begin{equation}
\label{normSH3}
[\vec S,H']\ \ll\  \vec S \times H'
\end{equation}  
(here in the sense of the $\normtwo{\cdot}$ norm) that the single spin coherent states $|\vec n_1\rangle$ are much more robust than quantum superpositions of macroscopically distinct coherent states $|\psi\rangle=c_1|\vec n_1\rangle+ c_2|\vec n_2\rangle$ with $\theta_{12}\gg\sqrt{j}$.
It is in the regime
\begin{equation}
\label{t1tt2a}
\tau_1\ll t\ll \tau_2
\end{equation} 
that coherent states behave as classical states (pointer states), and that one can use a semiclassical picture.

In general, a pure state of the form $|\psi\rangle=c_1|\vec n_1\rangle+ c_2|\vec n_2\rangle$ becomes a statistical mixture of $|\vec n_1\rangle$ and $|\vec n_2\rangle$ after a time of order $\tau_1$. 
But this decoherence timescale depends on the distance between the two states, i.e. of the angle $\theta$ between the vectors $\vec n_1$ and $\vec n_2$.
In the semiclassical regime when the spin is large $j\gg1$, and where this distance is small, but still large compared to the width of a coherent state, $1/\sqrt{j}\ll \theta\ll 1$, using \eq{lmaxthet}, \eq{ZlExp} and the scaling \eq{Mz1tlarge} discussed in \ref{z1tz1} ,  we have for typical decoherence timescale
\begin{equation}
\label{tdecthet}
\tau_{\mathrm{dec.}}(\theta)\sim \tau_1 /\sin(\theta/2)^2\quad\text{such that}\ \tau_1\ll\tau_{\mathrm{dec.}}(\theta)\ll \tau_2
\end{equation}
Thus the closer two coherent states are in phase space, the longer it takes for decoherence to wash out quantum correlations between these two states.
This is a well known effect for harmonic oscillator or free particles discussed  for instance  in
\cite{StrunzHaakeBraun2002, StrunzHaake2003, Zurek2003}.

Beyond the regime discussed here, i.e. for $t\gg\tau_2$, coherent states start to evolve, and in fact becomes statistical mixtures according to a quantum diffusion process that we discuss in  section~\ref{ssdiffus}.

\subsection{Illustrated examples of evolutions}
\label{sIllEv}
\subsubsection{The cases considered}
\label{ssIntrIl}
We have a complete explicit solution of the evolution of the $(l,m)$ components of the density matrix for the spin, starting from arbitrary initial conditions.
The $(l,m)$ components are nothing but the components of the decomposition of the Wigner transform of the density matrix in spherical harmonics. Therefore it is both tempting and easy to illustrate our results by explicitly plotting the time evolution of the Wigner transform for various initial states, and various choices of dynamics (given by the couplings $\Delta(l)$), and various values of the spin $j$.

In the following we take a large (but not tremendously large) value of the spin:  $j=20$.
We chose as coupling distributions the simplest case
\begin{equation}
\label{DistDel1}
\Delta(l)=0 \quad\text{unless}\ l=1
\end{equation}
and express the time evolution in units such that $\tau_0=1$. This corresponds for the various time scales to
\begin{equation}
\label{tau012}
\tau_0=\tau_1=1\quad,\qquad\tau_2=j/2=10
\end{equation}
We represent the Wigner transform $W(\vec n)$ which is a real function over the unit sphere $\mathcal{S}_2$ as a real function $W(z)$ over the complex plane $\mathbb{C}$ using the stereographic projection
\begin{equation}
\label{vecnexpl}
\vec n=(\sin\theta\cos\phi,\sin\theta\sin\phi,\cos\theta)\to z=x+\imath y=r\emath^{\imath\phi}\quad,\ r=2\arctan(\theta/2)
\end{equation}

\subsubsection{Evolution of random states}
\label{ssEvRaSt}
\begin{figure}[t]
\label{fRandSt}
\begin{center}
\ifincmov
\includemovie[
autoplay,toolbar,
mouse=true,
text=(included movie),
poster="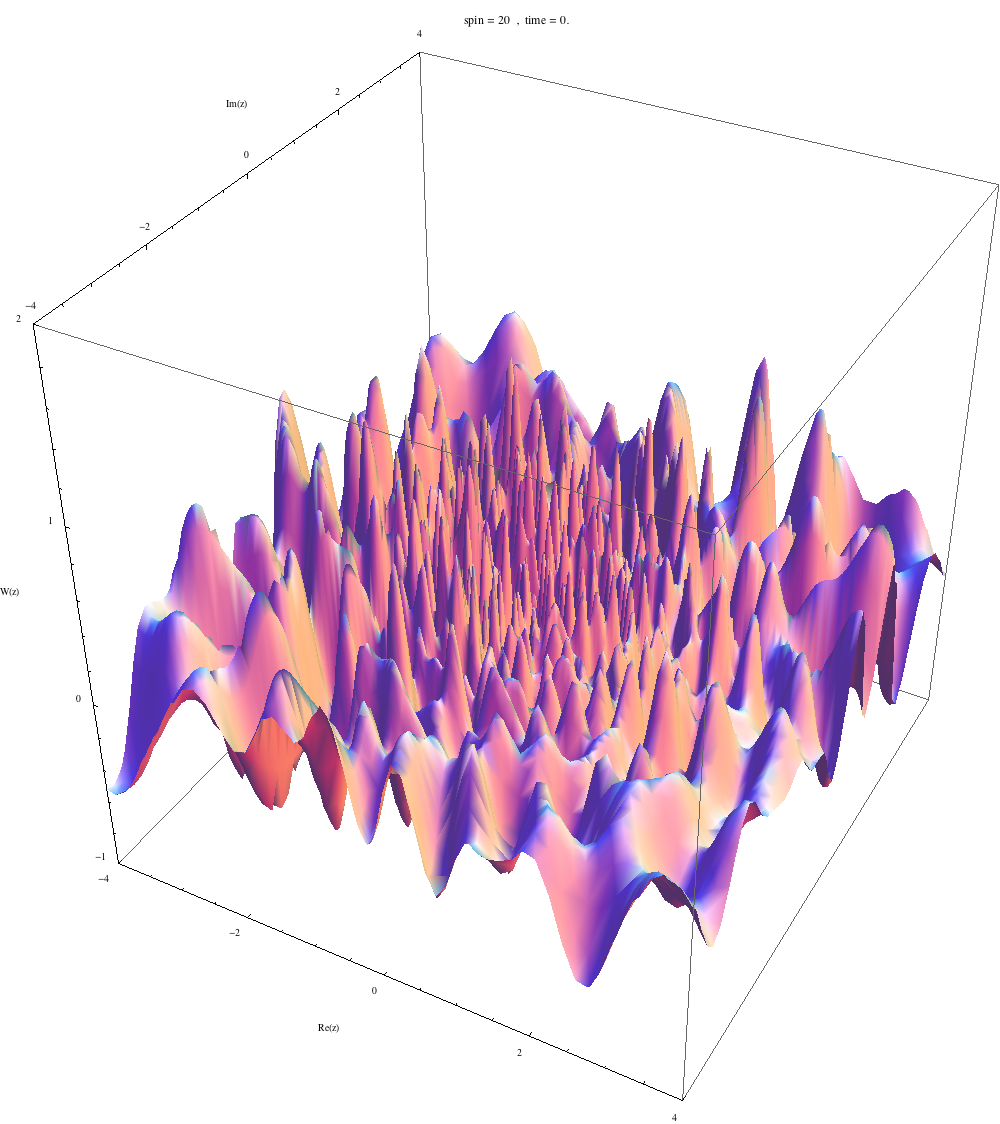"
%]{12cm}{12cm}{decoherence-9.mp4} 
]{12cm}{12cm}{anc/decoherence-9.mp4} 
\else
\includemovie[
autoplay,toolbar,
url,
mouse=true,
text=(online movie),
poster="WignerRandomState.png"
%]{12cm}{12cm}{decoherence-9.mp4} 
]{12cm}{12cm}{http://ipht.cea.fr/Pisp/francois.david/files/movies/decoherence-9.mp4} 
\fi
\medskip
\caption{Evolution of a random  state for spin $j=20$. The Wigner distribution $W(\vec n)$ is represented using stereographic projection of the complex plane (this is a .mp4 embedded file, use Adobe Reader® and click on the image to see the video).}
%\label{fRandSt}
\end{center}
\end{figure}

We start with a random pure state $|\psi\rangle\propto\sum\limits_{m=-j}^j a_m|m\rangle$, with the $a_m$ complex Gaussian independent random variables. In this case the initial $W^{(lm)}$'s are (not independent) random variables of the same order $(2j+1)^{-1}$, independently of $l$ and $m$
(except for $W^{(0,0)}=(2 j+1)^{-1/2}$).
For $0<t<\tau_1$ all the high angular frequency modes $W^{(lm)}(t)$ with $l\sim j$ undergo a fast decay, and for $t>\tau_1$ only the low frequency modes with $l\sim\sqrt{j}$ are still there, which decay at a much slower pace.
The evolution of the Wigner function is depicted in \fig{fRandSt} from $t=0$ to $t=8$, that is approximately for $0\le t\le \tau_2$. 
One sees indeed the behaviour dicussed above.

In other words, one starts from the pure quantum state $|\psi\rangle$ which can be written as a quantum superposition of  $\sim 2j+1$ semiclassical coherent states. This state evolves and  becomes for $t>\tau_1$ a statistical mixture of these coherent states, with approximately equal probabilities of order $1/(2j+1)$ (with small fluctuations of relative  order $1/\sqrt{2j+1}$ which are responsible for the long-lived remanent wiggles.

\subsubsection{Evolution of a coherent state}
\label{ssEvCoSt}
\begin{figure}[t]
\begin{center}
\ifincmov
\includemovie[
autoplay,toolbar,
mouse=true,
text=(included movie),
poster="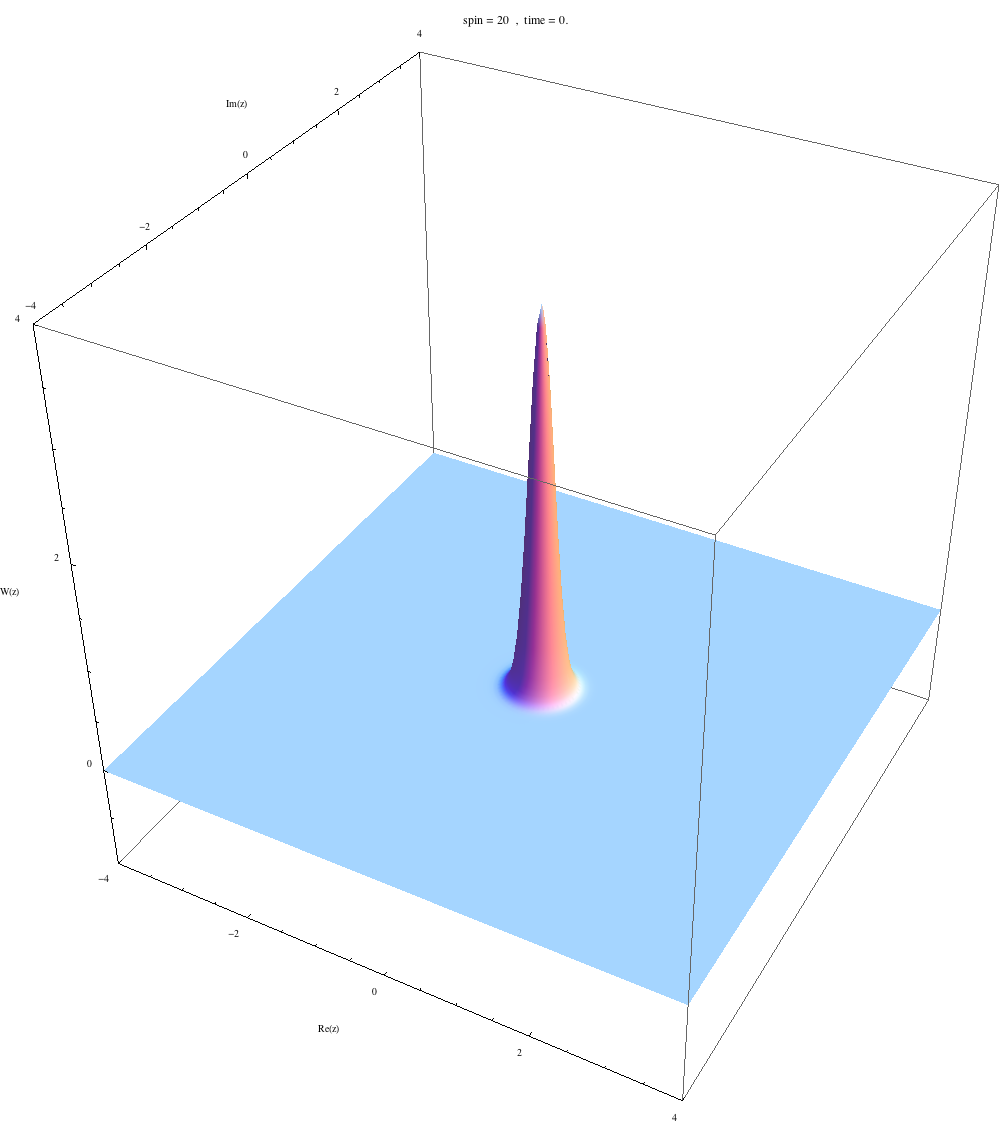"
]{12cm}{12cm}{anc/decoherence-12.mp4}
\else
\includemovie[
autoplay,toolbar,
url,
mouse=true,
text=(online movie),
poster="WignerCoherentState.png"
]{12cm}{12cm}{http://ipht.cea.fr/Pisp/francois.david/files/movies/decoherence-12.mp4}
\fi
\medskip
\caption{Evolution of a coherent state for spin $j=20$ (this is a .mp4 embedded file, use Adobe Reader® and click on the image to see the video).}
\label{fCohSt}
\end{center}
\end{figure}

On \fig{fCohSt} we plot the evolution of a coherent state $|\vec n\rangle=|\theta,\phi\rangle$, here the one centered at the origin
\begin{equation}
\label{Coher0}
|\vec e_z\rangle=|\theta=0,\phi\rangle=|j\rangle
\ .
\end{equation}
One sees immediately that its evolution in much slower and that it stays localized in phase space, as expected. 
We are working for finite $j$ so that $\tau_2$ is larger that $\tau_1$ but finite, which explain the fact that the Wigner transform widens with time.
Its evolution will be discussed in more details in \ref{ssdiffus}. 

\newpage
\subsubsection{Evolution of superpositions of coherent states}
\label{ssEvSupC}
We now look at the evolution of quantum superpositions of coherent states. 
These kind of states are often dubbed ``Schr\"odinger cat states'' or ``cat states'' in the literature.

\begin{figure}[t]
\begin{center}

\ifincmov

\includemovie[
autoplay,toolbar,
mouse=true,
text=(included movie),
poster="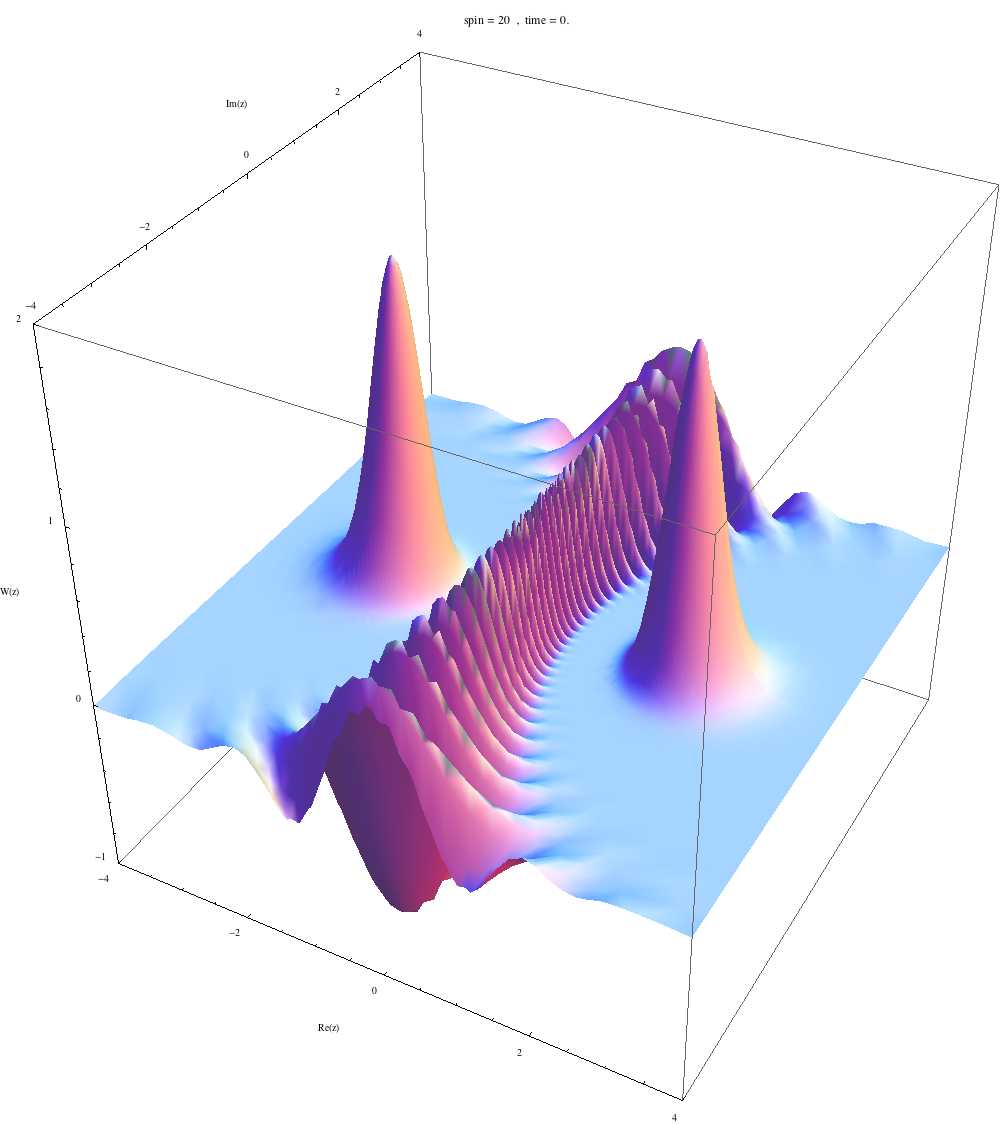"
]{12cm}{12cm}{anc/decoherence-5.mp4}

\else

\includemovie[
autoplay,toolbar,
url,
mouse=true,
text=(online movie),
poster="WignerCatTwo.png"
]{12cm}{12cm}{http://ipht.cea.fr/Pisp/francois.david/files/movies/decoherence-5.mp4}

\fi
\medskip
\caption{Evolution of a "2-states cat" for $j=20$ (this is a .mp4 embedded file, use Adobe Reader® and click on the image to see the video).}
\label{fCat2}
\end{center}
\end{figure}

\paragraph{2 state cats:}
On \fig{fCat2} we start from a simple superposition of the two opposite (hence orthogonal) coherent states. 
\begin{equation}
\label{2staCat}
|\psi\rangle={1\over \sqrt{2}}\left(|\pi/2,0\rangle+|\pi/2,\pi\rangle\right)
\end{equation}
At time $t=0$ the Wigner transform shows the two peaks associated to the two coherent states and the strong interference fringes along the $\phi=\pm\pi/2$ line (a great circle on the sphere), which contains the information about the superposition between the two coherent states.

As time increases the two coherent peaks evolves slowly, while the interference fringes disappear very quickly with a characteristic time of order $\tau_1$.
We thus see explicitely the decoherence induced by the couping of the spin with the environement, which transforms the quantum superposition of the two coherent states into a statistical mixture (here with equal probabilities) of these two states.

\begin{figure}[t]
\begin{center}

\ifincmov

\includemovie[
autoplay,toolbar,
mouse=true,
text=(included movie),
poster="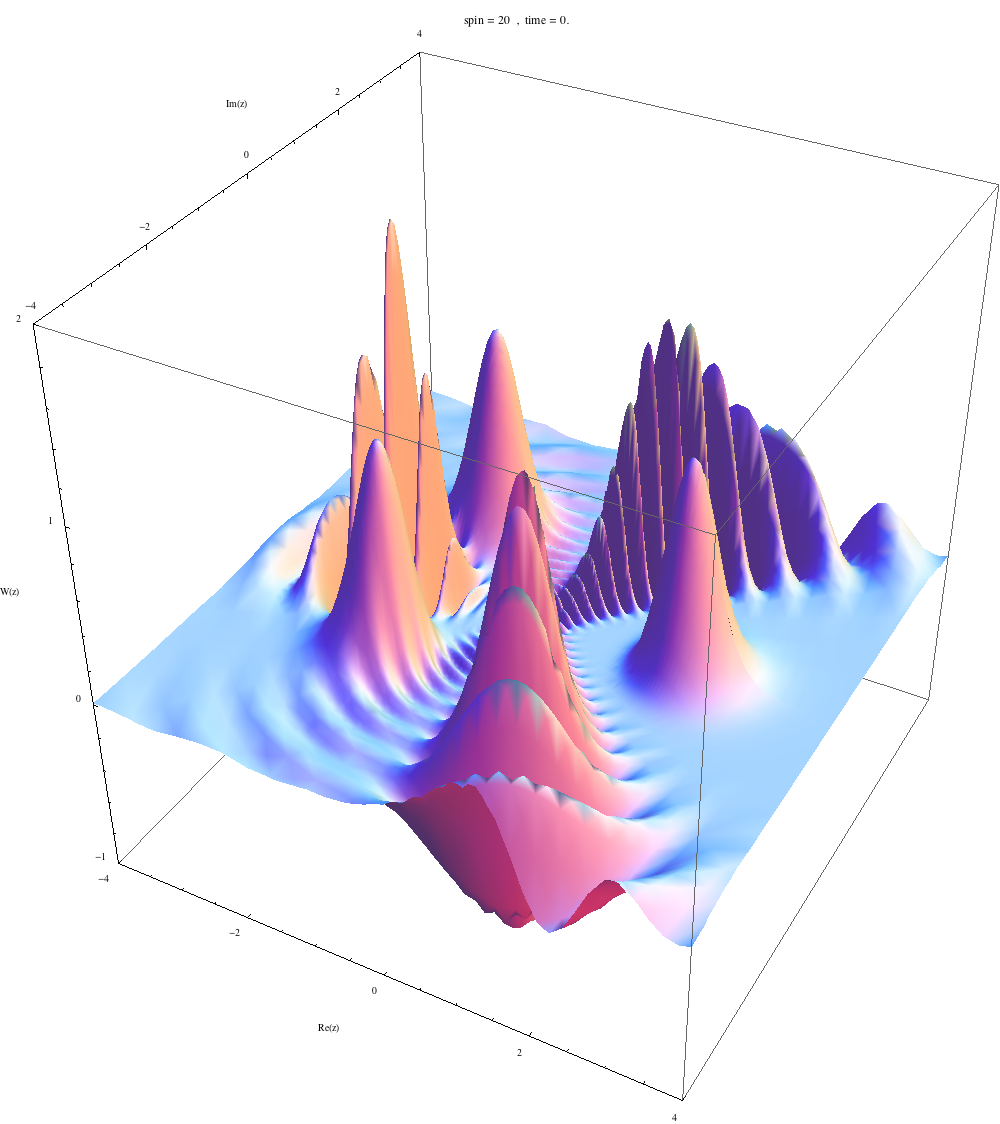"
]{12cm}{12cm}{anc/decoherence-10.mp4}

\else

\includemovie[
autoplay,toolbar,
url,
mouse=true,
text=(online movie),
poster="WignerCatThree.png"
]{12cm}{12cm}{http://ipht.cea.fr/Pisp/francois.david/files/movies/decoherence-10.mp4}

\fi 
\medskip
\caption{Evolution of a "3-states cat"  for $j=20$ (this is a .mp4 embedded file, use Adobe Reader® and click on the image to see the video).}
\label{fCat3}
\end{center}
\end{figure}

\paragraph{3 state cats:}
On \fig{fCat3} we start from a more complex superposition of three coherent states. 
\begin{equation}
\label{3staCat}
|\psi\rangle={1\over \sqrt{3}}\left( |\pi/2,0\rangle + |\pi/2,3\pi/4\rangle +|\pi/2,5\pi/4\rangle \right)
\end{equation}
At time $t=0$ the Wigner transform shows the tree peaks associated to the three coherent states and the interference fringes in between, which contains the information about the superposition between the three coherent states.
The rightmost peak corresponds to the first coherent state $|\pi/2,0\rangle$, and the leftmost pair of peaks corresponds to the second and third coherent state $ |\pi/2,3\pi/4\rangle$ and $|\pi/2,5\pi/4\rangle$. These two states are closer from each other than from the first one.

%\newpage

As time increases the three coherent peaks evolves slowly, and the interference fringes disappear quickly.
One sees that the coherences between the close second and thirst states decay more slowly that the coherences between these states and the first one, as expected from \eq{tdecthet}.

%\newpage
\subsection{Evolution of coherent states and quantum diffusion}
\label{ssdiffus}
We now discuss in more detail the dynamics of a single coherent state at large times $t>\tau_1$.
We concentrate on the semiclassical limit where $j\gg1$.
Then the modes for a single coherent state are such that $l\sim\sqrt{j}\ll j$, the corresponding function $Z(l)$ is always $Z(l)\simeq 1$ and we may approximate the evolution function $\widehat{\mathcal{M}}^{(l)}(t)=M(t,Z(l))$ by its universal form when $1-Z(l)$ is small
\begin{equation}
\label{PsiRap}
\widehat{{\mathcal{M}}}^{(l)}(t)\simeq \Psi\left(t\ \frac{4}{3\pi}\frac{D_0}{\tau_0}\frac{l(l+1)}{j(j+1)}\right)
\end{equation}
with $\Psi(t)$ given by \eq{Psi1} and \ref{Psi2}.

Using the asymptotics \eq{WlmCas} for the modes of a coherent state at large $j$, and starting at time $t=0$ from the coherent state $|j\rangle$
of \eq{Coher0}, the harmonics of the Wigner transform of the density matrix read explicitely
\begin{equation}
\label{WCSas}
W^{(l)}(t)=\,\frac{2l+1}{\sqrt{2j+1}}\exp\left({-\frac{l^2}{2j}}\right)
\Psi\!\!\left(\!t\, \frac{4}{3\pi}\frac{D_0}{\tau_0}\frac{l(l+1)}{j(j+1)}\right)
\end{equation}
There are two very different time regimes.

\subsubsection{Small time $t<\tau_2=\tau_{\mathrm{diff}}$ regime}
\label{ssReg2}
For the relevant modes $l\sim \sqrt{j}$, the variable $t'=t\, \frac{4}{3\pi}\frac{D_0}{\tau_0}\frac{l(l+1)}{j(j+1)}$ for $\psi$ is still very small, $t'\ll 1$, and  in  \eq{WCSas} $\Psi(t')\simeq 1$. 
  \begin{equation}
\label{ Wltt0}
W^{(l)}(t)=W^{(l)}(0)=\frac{2l+1}{\sqrt{2j+1}}\exp\left({-\frac{l^2}{2j}}\right)
\end{equation}
The coherent state does not evolve yet.
The Wigner distribution in phase space $W(\vec n,t)$ given by \ref{WWY} is a Gaussian distribution with width 
$$\Delta_{\theta}\simeq 1/\sqrt{j}\ .$$
When $t\sim\tau_2$  the two terms $W^{(l)}(0)$ and $\Psi(t')$ are of the same order and the Wigner transform $W(\vec n,t)$ starts to widen in phase space. This is the regime we are seeing in \fig{fCohSt} since $j$ is not very large.

\subsubsection{Large time $t\gg \tau_2=\tau_{\mathrm{diff}}$ regime} 
\label{ssReg3}
When $t\gg\tau_2$ and $l$ is small $l\sim\sqrt{j}$ the $\Psi(t')$ function dominates over the initial $\exp(-l^2/2j)$ term.
The coefficients of the Wigner distribution take the form
\begin{equation}
\label{Wltt2}
W^{(l)}(t)\simeq\frac{2l+1}{\sqrt{2j+1}}\  \Psi\left(t\ \frac{4}{3\pi}\frac{D_0}{\tau_0}\frac{l(l+1)}{j(j+1)}\right)
\end{equation}
This gives a Wigner distribution $W(\vec n,t)$ in phase space (the unit sphere) with a width
\begin{equation}
\label{DelThet}
\Delta_\theta(t) \simeq \sqrt{{t\over\tau_2\,j}}
\end{equation}
 %$$\Delta\theta(t) \simeq \sqrt{(t/\tau_2)}/\sqrt{j}$$
 The width of the distribution growths with the time $t$ as $\sqrt{t}$, thus the evolution of the spin in this semiclassical regime is a diffusion-like process.
 But as we shall see, the distribution in phase space is not exactly a Gaussian.
 
 \subsubsection{Final time $ t\gg \tau_{\mathrm{equ}}=j\tau_2$ regime}
 \label{ssReg4}
Finally, and as expected, there is a final equilibration time
\begin{equation}
\label{ }
\tau_3=j\tau_2=\tau_{\mathrm{equ}}
\end{equation} 
which corresponds to the equilibration time for the spin.
When $t\gg\tau_3=j\tau_2$, only the $l=0$ mode survives, the quantum diffusion has completely averaged the spin over the phase space and we obtain a rotationally invariant complete statistical mixture of states 
\begin{equation}
\label{Wfinal}
W(\vec n,t)={1\over \sqrt{4\pi\,(2j+1)}}
\end{equation}
It corresponds to the canonical distribution since the Hamiltonian for the spin is $H_\mathcal{S}=0$.
We remind that from Sect.~\ref{sWiHu} in our normalizations for the Wigner distribution  $$\int_{\mathcal{S}_2} d^2\vec n\  W(\vec n,t)\,=\sqrt{{4\pi\over 2j+1}}$$

 \subsection{Quantum diffusion and non-Markovianity}
 \label{ssNonMark}
 \subsubsection{The spin density profile}
 \label{ssSpinDens}
 We now look more closely at the diffusive regime $\tau_2\ll t\ll\tau_3$.
Since the spin $j$ is assumed to be large, a coherent state evolves into a mixed state whose Wigner distribution $W(\vec n,t)$ is still localized in phase space, with a width larger than the size of a quantum coherent state $\sqrt{j}$, but much smaller than the size of phase space, of order unity.

We perform the rescaling of  the spin components 
\begin{equation}
\label{rescS2}
\vec n=(\vec u,\sqrt{1-\vec u^2})\ ,\ \ \vec u=\sqrt{2j+1}\,\vec z
%\ ,\quad \vec n'=\simeq(x,y,1)=(\vec z,1)\quad\text{with}\ z=x+\imath y
%\quad\text{and}\ t\to t'=t/\sqrt{j}
\end{equation}
so that the curvature of phase space becomes negligible, and phase space becomes the flat complex plane $\mathbb{C}$ with coordinate $z=x+\imath y$ where $\vec z=(x,y)$.
We perform a similar rescaling for the time
\begin{equation}
\label{rescT}
t=t'\tau_2={\tau_2 j\over D_0}t'
\end{equation}
and we normalize the Wigner distribution so that
\begin{equation}
\label{ }
\int_\mathbb{C} d^2\vec z \ W(\vec z,t')=1
\end{equation}
In this large spin $j$ and not too large time $t$ regime, the phase space is similar to that of an harmonic oscillator and our results can be compared to those of decoherence models for an harmonic oscillator, see in particular 
\cite{BraunHaakeStrunz2001,StrunzHaakeBraun2002,StrunzHaake2003}.

From \ref{Wltt2} we find that in this limit the Wigner distribution  is simply the Fourier transform of the universal decoherence function $\Psi$
 \begin{align}
\label{Wquantum}
W(\vec z,t')&
=(2\pi)^{-2}\int d^2\vec\ell\ 
\Psi\left(
{8\over 3\pi} t'\,{|{\vec{\ell}\,}|}^2
\right)\,
\emath^{\imath\vec\ell\vec z}
\nonumber\\
&=
{3\over 8\pi} {1\over t'}\int_{-\pi/2}^{\pi/2}d\theta\,\cos(\theta)\,\exp\left(-{|\vec z|^2\over t'\,\cos(\theta)}{3\pi\over 16}\right)
\end{align}
This result is valid of course as long as the width of the distribution is large but smaller that $\sqrt{j}$ (the radius of the rescaled sphere), i.e. for time scales such that
\begin{equation}
\label{tt2t3}
\tau_2\ll t\ll j\tau_2=\tau_3
\end{equation}
In this quantum diffusive regime the Wigner distribution takes the universal scaling form
\begin{equation}
\label{WquaSca}
W(\vec z,t)={1\over t}W_{\mathrm{quantum}}(r)\ ,\ \ r=|\vec z]/\sqrt{t}
\end{equation}

\begin{figure}[t]
\begin{center}
\includegraphics[width=4in]{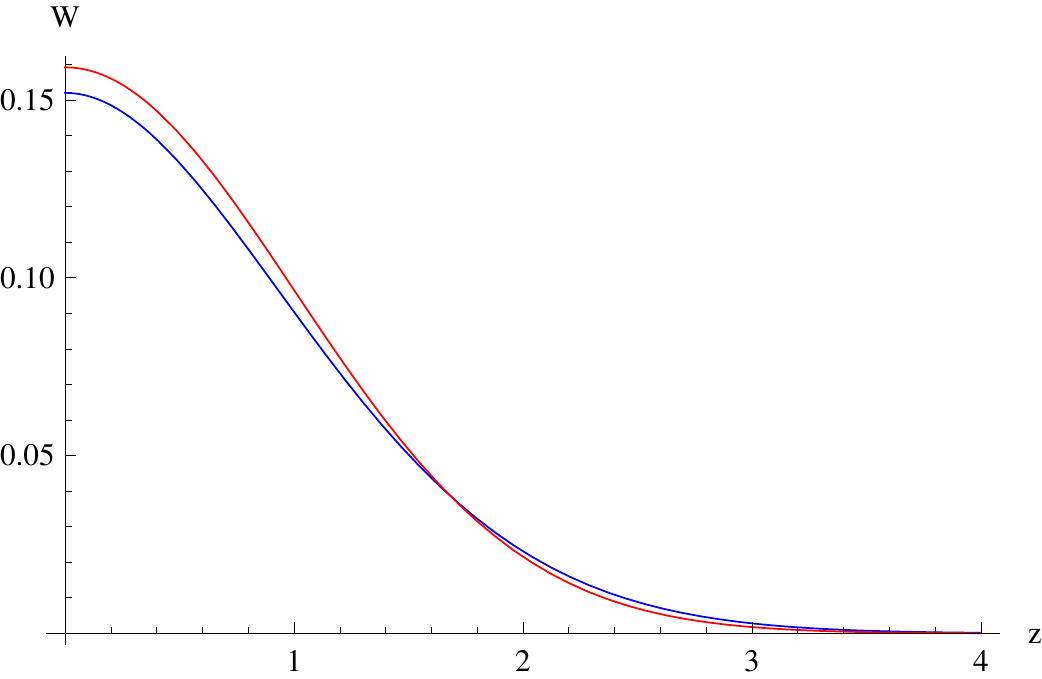}
\caption{The quantum phase space density profile $W_{\mathrm{quantum}}(z)$ {(blue curve)} versus the classical one $W_{\mathrm{classical}}(z) $ (red curve) in the large $j$ limit. Normalization is such that the variance is the same$\langle |z|^2\rangle=1$. 
There is a small but definite difference between the classical and the quantum cases.}
\label{WqvsWc}
\end{center}
\end{figure}

This universal function $W_{\mathrm{quantum}}$ is plotted on \fig{WqvsWc}, together with the standart Gaussian which represents the density profile associated to a classical Markovian local diffusion process (Wiener process) in phase space.
\begin{equation}
\label{WclaWie}
W_{\mathrm{classical}}(\vec z,t)=\frac{1}{2\pi\,t}\exp\left(-|\vec z|^2/(2t)\right)
\end{equation}

\subsubsection{Non-Markovianity}
\label{ssNonMarkov}
 It is interesting to notice that there is a small but definite difference between the two curves.
Thus, once the decoherence caused by the coupling of the spin to the external system $\mathcal{E}$ has absorbed the quantum correlations between the coherent states (and reduced in practice the quantum phase space to the semiclassical phase space, the sphere $\mathcal{S}_2$), the coupling between the spin $\mathcal{S}$ and $\mathcal{E}$ still affects dynamics of the spin.
This effective dynamics is a quantum  diffusion process, not exactly a classical diffusion. Non trivial quantum correlations and quantum memory effects induced by the dynamics of $\mathcal{E}$ and the couplings are present at all time scales, resulting in the non trivial and non classical diffusion profile given by \eq{Wquantum}.

Indeed, the self-similar profile $W(\vec z,t)$ given by \eq{Wquantum} can \emph{never} be the solution of a homogeneous evolution equation local in time of the form
\begin{equation}
\label{ }
{d\over dt}W(\vec z,t)\ = \mathcal{M}_{\vec z}\,W(\vec z,t)\quad,\qquad\mathcal{M}_{\vec z}\  \text{differential operator in}\ \vec z
\end{equation}
(as would be the case if it could be written (at least in this limit $t\gg\tau_2$) as a quantum master equation of a Lindblad form).
The evolution of the spin at times larger than the decoherence time $\tau_1=\tau_{\mathrm{dec}}$ is described by a universal self similar but non-Markovian quantum diffusion process.
As will be discussed more in section~\ref{ssCompLitt}, such a non Markovian behavior is not so surprising. Non Markovian relaxation processes have been already observed and discussed in two level systems.

This diffusion is universal in the following sense. It does not depend on the precise choice of the parameters $\Delta(l)$ for the amplitude of the couplings in the different spin channels $l$, only from the fact that the couplings belong to a SU(2)$\times$U($N$) GUE ensemble, and that they are important only for small $l$'s (in our case for a finite number of channels $l\le l_0$). 
However the diffusion process cannot really depend on the internal dynamics of the environment $\mathcal{E}$ (since this dynamics is given by the $l=0$ channel hence is GUE), nor of the initial state taken for $\mathcal{E}$, since we take a random state $|\psi_{\mathrm{random}}\rangle\in\HaE$.

%Here the diffusion does not depend either of the internal dynamics of the environment $\mathcal{E}$ (since this dynamics is given by the $l=0$ channel hence is GUE), or of the initial state taken for $\mathcal{E}$, since we take a random state $|\psi_{\mathrm{random}}\rangle\in\HaE$.

%As will be discussed more in section~\ref{ssCompLitt}, such a non Markovian behavior is not so surprising. Non Markovian relaxation processes have been already observed and discussed in two level systems.

As we shall show in section~\ref{ssfast}, a Markovian behavior emerges only in the limit where the dynamics for $\mathcal{E}$ becomes very fast, and when the initial state is chosen to be an energy eigenstate of $H^{(0)}$ for $\mathcal{E}$.

\subsubsection{Spin relaxation}
\label{ssSpinRelax}
Finally, as a simple application, let us compute the long time evolution of the spin component along one axis (here the axis $z$), assuming one starts from an initial spin state $|\psi_\mathcal{S}\rangle$ such that $S_z(0)=\langle \psi_\mathcal{S}| S^3|\psi_\mathcal{S}\rangle\neq 0$. From
\begin{equation}
\label{Szoft}
S_z(t)=\langle S^3\rangle(t)= \tr\left(\rho_\mathcal{S}(t) S^3\right)
\end{equation}
and averaging over the $H_\mathrm{int}$ ensemble we get simply the decoherence function for the $l=1$ mode
\begin{equation}
\label{Szoftav}
\overline{S_z}(t)=\widehat{\mathcal{M}}^{(1)}(t)\ S_z(0)
\end{equation}
where (for large $j$)
\begin{equation}
\label{M1oft}
\widehat{\mathcal{M}}^{(1)}(t)
\simeq \Psi\left(t{4\over 3 \pi}{D_0\over \tau_\mathrm{dyn}}{2\over j(j+1)}\right)\simeq
\Psi\left({t\over \tau_\mathrm{equ}}{8\over 3\pi}\right)
\end{equation}
with $\Psi(t)$ the function defined in section~\ref{z1tz1}.
So the magnetization relaxes to its equilibrium value $0$ with a typical time scale $\tau_{equ}=\tau_3$, as expected. However, the relaxation is not exponential (as this would be the case in the quantum diffusion process were Markovian), but only algebraically as
\begin{equation}
\label{Szdecay}
\overline{S_z}(t)\ \propto\ {1\over t^3}\quad,\qquad t\to\infty
\end{equation}

\section{External dynamics for $\mathcal{E}$}
\label{sExtDyn}
Up to now, and for simplicity, we have treated the internal dynamics of $\mathcal{E}$ exactly on the same footing than the coupling between $\mathcal{S}$ and $\mathcal{E}$. In particular the Hamiltonian $H^{(0)}$ (corresponding to the $l=0$ sector of $H$) was chosen to be a random Hamiltonian in a  GUE ensemble, its distribution had the same $U(N)$ invariance than for the $l>0$ sectors. 
No particular attention was given to the initial state $|\psi_{\mathrm{init}}\rangle$ for $\mathcal{E}$, which was taken to be a random state.
 
\subsection{External Hamiltonian for $\mathcal{E}$}
\label{sExtHam}
Now the Hamiltonian $\HaE$ for the system $\mathcal{E}$, i.e. the $l=0$ component  $H^{(0)}$ of the Hamiltonian $H_{\mathcal{SE}}$
\footnote{\ Strictly speaking $H^{(0)}=\mathbf{1}_{\mathcal{S}}\otimes\HaE$}, is taken to be a fixed (not necessarily GUE) Hamiltonian with a regular normalized state density in the large $N$ limit. We define the density of states $\rho(E)$ and the normalized density of states $\nu(E)$ by
\begin{equation}
\label{nu2E}
\nu(E)=\lim_{N\to\infty}\frac{1}{N}\rho(E)
\quad,\qquad \rho(E)=\sum_{a=1}^N \delta(E-E_\alpha)
\end{equation}
and assume that $\nu(E)$ is finite and regular.
$E_\alpha$ are the eigenvalues of the Hamiltonian $H^{(0)}$, considered as a $N\times N$ matrix acting on $\mathcal{H}_{\mathcal{E}}=\mathbb{C}^N$. 
In the calculations we choose as a basis $\{|\alpha\rangle,\,\alpha=1,\cdots N\}$ of $\mathcal{H}_{\mathcal{E}}$ the eigenstates 
%$|a\rangle=|E_a\rangle$
 of $H^{(0)}$.
\begin{equation}
\label{Ealpha}
|a\rangle=|E_a\rangle
\quad,\qquad H^{(0)}|E_\alpha\rangle=E_\alpha |E_\alpha\rangle
\end{equation}
For the $\mathcal{S}+\mathcal{E}$ interaction (the $l\neq 0$ sectors) we keep a random Hamiltonian $H'$ in the SU(2)$\times$U($N$) GUE ensemble, characterized by the 
variances
\begin{equation}
\label{Deltbpr}
\bold{\Delta'}=\{\Delta(l),\,l=1,\cdots 2j+1\}
\end{equation} 
 
 \subsection{General form of the influence functional}
 \label{sInfFEx}
We can then repeat the calculation for the influence functional.
We shall be interested in the evolution of the spin  $\mathcal{S}$, starting from a initial separable state of the form
\begin{equation}
\label{rhoE}
\rho(0)=\rho^{\mathcal{S}}(0)\otimes |E_\alpha\rangle\langle E_\alpha|
\end{equation}
i.e. an arbitrary spin state for $\mathcal{S}$ times a given pure energy eigenstate for $\mathcal{E}$.
One can choose a more general fixed initial state $|E\rangle$ for $\mathcal{E}$, neither random nor a pure eigenstate of $\HaE$, but thanks to the $U_N$ invariance of the ensemble of the $\HaSE$ it is in fact sufficient to consider the previously introduced case.

The influence functional is now a tensor $\mathcal{M}(t,E)$ depending on $t$ and $E$
\begin{equation}
\label{M2tE}
%\mathcal{M}(t,E)
%\quad,\qquad
\rho^{\mathcal{S}}_{ru}(t)=
\overline{\tr_{\mathcal{E}}\left( e^{-\imath t H}(\rho^\mathcal{S}(0)\otimes |\alpha\rangle\langle \alpha|) e^{\imath t H}\right)}_{ru}
=\mathcal{M}_{ru,st}(t,E_\alpha)\rho^\mathcal{S}_{st}(0)
\end{equation}
with $H=H^{(0)}+H'$.
We shall compute this functional in the large $N$ planar limit.

We better consider the double resolvent
\begin{equation}
\label{G2rsab}
\mathcal{G}_{ru,st}^{\alpha,\beta} (x,y) = \overline{\langle r\alpha|\frac{1}{x-H}|s\beta\rangle\langle t\beta |  \frac{1}{y-H} | u \alpha\rangle}
\end{equation}
and the single resolvent
\begin{equation}
\label{H2rsab}
\mathcal{H}_{r,s}^{\alpha,\beta} (x) = \overline{\langle r\alpha|\frac{1}{x-H}|s\beta\rangle}
\end{equation}
Obviously, when there is no coupling between $\mathcal{S}$ and $\mathcal{E}$,  this resolvent is simply
\begin{equation}
\label{Hrsab0}
{\langle r\alpha|\frac{1}{x-H^{(0)}}|s\beta\rangle}=\delta_{r,s}\,\delta_{\alpha,\beta} \, \frac{1}{x-E_\alpha}
%H^{(0)}_\alpha(x)
\end{equation}

When there is a coupling, i.e. when the $\Delta(l)$'s are non zero, thanks to the SU(2) invariance of the distribution ensemble for the $H'$ and the fact that the $|\alpha\rangle$ are eigenstates of $H^{(0)}$
we can show that $\mathcal{H}_{r,s}^{\alpha,\beta} (x)$ is still of this form
\begin{equation}
\label{HrsabC}
\mathcal{H}_{r,s}^{\alpha,\beta} (x)=\delta_{r,s}\,\delta_{\alpha,\beta} \, \widetilde C_\alpha(x)
\end{equation}
In the large $N$ limit, $\widetilde C_\alpha(x)$ satisfies the recurrence equation (which generalizes the recurrence equation \ref{Heq} for $\mathcal{H}(x)$ )
\begin{equation}
\label{CtaxEq}
\widetilde C_\alpha(x)=\frac{1}{x-E_\alpha}+\frac{\widehat{D}}{x-E_\alpha}\widetilde C_\alpha(x)\,\left(\sum_{\beta=1}^N\widetilde C_\beta(x)\right)
\end{equation}
where 
\begin{equation}
\label{DhatRec}
%\widehat{\Delta}(0)=
\widehat{D}=\sum_{l=1}^{2j} \frac{2l+1}{2j+1}\,\Delta(l)
%\quad,\qquad
%\tilde\Delta(l)=N \Delta(l)
\end{equation}

Thanks to the SU(2) invariance, we can easily show that the double resolvent  $\mathcal{G}$ is diagonal. 
Its double Wigner transform decomposition takes the form
\begin{align}
\label{GnewHat}
\left.{W_{\!\widehat{\mathcal{D}}}}\right._{(l_1,m_1),(l_2,m_2)}^{\alpha,\beta}&=
\delta_{l_1,l_2}\,\delta_{m_1+m_2,0}\, {(-1)}^{m_1}\,  \left.{\widehat{\mathcal{G}}}\right._{(l_1)}^{\alpha,\beta}(x,y)
\end{align}
$\left.{\widehat{\mathcal{G}}}\right._{(l)}^{\alpha,\beta}(x,y)$ is given by a planar recursion equation similar to  \eq{RecGl} for $\left.{\widehat{\mathcal{G}}}\right._{(l)}(x,y)$. 
Its solution is
\begin{equation}
\label{GnHatExp}
\left.{\widehat{\mathcal{G}}}\right._{(l)}^{\alpha,\beta}(x,y)\ =\ \delta_{\alpha,\beta}\, \widetilde{C}_\alpha(x)\,\widetilde{C}_\alpha(y)
+\frac{\widehat{D}(l)\ \widetilde{C}_\alpha(x)\widetilde{C}_\alpha(y)  \widetilde{C}_\beta(x)\widetilde{C}_\beta(y)}{1-\widehat{D}(l)
\left(\sum\limits_{\gamma=1}^{N}\widetilde{C}_\gamma(x)\widetilde{C}_\gamma(y)\right)}
\end{equation}
where the $\widehat{D}(l)$ are defined, as in \eq{HatDelOfL} (same SU(2) structure) by
\begin{equation}
\label{DhatN}
\widehat{D}(l_1)=\sum_{l'=1}^{2j}\Delta(l') (2l'+1) (-1)^{2j+l'+l_1}
\sixj{j}{j}{l'}{j}{j}{l_1}
\end{equation}
Note however that in \eq{DhatN} the sum over $l$ excludes the $l=0$ case, contrarily to the sum in \eq{HatDelOfL} which defines the $\widehat{\Delta}(l)$'s.

Since we are only interested in taking the trace over the final states $|\beta\rangle$ we simply have to consider
\begin{equation}
\label{FalxyEq}
\left.{\widehat{\mathcal{F}}}\right._{(l)}^{\alpha}(x,y)\ =
\sum_{\beta=1}^N\left.{\widehat{\mathcal{G}}}\right._{(l)}^{\alpha,\beta}(x,y)\ =\ 
\frac{\widetilde{C}_\alpha(x)\widetilde{C}_\alpha(y)  }{1-\widehat{D}(l)
\left(\sum\limits_{\gamma=1}^{N}\widetilde{C}_\gamma(x)\widetilde{C}_\gamma(y)\right)}
\end{equation}
It is of course natural in the large $N$ limit to rexpress the sum over states of $\mathcal{E}$ as a continuum integral over the spectrum  of $H^{(0)}$.
\begin{equation}
\label{ContSpectLim}
\sum_{\alpha=1}^N\ \to\ N\int dE \ \nu(E)
\quad,\quad
\widetilde{C}_\alpha(x)\ \to\ \widetilde{C}(x,E_\alpha)
\quad,\quad
\widetilde{F}^\alpha_{(l)}(x,y)\ \to\ \widetilde{F}_{(l)}(x,y,E_\alpha)
\end{equation}
\eq{CtaxEq} becomes
\begin{equation}
\label{CCtilde}
\widetilde{C}(x,E)=\frac{1}{x-E-\widehat{\Delta}'\,\widetilde{C}(x)}
\quad,\qquad
\widetilde{C}(x)=\int dE\ \nu(E)\,\widetilde{C}(x,E)
\end{equation}
with
\begin{equation}
\label{DeltPrime}
\widehat{\Delta}'=N\,\widehat{D}=\sum_{l=1}^{2j} \frac{2l+1}{2j+1}\,\widetilde{\Delta}(l)
\quad,\qquad
\widetilde\Delta(l)=N \Delta(l)
\end{equation}
while \eq{FalxyEq} becomes
\begin{equation}
\label{FElxyEq}
 \widetilde{\mathcal{F}}_{(l)}(x,y,E)\ =\ \frac{\widetilde{C}(x,E)\widetilde{C}(y,E)}{1-\widehat{\Delta}'(l)\int dE'\, \nu(E')\,\widetilde{C}(x,E')\widetilde{C}(y,E')}
 \end{equation}
 where
 \begin{equation}
\label{FElxyEq2}
% \ ,\quad
 \widehat{\Delta}'(l)=N\,\widehat{D}(l)
 =\sum_{l'=1}^{2j}\widetilde\Delta(l') (2l'+1) (-1)^{2j+l'+l_1}
\sixj{j}{j}{l'}{j}{j}{l_1}
\end{equation}
The notation $\widehat\Delta'$ and $\widehat\Delta'(l)$ (with a tilde) in the definitions of \eq{DeltPrime} and \eq{FElxyEq2} is here to recall that there is no $l=0$ contribution in the sum over the $l$'s, contrary to the definition for $\widehat\Delta(l)$ and $\widehat\Delta(0)$ given by \eq{HatDelOfL} and \eq{HatDelOf0}.
We have obviously
\begin{equation}
\label{WhDp0}
 \widehat{\Delta}'(0)= \widehat{\Delta}'
\end{equation}

As in the previous situation, the evolution of the reduced density matrix factorizes into each $(l,m$) sector, and the influence functional becomes a single function of the time $t$, of the angular momentum $l$, and now of the initial state energy $E$. 
Thanks to the SU(2) invariance it is still independent of $m$.
It is given by the integral
\begin{equation}
\label{MhatTE}
\widehat{\mathcal{M}}^{(l)}(t,E)=\oint\frac{dx}{2\imath\pi}\oint\frac{dy}{2\imath\pi}\,\emath^{-\imath t (x-y)}\,  \widetilde{\mathcal{F}}_{(l)}(x,y,E)
\end{equation}
This can be rewritten in a simpler form by the change of variables (and its inverse)
\begin{equation}
\label{WofXofW}
w=W(x)=x-\widehat{\Delta}'\widetilde{C}(x)
\quad,\qquad
x=X(w)
\end{equation}
Indeed \eq{CCtilde} becomes
\begin{equation}
\label{CtildexxE}
\widetilde{C}(x)
=\int dE\,\frac{\nu(E)}{w-E}
\quad,\qquad
\widetilde{C}(x,E)=\frac{1}{w-E}
\end{equation}
so that $\widetilde{C}(X(w))$ is the Hilbert transform of $\nu(N)$, i.e. the resolvent of $H^{(0)}$.
Then in \eq{FElxyEq} we rewrite the integral
\begin{align}
\label{intnucc}
\int dE'\, &\nu(E')\,\widetilde{C}(x_1,E')\widetilde{C}(x_2,E')=\int dE'\,\nu(E'){1\over W(x_1)-E'}{1\over W(x_2)-E'}
\nonumber\\
&=-{\widetilde{C}(W(x_1))-\widetilde{C}(W(x_2))\over W(x_1)-W(x_2)}=
{1\over \widehat{\Delta}'}\left(-{x_1-x_2\over W(x_1)-W(x_2)}+1\right)
\end{align}
and after some algebra the evolution kernel $\widehat{\mathcal{M}}^{(l)}(t,E)$ is written in a form similar to \eq{Mint1} as an integral representation involving  the variances $\Delta(l)$'s  for the interaction Hamiltonians $H_{SE}=\sum H^{(l,m)}$ and the Hamiltonian for E, $H_E=H^{(0)}$, through the function $W(x)$ (related to the inverse of the resolvent for $\HS$), and a parameter $Z'(l)$ 
\begin{equation}
\label{MltEexpl}
\widehat{\mathcal{M}}^{(l)}(t,E)=\oint\frac{dx_1}{2\imath\pi}\oint\frac{dx_2}{2\imath\pi}\frac{\emath^{-\imath t(x_1-x_2)}}{(W(x_1)-E)(W(x_2)-E)}
\frac{1}{(1-{{Z}^{\prime}}(l))+{{Z}^{\prime}}(l) \left(\frac{x_1-x_2}{W(x_1)-W(x_2)}\right)}
\end{equation}
where
\begin{equation}
\label{ZprimeL}
Z'(l))=\frac{ \widehat{\Delta}'(l)}{ \widehat{\Delta}'(0)}
\end{equation}
with $\widehat{\Delta}'(l)$ defined through \eq{DhatN} and  \eq{FElxyEq}.

\subsection{Application to the Wigner ensemble}
\label{sAppWign}
\subsubsection{General form of the solution}
\label{ssGenFoW}
As a simple and illustrative ensemble, let us treat the case where the density spectrum of $\HaE$ is the Wigner semi-circle distribution with width $2E_0$. 
\begin{equation}
\label{nuWigner}
\nu(E)={2\over\pi\ E_0^2}\sqrt{E_0^2-E^2}
\end{equation}
This corresponds to take for $H^{(0)}$ a sample in a GUE distribution, as in the previous section, with the identification
\begin{equation}
\label{widthWig}
E_0=2\,\sqrt{\bar{\Delta}(0)}
\quad\text{where}\quad
\bar{\Delta}(0)={N\over 2j+1}\Delta(0)=\widehat{\Delta}_{\mathrm{av}}
\end{equation}
But now we are able to study the dependence of the final state on the choice of initial state $|E\rangle$.

To compare our calculations with the results of the previous section, it is convenient to perform the same rescalings.
The first  time scale $\tau_0$ is now
\begin{equation}
\label{t0New}
\tau_0={1\over \sqrt{\widehat{\Delta}(0)}}
\quad,\quad
\widehat{\Delta}(0)=
\widehat{\Delta}_{\mathrm{av}} + \widehat{\Delta}'(0)= {E_0^2\over 4} + \sum_{l=1}^{2j} \frac{2l+1}{2j+1}\,\widetilde{\Delta}(l)
\end{equation}
The $Z(l)$ parameters of \eq{Z2l} (not to be confused with the $Z'(l)$ of \eq{ZprimeL}) and the average $Z_{\mathrm{av}}$ of \eq{ZAv} are
\begin{equation}
\label{ZlNew}
Z(l)={\Deltaav+\widehat{\Delta}'(l)\over \Deltaav+\widehat{\Delta}'(0)}
\quad,\qquad
Z_{\mathrm{av}}={{\widehat{\Delta}}_{\mathrm{av}}\over  \widehat{\Delta}(0)}
={\Deltaav\over \Deltaav+\widehat{\Delta}'(0)}
\end{equation}

After performing the rescaling 
\begin{equation}
\label{bartN}
\bar t= t/\tau_0
\quad,\qquad
\bar E= E\,\tau_0
\end{equation}
%$
%t\to t/\tau_0$, $E\to E\,\tau_0
%$,
the evolution kernel $\widehat{\mathcal{M}}^{(l,m)}(t,E)$ takes the form
\begin{equation}
\label{MhtEN}
\widehat{\mathcal{M}}^{(l,m)}(t,E)=M(\bar t,\bar E,Z(l),Z_{\mathrm{av}})
%=M(t/\tau_0,E\tau_0,Z(l))
\end{equation}
The function $M(t,E,Z,Z_{\mathrm{av}})$ is given by the integral representation, which generalizes \eq{Mint1}
\begin{align}
\label{MtEZZ}
 M(\bar t,\bar E,Z(l),Z_{\mathrm{av}})   =&   \oint \frac{dX_1}{2\imath\pi}  \oint \frac{dX_2}{2\imath\pi}
%\emath^{-\imath t(X_1-X_2)} 
\frac{\emath^{-\imath \bar t(X_1-X_2)}}{(W_1-\bar E)(W_2-\bar E)}
 \frac{1-Z_{\mathrm{av}}H_1H_2} {1-Z(l) H_1 H_2}
%\nonumber\\
  %  &\quad \frac{1-Z_{\mathrm{av}}H_1H_2} {1-Z(l) H_1 H_2}
\end{align}
where as above $H_{1,2}$ are given by the function $H(X)$ (the resolvent for the Wigner ensemble)
\begin{equation}
\label{H2XN}
H_{1,2}=H(X_{1,2})
\quad,\qquad
H(X)={1\over 2}(X-\sqrt{X^2-4})\quad\text{i.e.}\ X=H+{1\over H}
\end{equation}
while  $W_{1,2}$ are given by the function $W(X)$ defined from $H(X)$ and $Z_{\mathrm{av}}$ through
\begin{equation}
\label{W2XN}
W_{1,2}=W(X_{1,2})
\quad,\qquad
\sqrt{Z_\mathrm{av}}H(X)=H\left({W(X)\over\sqrt{Z_\mathrm{av}}}\right)
\quad\text{i.e.}\ W(X)=Z_{\mathrm{av}}H(X)+{1\over H(X)}
\end{equation}

As in the previous section, $H(X)$ is analytic in $\mathbb{C}\backslash{[-2,2]}$, with the cut along the interval $X\in I_1=[-2,2]$ corresponding to the spectrum of the total Hamiltonian  $H=\HaE+\HaSE$.
Away from this cut, one has $X\in\mathbb{C}\backslash I_1\implies |H(X)|>1$.
In addition $H(W)$ is analytic in $\mathbb{C}\backslash{[-2\sqrt{Z_\mathrm{av}},2\sqrt{Z_\mathrm{av}}]}$, with the cut along the interval $W\in I_2={[-2\sqrt{Z_\mathrm{av}},2\sqrt{Z_\mathrm{av}}]}$, included in the previous cut since $Z_\mathrm{av}<1$.
Away from this cut, one has $W\in\mathbb{C}\backslash I_2\implies |H|>1\sqrt{Z_\mathrm{av}}$.
Remember also that $\bar E$ is the (rescaled) energy of the initial state $|E\rangle\in\HE$, chosen to be an eigenstate of $\HaE$, hence
$|\bar E|\le 2\sqrt{Z_\mathrm{av}}$.
Therefore the integral form for $M(\bar t,\bar E,Z(l),Z_{\mathrm{av}}) $ given in \eq{MtEZZ} is defined for a contour in $X_1$ and $X_2$ encircling the cut $I_1=[-2,2]$ and the poles $W(X)=\bar E$, or equivalently for a contour in $H_1$ and $H_2$ such that $|H|>1/\sqrt{Z_{\mathrm{av}}}$.

\subsubsection{Discussion of the solution}
\label{ssDiscSol}
Altough it depends on more parameters, the function $M(\bar t,\bar E,Z,Z_{\mathrm{av}})$ has a large $t$ behavior similar to the behavior of $M(t,Z)$ depicted in \fig{f:MtZ}.
As long as $Z<1$ it decays when $t\to\infty$ (fast decoherence).

In particular, if we choose
$Z_\mathrm{av}=0$ and $E=0$ we recover the function $M(t,z)$ given by \eq{Mint1}, depicted on \fig{f:MtZ}  and studied above.
\begin{equation}
\label{MtE0ZZ}
M(t,E=0,Z(l),Z_\mathrm{av}=0)=M(t,Z(l))
\end{equation}

\begin{figure}[h]
\begin{center}
\includegraphics[width=5in]{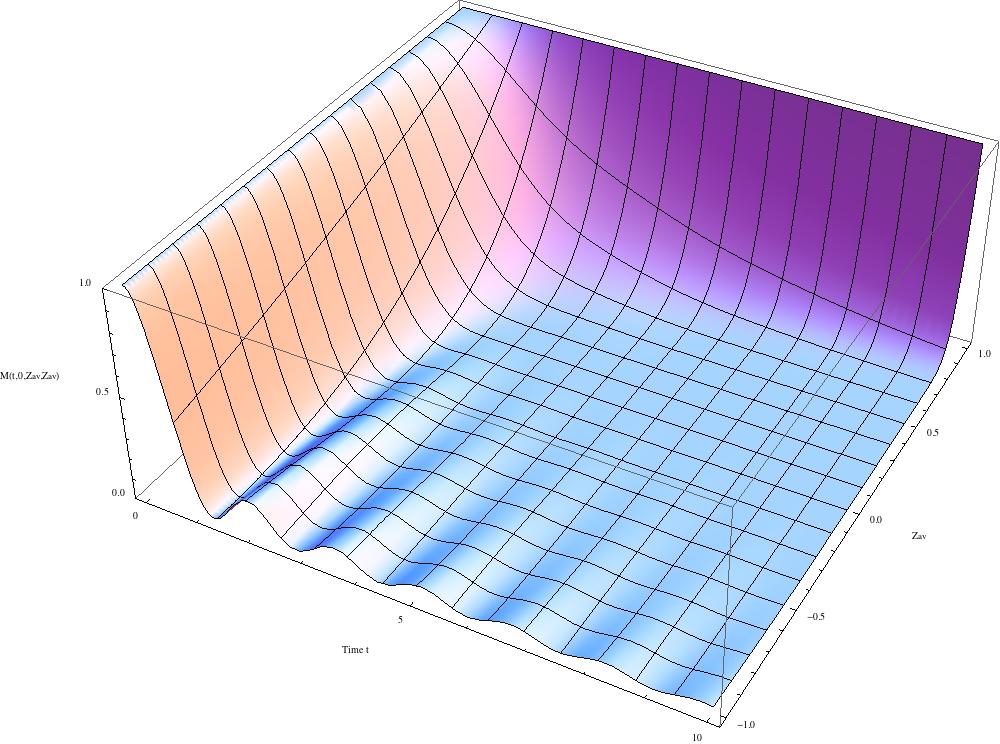}
\caption{The decoherence function $M(t,E,Z,Z_\mathrm{av})$ plotted as a function of $t$ and of $Z$ for $E=0$ and $Z=Z_\mathrm{av}$. Its behavior as a function of time $t$ and of $Z_\mathrm{av}$ is similar to that of the function $M(t,Z)$.}
\label{f:Nfunction}
\end{center}
\end{figure}

For typical $l$'s (i.e. $l\sim j$) $Z(l)\sim Z_{\mathrm{av}}$. 
If we choose
\begin{equation}
\label{ZZavE0}
Z(l)=Z_{\mathrm{av}}\quad\text{and}\quad \bar E=0
\end{equation}
we get
\begin{equation}
\label{MZavE0}
M(\bar t,0,Z_{\mathrm{av}},Z_{\mathrm{av}})=\left|N(t,Z_\mathrm{av})\right|^2
\quad,\qquad
N(t,Z_\mathrm{av})=\oint {dX\over 2\imath\pi}{\emath^{-\imath t X}\over W(X)}
\end{equation}
The function $N(t,Z)$ is easily calculated, for instance from its small $t$ and small $Z$ series expansion.
The  function $M(t,0,Z,Z)$ is depicted in \fig{f:Nfunction}.
Although not exactly the same as the decoherence function $M(t,Z)$ obtained in the previous section (see \eq{Mhat2M}) and  depicted on \fig{f:MtZ}, one sees that it is very similar. 
In particular one sees that when $Z_\mathrm{av}$ is close to 1, the decoherence becomes very slow. 

\subsubsection{Large $t$ asymtotics}
\label{ssLtWig}
This is a general feature: for fixed $Z_\mathrm{av}<1$ and fixed $E$, when $l$ is small (i.e. $l\ll j$), $Z(l)\simeq 1$ and the decoherence becomes  always very slow. 
This can be shown by reexpressing the integral representation \ref{MtEZZ} for the function $M(\bar t,\bar E, Z(l),Z_{\mathrm{av}})$ by a double contour integral over $H_1$ and $H_2$. 
The contours can be choosen to be the unit circle $H_{1,2}=\emath^{\imath \theta_{1,2}}$, and at large time $\bar t$ the integral can be evaluated by the saddle point method. The relevant saddle points are the extrema of the functions $X(H)=H+1/H$, situated at  $H_1$ and $H_2=\pm 1$.
We shall not detail the full calculation here, but one can show that the final large $\bar t$ behavior of $M(\bar t,\bar E, Z(l),Z_{\mathrm{av}})$ is of a similar form than for the function $M(t,Z(l)$ given by \eq{Mtztinf}.
\begin{equation}
\label{MtEzztinf}
M(\bar t,\bar E, Z(l),Z_{\mathrm{av}})= t^{-3}\left(A-B\sin(4t)\right)\left(1+\mathcal{O}(t^{-1})\right)
\end{equation}
The coefficients $A=A(\bar E, Z(l),Z_{\mathrm{av}})$ and $B=B(\bar E, Z(l),Z_{\mathrm{av}})$ depend in a complicated way of the spin $l$ through $Z(l)$, but now also of the energy $\bar E$ and of $Z_{\mathrm{av}}$.

Let us just give the form of these coefficients in the limit of small angular momentum $l$. This is the limit where the decoherence becomes slow and where coherent states emerges as semiclassical states. In this case we have seen that $Z(l)$ is very close to 1, and  we find that in this limit
\begin{equation}
\label{Acoeff1}
A\ \simeq\  {1\over \pi}\,{(1+Z_{\mathrm{av}})^2+E^2 \over (1+Z_{\mathrm{av}})^2-E^2} \  {(1-Z_{\mathrm{av}})\over(1-Z(l))^3}
%\ +\ \mathcal{O}((1-Z(l))^{-2})
\end{equation}
\begin{equation}
\label{Bcoeff1}
B\simeq 
%{1\over \pi}{(1+Z_{\mathrm{av}})^2+E^2 \over (1+Z_{\mathrm{av}})^2-E^2}  {(1-Z_{\mathrm{av}})\over(1-Z(l))^3}
\mathcal{O}(1)
\end{equation}
One sees in particular that in this regime the function is a simple scaling function of the variable $t(1-Z(l))$.

\subsubsection{Small $Z(l)$ scaling}
\label{sssmalZl}
One can study the regime (similar to the regime of \ref{z1tz1}) where 
\begin{equation}
\label{ZlScaling}
Z(l)\to 1\ ,\quad \bar t (1-Z(l))=t'=\mathcal{O}(1)
\end{equation}
In this limit the dominant contribution in the integral \ref{MtEZZ} is given by $H_2\simeq 1/(Z(l) H_1)$. Noting $H_1=\emath^{-\imath\theta}$ and integrating over $H_2$ we find the scaling function
\begin{equation}
\label{Mscaling1}
M_{\mathrm{scaling}}(\bar t,\bar E, Z(l),Z_{\mathrm{av}})=\int_{0}^{\pi} {d\theta\over \pi}\, {(1-\cos(2\theta))(1-Z_{\mathrm{av}})\over{( (1+Z_{\mathrm{av}})\cos\theta-\bar E)^2+(1-Z_{\mathrm{av}})}^2(\sin\theta)^2}\ \emath^{-2t'\sin\theta}
\end{equation}
For large $t'$ this integral is dominated by the contribution of the endpoints $\theta=0,\pi$, and this function scales indeed as
\begin{equation}
\label{algDec}
M_{\mathrm{scaling}}\ \simeq\  {1\over \pi}\,{(1+Z_{\mathrm{av}})^2+E^2 \over (1+Z_{\mathrm{av}})^2-E^2} \  {(1-Z_{\mathrm{av}})\over{t'}^3}
\end{equation}
However, when $1-Z_{\mathrm{av}}$ is small there is an interesting regime where this integral is dominated by the point $\theta_c$ where the denominator is small, namely
\begin{equation}
\label{theta2E}
(1+Z_{\mathrm{av}})\cos\theta_c-\bar E=0
\end{equation}
The integral can be approximated by a Lorentzian integral and one finds that the function $M_{\mathrm{scaling}}$ scales as
\begin{equation}
\label{expDec}
M_{\mathrm{scaling}}\ \mathop{\simeq}_{Z_{\mathrm{av}}\to 1}\  \exp\left({-t'\,\sqrt{4-\left.{\bar{E}}\right.^2}}\right)
\end{equation}
The crossover between the exponential decay \ref{expDec} (valid for $t'<t'_{\mathrm{cross}}$) and the algebraic decay \ref{algDec} (valid for $t'>t'_{\mathrm{cross}}$) occurs for
\begin{equation}
\label{tcrossing}
t'_{\mathrm{cross}}\sim\log(1/(1-Z_{\mathrm{av}}))=\log({\tau_1/\tau_0})
\end{equation}
%\textit{Give details? }

\subsection{The limit of fast $\mathcal{E}$ dynamics ($\tau_1\gg\tau_0$)}
\label{ssfast}
\subsubsection{Fast dynamics}
\label{ssFDyn}
\begin{figure}[h]
\begin{center}
\includegraphics[width=4in]{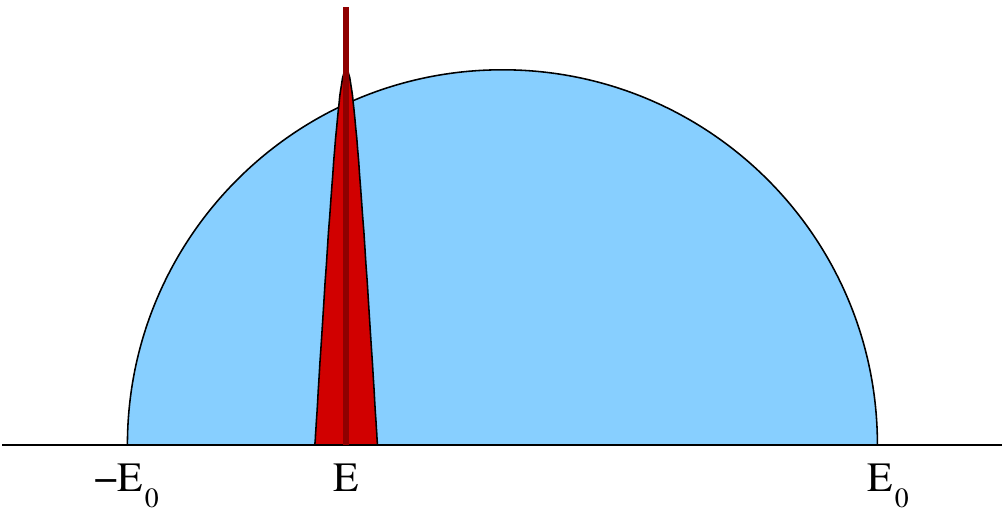}
\caption{The case of fast $\mathcal{E}$ dynamics. The energy spectrum density $\nu(E)$ of $\HaE$ is depicted in blue. $E$ is the energy of the initial energy eigenstate $|E\rangle$ of $\mathcal{E}$ that we couple with the spin $\mathcal{S}$ at time $t=0$. The energy perturbation of this state and its mixing with other energy eigenstates $|E'\rangle$ of $\mathcal{E}$ induced by the coupling Hamiltonian $\HaSE$ is represented in red.}
\label{fExtDyn1}
\end{center}
\end{figure}

These results can be used to study the regime 
where the dynamics of $\mathcal{E}$ is much faster than the dynamics on $\mathcal{S}$ induced by the coupling $\mathcal{S}+\mathcal{E}$. 
In the situation considered here (Wigner ensemble for $\HaE$), this occurs if the width of the spectrum of $\HaE$, namely $2 E_0$, is much larger than the width of the spectrum of $\HaSE$, which is of the order of $\sqrt{\widehat{\Delta}'(0)}$
\begin{equation}
\label{D0ggDh0}
E_0=2\sqrt{\Deltaav}\ \gg\ \sqrt{\widehat{\Delta}'(0)}
\end{equation} 
This situation is depicted on \fig{fExtDyn1}.

In this case we use the results of  Section \ref{ssTS}). 
The time scale $\tau_0$ for  the internal dynamics of the whole system is from \eq{t0New}
\begin{equation}
\label{tau0E0}
\tau_0={1\over\sqrt{\widehat{\Delta}(0)}}\simeq{1\over\sqrt{\Deltaav}}={2\over E_0}
\end{equation}
The decoherence time scale $\tau_1$ for a typical spin state is from \eq{t1def}
\begin{equation}
\label{ }
\tau_1={\tau_0\over 1-Z_{\mathrm{av}}}\simeq \tau_0{\Deltaav\over{\widehat{\Delta}}'(0)}
\end{equation}
hence it is much larger that $\tau_0$.
\begin{equation}
\label{t1ggt0bis}
\tau_1\gg\tau_0
\end{equation}
We remind that from \eq{t0New} $\widehat{\Delta}'(0)$ is defined by
\begin{equation}
\label{Dprime0}
\widehat{\Delta}'(0)=\sum_{l=1}^{l_0} (2l+1)\,\bar\Delta(l)
\end{equation}
where the $\bar\Delta(l)$ are of order $\mathcal{O}(1)$.
The factor $Z(l)$ behaves for small $l$  according to \eq{ZlExp} 
\begin{equation}
\label{Zlapprox}
Z(l)=1-{l(l+1)\over j(j+1)}{D_0\over 4}+\cdots
\end{equation}
with now
\begin{equation}
\label{D0expl2}
D_0\ \simeq\ {1\over \Deltaav}\sum_{l'=1}^{l_0} \bar\Delta(l)\, l (l+1) (2l+1)
\ \ll\ 1
\end{equation}
We have seen that the decoherence time scale for a coherent state is from \eq{t2def}
\begin{equation}
\label{tau2rapp}
\tau_2=\tau_0 {j\over D_0}
\end{equation}
Therefore when the spin $j$ is large it is always much larger than $\tau_1$ since
\begin{equation}
\label{tau2over1}
{\tau_2\over\tau_1}={j\sum_{l=1}^{l_0} (2l+1)\bar\Delta(l)\over \sum_{l=1}^{l_0}l(l+1)(2l+1)\bar\Delta(l)}\sim{j\over l_0^2}\gg1
\end{equation}

\subsubsection{Evolution of coherent states}
\label{ssEvCSF}
In the limit of large $E_0$, from \eq{D0ggDh0} and \eq{ZlNew} we have in any case
\begin{equation}
\label{ZavZl21}
Z_{\mathrm{av}}\simeq 1\quad\text{and}\quad Z(l)\simeq 1
\end{equation}
and we can apply the results of the section \ref{sssmalZl}.
Let us consider as an initial state $|\Psi\rangle$ of the system $\mathcal{S}+\mathcal{E}$ the single spin coherent state $|\vec e_z\rangle$ times the $\HaE$ eigenstate $|E\rangle$.
\begin{equation}
\label{InStaE}
|\Psi(t=0)\rangle=|\vec e_z\rangle\otimes|E\rangle
\end{equation}
We study the evolution of the reduced spin matrix density for times $t\gg\tau_0$, as was done in Section \ref{ssdiffus} for the case of a spin coherent state $|\vec e_z\rangle$ times a random $\mathcal{E}$ state $|\psi\rangle$.
The harmonics of the Wigner transform of the density matrix are now, using \eq{expDec} for the scaling of the evolution functional $M(t,E,Z(l),Z_{\mathrm{av}})$
\begin{equation}
\label{WltE}
W^{(l)}(t)={2l+1\over\sqrt{2j+1}}\ \exp\left(-{l^2\over 2j}\right)\,\exp\left(-{t}\,{l(l+1)\over j(j+1)}{D_0\over 4}\sqrt{E_0^2-E^2}\right)
\end{equation}
to be compared with \eq{WCSas}.
This gives a Wigner distribution $W(\vec n,t)$ on the unit sphere with width proportional to $\sqrt{t}$ when $t\gg\tau_2$, hence the evolution of the spin is again a quantum diffusion process.

\subsubsection{Markovian limit for quantum diffusion in phase space}
\label{ssMarkDiff}
For fast $\mathcal{E}$ dynamics, initial state $|E\rangle$ and large spin $j\gg1$, the spin diffusion at $t\gg\tau_2$ becomes Markovian, with no memory effects. 
Indeed, at not too large times, whan we can approximate the phase space $\mathcal{S}_2$ by the tangent complex plane $\mathbb{C}$,
%we can perform a similar rescaling in phase space than in \ref{rescS2} and \ref{rescT} to approximate the unit sphere by the complex plane.
%But now 
the Wigner distribution $W(\vec u,t)$, which is the Fourier transform of the decoherence function 
$$M^{(l,m)}(t)=\exp\left(-{t}\,{l(l+1)\over j(j+1)}{D_0\over 4}\sqrt{E_0^2-E^2}\right)$$
is exactly a Gaussian distribution of the form \ref{WclaWie},  given by \eq{Wquantum}.
\begin{equation}
\label{WClass3}
W(\vec u,t; D)={1\over 4\pi D\, t}\exp(-|\vec u|^2/(4D\,t))
\ ,
\end{equation}
instead of the non-classical distribution function $W_{\mathrm{quantum}}$ of \eq{Wquantum}.
This corresponds to a classical diffusion process (Wiener process) on the unit sphere\footnote{\ Important: we do not perform the rescaling \ref{rescS2} by a factor $\sqrt{2j+1}$ on the spin here.}
$ \mathcal{S}_2$ with a diffusion coefficient $D$ which depends on the energy $E$ of the initial state through
\begin{equation}
\label{DClasExpl}
D={D_0\sqrt{E_0^2-E^2}\over 4 j (j+1)}
\end{equation}
This Markovian regime where quantum diffusion is classical is expected to hold only when the exponential scaling \ref{expDec} of  $M_{\mathrm{scaling}}$ is valid. Since it holds only for  $t(1-Z(l)) < t'_{\mathrm{cross}}$ given by \ref{tcrossing}, we can show that non Markovian deviations to the  
distribution function $W(\vec Z,t)$ given by  \ref{WClass3} occurs when (this is a very rough estimate)
\begin{equation}
\label{MarkBreak}
{{|\vec u\hskip 1.pt|}^2\over t}> 4D \log(\tau_1/\tau_0)
\end{equation}
This means that departure to Markovian behavior for finite $\tau_1/\tau_0$ are more easily observable in the large distance/small time diffusive regime.

\subsubsection{A Golden Rule formula for the diffusion coefficient}
\label{ssGoldRu}
It is interesting to note that the diffusion coefficient $D$ can be rewritten in term of the normalized density of states $\nu(E)$ for the system  $\mathcal{E}$, given by
\eq{nuWigner}
\begin{equation}
\label{nuexpl}
\nu(E)={2\over \pi\,E_0^2}\sqrt{E_0^2-E^2}
\end{equation}
and of the norm of the commutator
\begin{equation}
\label{CSH}
\vec C=\imath[\vec S,\HaSE]
\end{equation}  between the spin operator $\vec S$ and the coupling Hamiltonian $\HaSE$. Indeed, using 
\eq{D0expl2} for $D_0$, \eq{widthWig} for $E_0$, and the eqs.~(\ref{t2def}-\ref{normSH3}) of sect.~\ref{sstau2} we have
\begin{equation}
\label{DofNu}
D=2\pi\,\nu(E)\ {1\over N (2j+1)^2}\tr\left({\vec C}^2\right)=2\pi\,\nu(E)\ {1\over 2j+1}{|\!|\,{\vec C}\,|\! | }^2
\end{equation}
Using \eq{nu2E} we can rewrite the diffusion coefficient $D$ in term of the full density of states $\rho(E)=N\nu(E)$ for the Hamiltonian $\HaE$, acting on the full system, and of the average squared norm of a matrix element of the commutator $\vec C$
\begin{equation}
\label{ExpC2}
\mathbb{E}\left[{\left | \langle \Psi | \vec C | \Psi'\rangle \right |}^2\right]={1\over N(2j+1)}{|\!|\,{\vec C}\,|\! | }^2={1\over {(N(2j+1))}^2} \sum_{\Psi,\Psi'}{\left | \langle \Psi | \vec C | \Psi'\rangle \right |}^2
\end{equation}
where $\{|\Psi\rangle\}$ is a basis of $\HaSE=\mathbb{C}^{N(2j+1)}$. Indeed we rewrite \eq{DofNu} as the general form for the spin diffusion coefficient $D$
\begin{equation}
\label{DGolden}
D\ =\ 2\pi\, \rho(E)\  \mathbb{E}\!\left[{ | \langle \Psi | \vec C | \Psi'\rangle  |}^2\right]
\end{equation}
%{\large{\emph{Check! Recheck!}}}

This form for the diffusion coefficient for the dynamics of coherent states, as a function of the energy $E$ of the initial state for the environment, is similar to the Fermi Golden Rule for the transition rate of a state $\psi$ of a quantum system into a continuum of states $\Psi'$ (with density of final states $\rho_\mathrm{final}$), induced by a perturbation Hamiltonian $H_{\mathrm{pert}}$.
\begin{equation}
\label{FGR}
T_{\Psi\to\Psi'}\ =\ 2\pi\,\rho_\mathrm{final}\,\left|\langle\psi|H_{\mathrm{pert}}|\Psi'\rangle\right|^2
\end{equation}
This is not so surprising since we are in a regime where effective dynamics is Markovian, and it is known that there is a relation between the Markov approximation and the Fermi golden rule for simple systems coupled to a bath (see \cite{Aliki77} and references therein).
The form \ref{DGolden} for the diffusion coefficient can therefore be viewed as a fluctuation dissipation relation for the quantum dynamics of the spin in the regime $t\gg\tau_2$. What is interesting is that it involves the matrix elements of the commutator $\vec C$ between the spin operator $\vec S$ and the interaction Hamiltonian $\HaSE$.

We therefore expect that the relation \ref{DGolden} for $D$ is general and independent of the the explicit form of the density of states $\rho(E)$ for the  external Hamiltonian $\HaE$.
It should be easy to study the limit $\tau_1\gg\tau_2$ using the general form \ref{MltEexpl} for the evolution functional $M(t,E,Z(l),Z_{\mathrm{av}})$ instead of the explicit form \ref{MtEZZ} for a semicircle state density law. 
In particular, we remark that the derivation of the exponential scaling \ref{expDec} starting from the general form \ref{Mscaling1} of the scaling function $M_{\mathrm{scaling}}$ at small $l$ uses a Lorentzian approximation very similar to the one used in the textbooks derivation of the Fermi Golden Rule.

\subsubsection{Initial state $|\psi_{\mathcal{E}}\rangle$ dependence and randomized Markovian processes}
It remains to understand why, even in the limit of fast dynamics  $\tau_1\gg\tau_0$, if for $\mathcal{E}$ as initial state $|\psi_{\mathrm{init}}\rangle=|\psi(0)\rangle$ in $\HaE$   one starts from a random state $|\psi_{\mathrm{random}}\rangle$, in the quantum diffusion regime the distribution profile in phase space $W(\vec u,t)$ for the spin is not a Gaussian, but is given by the non-Gaussian quantum distribution $W_{\mathrm{quantum}}$ of \eq{Wquantum} given by the Fourier transform of the scaling function $\Psi$ obtained through \eq{Psi1}.

This can be understood easily. 
First let us come back to the case where $\HaE$ is a GUE random hamiltonian, so that its normalized density of states $\nu(E)$ is given by the Wigner semi-circle law \ref{nuexpl}
\begin{equation}
\label{nuexpl2}
\nu(E)={2\over \pi\,E_0^2}\sqrt{E_0^2-E^2}
\end{equation}
In the fast $\mathcal{E}$ dynamics limit $\tau_1/\tau_0\to\infty$ and large spin limit $j\to\infty$, if one starts form a separable initial state product of a coherent spin state $|\vec n\rangle$ and of a random $\mathcal{E}$ state $|\psi_{\mathrm{random}}\rangle$, in the quantum diffusive regime $\tau_2\ll t\ll j \tau_2$ the probability distribution for the spin \ref{Wquantum} can be written as
\begin{equation}
\label{WnuWGa}
W(\vec u,t)=\int dE\,\nu(E)\,
W(\vec u,t;D(E))
%\quad,\qquad   W(\vec u,t;D(E) = {1\over 4\pi D(E)\,t}\exp\left(-{1\over 4 D(E)}{| \vec u |^2\over t}\right)
\end{equation}
with $\nu(E)$ given by \ref{nuexpl2}, $W(\vec u,t;D)$ the Gaussian density profile \ref{WClass3}  and $D(E)$ the spin diffusion coefficient \ref{DClasExpl} when starting from an initial energy eigenstate $|E\rangle$.
\begin{equation}
\label{WGausExp}
W(\vec u,t;D(E) = {1\over 4\pi D(E)\,t}\exp\left(-{1\over 4 D(E)}{| \vec u |^2\over t}\right)
\quad,\quad 
D(E)={D_0\sqrt{E_0^2-E^2}\over 4 j (j+1)}
\end{equation}

Eq.~\ref{WnuWGa} has a simple probabilistic interpretation. The spin diffusion process is simply a statistical superposition, with probability distribution 
$\nu(E)$, of the Wiener processes with the different  diffusion constants $D(E)$.
In other word, starting from a random state for $\mathcal{E}$, to estimate the probability (density) to observe the spin at position $\vec u$ in phase space at time $t$, for all practical purpose we may assume that:
\begin{description}
  \item[1 -] at time $0$ we choose at random an energy eigenstate $|E\rangle$ for $\mathcal{E}$ (therefore with probability law given by the density of states $\nu(E)$),
  \item[2 -]  then we let the spin diffuse according to a Wiener process with diffusion constant $D(E)$. 
\end{description}
This kind of non-Markovian process is build by  the operation of \emph{randomization} of a family of Markovian processes (see the mathematical litterature, for instance \cite{Feller71}). Here the family of Wiener processes depending on the parameter $E$ through the diffusion constant $D(E)$ is randomized with the probability distribution $\nu(E)$.

To understand the physical origin of this ramdomization,  we remark that a random state $|\psi_{\mathrm{random}}\rangle$ of $\mathcal{E}$ is nothing but a quantum superposition of energy eigenstates of $\HaE$ with random complex coefficients $c_E$, choosen independently according to the Gaussian normal distribution.
\begin{equation}
\label{ }
|\psi_{\mathrm{random}}\rangle=\sum_E c_E\, |E\rangle
\end{equation}
Hence the initial step {\bf{1}} amounts to project the initial state $|\psi_{\mathrm{random}}\rangle$ onto a random energy eigenstate $| E\rangle$, i.e. to perform an ideal (von Neumann) measurement of the energy of the external system $\mathcal{E}$.

This is to be expected.
Indeed, we are in the regime where the interaction Hamiltonian $\HaSE$ is a small perturbation compared to the internal $\HaE$ Hamiltonian for $\mathcal{E}$. But we are also in the large spin regime where $j\gg 1$, so the spin $\mathcal{S}$ can be considered as a large quantum system, with many states, weakly coupled to $\mathcal{E}$.
%In the same way than spin decoherence is governed by the time scales $\tau_1$ and $\tau_2$ with (from \ref{t0defa}, \ref{t1def} and \ref{t1overt2})
In the same way than spin decoherence occurs by the time scales $\tau_1$  with (from \ref{t0defa} and \ref{t1def})
\begin{equation}
\label{ }
{\tau_0\over\tau_1}={\left({\|  \HaSE  \|  \over \| H \|}\right)}^2 \simeq{\left({\|  \HaSE  \|  \over \| \HaE \|}\right)}^2\ \ll 1
\end{equation}
with $H=\HaSE + \HaE$, we expect that the decoherence between (not too close on the spectrum $E$ line) energy eigenstates $|E\rangle$ of $\HaE$ occurs at  a time scale $\tau_{\scriptstyle{e}}$ given by
\begin{equation}
\label{ }
{\tau_0\over\tau_e}= {\left(  { \| [ H ,  \HaE ] \| \over \| H \|  \cdot  \| \HaE \|  }  \right) }^2
\simeq  {\left(  { \| [ \HaSE ,  \HaE ] \| \over \| \HaE \|^2  }  \right) }^2\ \ll \ 1
\end{equation}
In other words, the interaction between $\mathcal{S}$ and $\mathcal{E} $ induces both decoherence for the spin and decoherence in energy for the external system.
As a consequence, at the time scale $\tau_2$, where quantum diffusion starts, the initial coherent state $|\vec n\rangle$ for the spin can be considered as a classical spin state $\vec n$, while the initial random state $|\psi_{\mathrm{random}}\rangle$ for $\mathcal{E}$ can be considered as a classical equidistributed statistical mixture of energy states $E$.

This suggests the general form of the distribution profile $W(\vec n,t)$ as a function of the $\mathcal{E}$ initial state $|\psi_{\mathrm{init}}\rangle$.
We consider a general $\HaE$ with a continuous density of states $\rho(E)= N\nu(E)$, and we normalise the energy eigenstates $|\mathbf{E}\rangle=(1/\sqrt{\rho(E)})|E\rangle$ so that
\begin{equation}
\label{ }
\langle \mathbf{E}|\mathbf{E'}\rangle=\delta(E-E')
\end{equation}
If we start from a general initial state $|\psi\rangle$ for $\mathcal{E}$ which has a smooth decomposition in energy eigenstates, that is to say
\begin{equation}
\label{ }
\langle \mathbf{E}|\psi\rangle =\psi(E)\quad\text{is a regular function of}\ E
\end{equation}
at least when averaged over small energy intervals $\Delta E\simeq 1/\tau_2$,
then the probability distribution for the spin at large time $t\gg\tau_\mathrm{diff}$ should be given by
\begin{equation}
\label{ }
W(\vec u,t;\psi)=\int dE\, 
{\left\vert{\langle \mathbf{E}|\psi\rangle}\right\vert}^2\,{1\over 4\pi D(E)\, t}\,\exp\left(-{| \vec u |^2\over 4\,D(E)\,t}\right)
\quad .
\end{equation}
with the effective diffusion constant $D(E)$ given by \eq{DGolden}
\begin{equation}
\label{ }
D(E)=2\pi\, \rho(E)\, \mathbb{E}
\quad,\qquad
\mathbb{E}= \mathbb{E}\left[\left| \langle \Psi|[\HaSE,\vec S]|\Psi'\rangle\right|^2\right] ={\tr\left((\imath [\HaSE,\vec S])^2\right) \over (N(2j+1))^2}
\end{equation}

\section{Discussion and conclusions}
\label{sConc}
\subsection{Summary}
\label{ssSummary}
We have proposed and solved a model of a general quantum spin $j$ interacting with a large external system (with $N$ states).
The interaction Hamiltonian $H_\mathrm{int}$ is described by a new random matrix model, relying on a GU$_{2\times N}$E ensemble, which takes into account the interactions in all possible angular momentum channels.
The only constraint is that the probability distribution for $H$ is Gaussian and invariant under the spin symmetry SU(2) and the external U($N$) group.
The dynamics if the model is solved exactly in the large $N$ limit, in the case where the internal dynamics of the spin is trivial.
%We obtain explicit expressions for the evolution functional of the density matrix for the spin.
We can thus study in full generality the decoherence of spin states in various dynamical regimes, and as a function of the initial condition and of the dynamics of the external system. 

In the semiclassical $j\to\infty$ limit, we are able to characterize quantitatively the interaction Hamiltonians 
%(given by the variances $\Delta(l)$ of the GU$_{2\times N}$E ensemble) 
which are such that coherent spin states have a much longer coherence time than any other states, and emerge effectively as the semiclassical ``pointer states'' of a classical spin.
The large time dynamics of these coherent states is found to be a quantum diffusion process, in general non-Markovian.

When the dynamics of the external system is much faster than the dynamics induced by its coupling with the spin, and when the initial state of the external system is an energy eigenstate, the quantum spin diffusion is found to be Markovian.
The diffusion coefficient takes a golden rule form, but involving matrix elements of the commutator between the spin and the interaction hamiltonian.
When the initial external state is a general state, the quantum spin diffusion is described by a randomization of Markov processes.  We argue that this is explained by the decoherence of energy states (for the external system) induced by the coupling with the spin.

\subsection{Comparison with previous studies of decoherence in spin systems}
\label{ssCompLitt}
We finally discuss the relation of the present work with some previous studies of open quantum spin systems.

In the pionnering work  \cite{TakahashiShibata1975}, the relaxation of a large spin $j$ in an external magnetic field  and coupled to a heat bath is already studied. 
A coherent state representation for the spin states different from ours is used.
The Born approximation and some methods of \cite{Agarwal1971} are used. 
%The Born aproximation and the methods of \cite{Agarwal1971} are used. 
Therefore their results are only valid for large time.
A relaxation towards a classical distribution is obtained and the corresponding diffusion coefficients are calculated.
In \cite{ShibataSaito1975}  the spin relaxation in the specific case of the $j=1/2$ spin is studied within the Markovian approximation.
In both papers the spin has a non trivial dynamics since it is coupled to an external field, this in not taken into account in our study.

In 
\cite{MelloPereyraKumar1988} the relaxation of a $j=1/2$ spin coupled to an external large system (bath) is studied. 
The spin has no internal dynamics ($H_\mathrm{spin}=0$), the bath has its own dynamics characterized by some $H_\mathrm{B}$ with its density of states.
The interaction is of the simple form $H_\mathrm{int}=\sigma_x\otimes V_\mathrm{ext}$, where $V_\mathrm{ext}$ is a random operator for the external system, chosen in a GOE ensemble local in energy $E$ (for instance the matrix elements of $V_\mathrm{ext}$ are non zero only between energy eigenstates such that the energy difference is finite $|E_1-E_2|\le\Delta E$, thus it is a finite width band GOE). They use the RMT methods which shall be used in most subsequents studies based on RMT, including the present one. With the initial condition considered they are able to study the relaxation of the spin at all time, and find an exponential relaxation (with already a Golden rule form for the relaxation time).

In \cite{LebowitzPastur2004} and \cite{LebowitzLytovaPastur2007} the study of relaxation for this spin $j=1/2$ model with a band GOE $H_\mathrm{int}$ has been extended to the non trival case where the spin has its own dynamics  $H_\mathrm{spin}\propto \sigma_z$.
Special attention is paid to the case where the external bath is formed of large number of independent large subsystems, in some particular limits where an effective temperature can be associated to the large bath.
Explicit expressions are obtained for the evolution of the spin $1/2$ density matrix, but in practice the relaxation regime towards an equilibrium distribution is studied, in particular in the so called van Hove limit where the coupling spin-bath is small, which corresponds to our limit $\tau_\mathrm{dyn}\ll\tau_\mathrm{dec}$. Thus the study of  \cite{LebowitzPastur2004,LebowitzLytovaPastur2007} is more limited than our work, since only the spin $j=1/2$ case is considered and the GOE ensemble for $H_\mathrm{int}$ is not SU(2) invariant, but also more general, since $H_\mathrm{int}$ is a band matrix in the energy states for the bath, and, most importantly, the spin is coupled to an external magnetic field, so that relaxation and thermalization effects can be studied.

Another study involving RMT techniques is  \cite{LutzWeidenmuller1999}. 
General system+bath couplings with band random matrix Hamiltonians are studied, both for a quantum harmonic oscillator and for a two level system. 
In the case studied, Markovian equations can be derived.

In \cite{EspositoGaspard2003} an extensive study of the relaxation of a two level system (spin $1/2$) coupled to a bath is performed. 
The coupling is given by a $H_\mathrm{int}=\sigma_x\otimes V_\mathrm{bath}$ Hamiltonian with $V_\mathrm{bath}$ a GUE random matrix,
 $H_\mathrm{bath}$ also given by a random matrix, and $H_\mathrm{spin}\propto\sigma_z$.
Most of the study is by numerical simulations. The finite $N$ effects, where $N$ is the size (number of states) of the bath, the statistics of the eigenvalues, the crossover between the Poisson and Wigner behavior for the spacing of the eigenvalues are very thoroughly studied.
Explicit checks of the self averaging property for different realization of the random Hamiltonian are done. 
The relaxation and equilibrium effects are considered. 
Interesting analytical results are obtained in the regime of weak coupling (which corresponds again to the regime $\tau_\mathrm{dec}\gg\tau_\mathrm{dyn}$ in the present paper). In particular the fact that the relaxation can be non-Markovian is observed and discussed.
The fact that in the regime of strong coupling the decay of the magnetization (evolution of  $\langle\sigma_z\rangle$) with time can be non-exponential but algebraic with some oscillations is also observed.

The relaxation of a two level system (and systems with a small but > 2 number of levels) coupled by RMT Hamiltonians to a bath have been also studied numerically and by the so called HAM method in  \cite{BorovskyGemmerMahler2003,GemmerMichel2006}. 
The two level system has been studied more extensively and precisely by the TCL method in 
\cite{BreuerPeterMichel2003}. Here also, it is found that depending on the initial states for the system and the bath, the relaxation dynamics can be non-Markovian.

Finally, let us quote some papers which discuss more specifically the problem of decoherence in spin systems.
There are of course many studies of the dynamics of several coupled spins $1/2$, and of their equilibration dynamics, see for instance
\cite{SaitoTakesueMiyashita1996}.
In a recent numerical study 
\cite{YuanKatsnelsonDeRaedt2009}
it is confirmed for instance that decoherence is a much faster process than the thermal relaxation.

In 
\cite{BraunHaakeStrunz2001}
\cite{StrunzHaakeBraun2002}
\cite{StrunzHaake2003}
decoherence is studied for various systems (mostly a particle or an oscillator) coupled to a bath through a simple $H_\mathrm{int}=Q_\mathrm{syst}\otimes B_\mathrm{bath}$ Hamiltonian (sometimes two), in the so-called interaction dominated regime, which should correspond in the present paper to the regime $\tau_\mathrm{syst}\ll\tau_\mathrm{dec}$.
Of particular interest for the present work is the section~VII of \cite{StrunzHaakeBraun2002}, where the system considered is a large spin $j$, studied by coherent states techniques.
However the difference is that the coupling agent is just $Q_\mathrm{syst}=\sigma_x$, while the bath Hamiltonian $B_\mathrm{bath}$ is a sum of independant $B_i$'s for small independant sub-baths (so that random limit theorems may be applied).
Thus it is not clear how to compare this model to our model.
Some small time expansion are used and with these approximation the decay of the off-diagonal elements of the density matrix is obtained (decoherence) but with a time dependence which is very different from the exact results that we have obtained.

\subsection{Generalizations and open questions}
\label{ssGeneral}
Many interesting questions remain to be addressed for this kind of models of a spin coupled to a large external system.
Firstly, we have seen that the non-Markovian dynamics that we obtain for large spin $j$ has been already observed in some two level systems.
In order to better understand this effect, and how the non-Markovianity is related to the dynamics of the coupling and to the initial states, one must generalize the random interaction Hamiltonians that we have considered to the more general case of an gaussian ensemble with a law depending on the initial and final energy states for the external system, as in 
\cite{MelloPereyraKumar1988,LebowitzPastur2004}.
It should not be difficult to extend the study of section~\ref{sExtDyn} to this more general case, but this will be discussed in a further work.

In order to compare more precisely our results with those of previous studies, one would like to study simpler coupling Hamiltonians of the form
$H_\mathrm{int}=Q_\mathrm{syst}\otimes B_\mathrm{bath}$ with for instance $Q_\mathrm{syst}=S_x$.
However, our model is solvable and its solution takes a simple form precisely because the random interaction Hamiltonian belongs to a ensemble $H_\mathrm{int}$ which is SU(2) invariant. This is clearly not the case for the simple $H_\mathrm{int}$'s considered previously, since there is a privileged $x$ direction.
Another simplification would be to consider a coupling of the large spin $j$ with a large collection of independent systems, as done in
 \cite{StrunzHaakeBraun2002}. However, here again, to do so while preserving the SU(2) invariance of the ensemble of random interaction Hamiltonian s is not that easy and does not lead to models simple enough to be discussed here.
 
We have not studied the case where the internal dynamics for the spin is non-trivial ($\HaS\neq 0$). This has been done analytically for the spin $j=1/2$ case in \cite{LebowitzPastur2004,LebowitzLytovaPastur2007}, and they have obtained explicit expressions (of course more complex than our results for $j=1/2$).
It should not be difficult to extend their results for our class of interaction Hamiltonian in the specific $j=1/2$ case (then only the $l=1$ sector contributes).
However the problem becomes increasingly complicated as the spin $j$ increases, and no general solution valid for generic spin $j$ has been obtained yet.
This is clearly an interesting problem, relevant for studying the interplay between decoherence and dissipation. 
If the internal spin dynamics is slow, i.e. when $\HaS$ is small, it should be possible to study the dynamics by standard approximation methods (short time expansion, TCL).

We have studied the matrix model in the planar large $N$ limit, where $N$  corresponds to the size (number of states) of the external system. 
It should be interesting to study the case of finite but large $N$, as has been done already for $j=1/2$ in \cite{EspositoGaspard2003} . 
Especially interesting should be the case where $N$ (the dimension of the external system) is of the same order than $j$ (the spin).

In our model we have considered a random $M\times M$ matrix model of the GUE type, but such that the distribution probability is not invariant under the whole U($M$) group, but only under the action of a subgroup $G$ acting on the space (in our case $G=$SU($2$)$\times$U($N$) with $M=(2j+1)N$).
This class on models belongs of course to the very general class of block random matrix models, 
but in our case the presence of the symmetry group $G$ and of its spin representation adds a lot of structure and of simplifications.
It  should be interesting (at least mathematically) to study these kind of models for more general representations of the group $G=$SU($2$)$\times$U($N$), and for general groups and representations.

Finally, the model presented here is very idealized and  is mostly of pedagogical and mathematical value. 
But it should be a first step in studying general but more realistic models of spin decoherence, in particular for systems where the classical macroscopic degrees of freedom do not come from a per se large quantum spin, but emerges from the interaction between many small quantum spins.

\section*{Acknowledgements}
I am very indebted to Michel Bauer for his interest, his insights and his patience during numerous discussions.
I am also very grateful to Roger Balian, Philippe Biane, Philippe di Francesco,  Alice Guionnet, Stéphane Nonnemacher,  Olivier Parcollet, Vincent Pasquier and André Voros for useful discussions and suggestions at various stages of this work.
I thank the referees for their useful comments and suggestions.
This work is supported by the ANR Project GranMa ``Grandes Matrices Aléatoires'' (ANR-08-BLAN-0311-01). 

%\bibliographystyle{plain}
%\bibliographystyle{alpha}

%\bibliographystyle{apalike}
%\bibliography{biblidecoherence}

\newpage
\begin{appendix}
\section{The function $Z(l)$ for various coupling distributions $\Delta(l)$ and spin $j$}
\label{app:fig}

\newlength{\Zplotwidth}
\setlength{\Zplotwidth}{2.8in}

\begin{figure}[h]
\begin{center}
\includegraphics[width=\Zplotwidth]{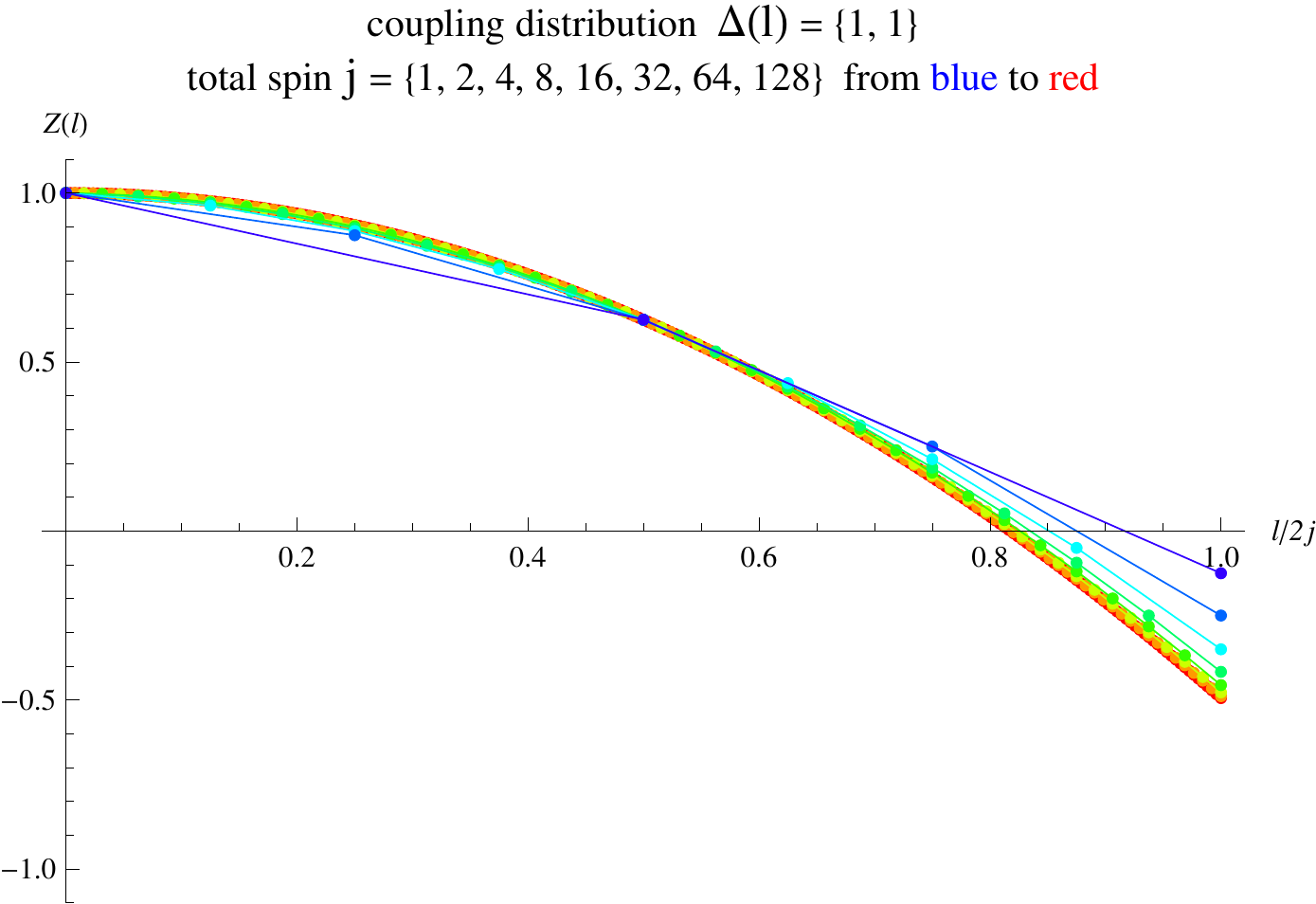} 
\includegraphics[width=\Zplotwidth]{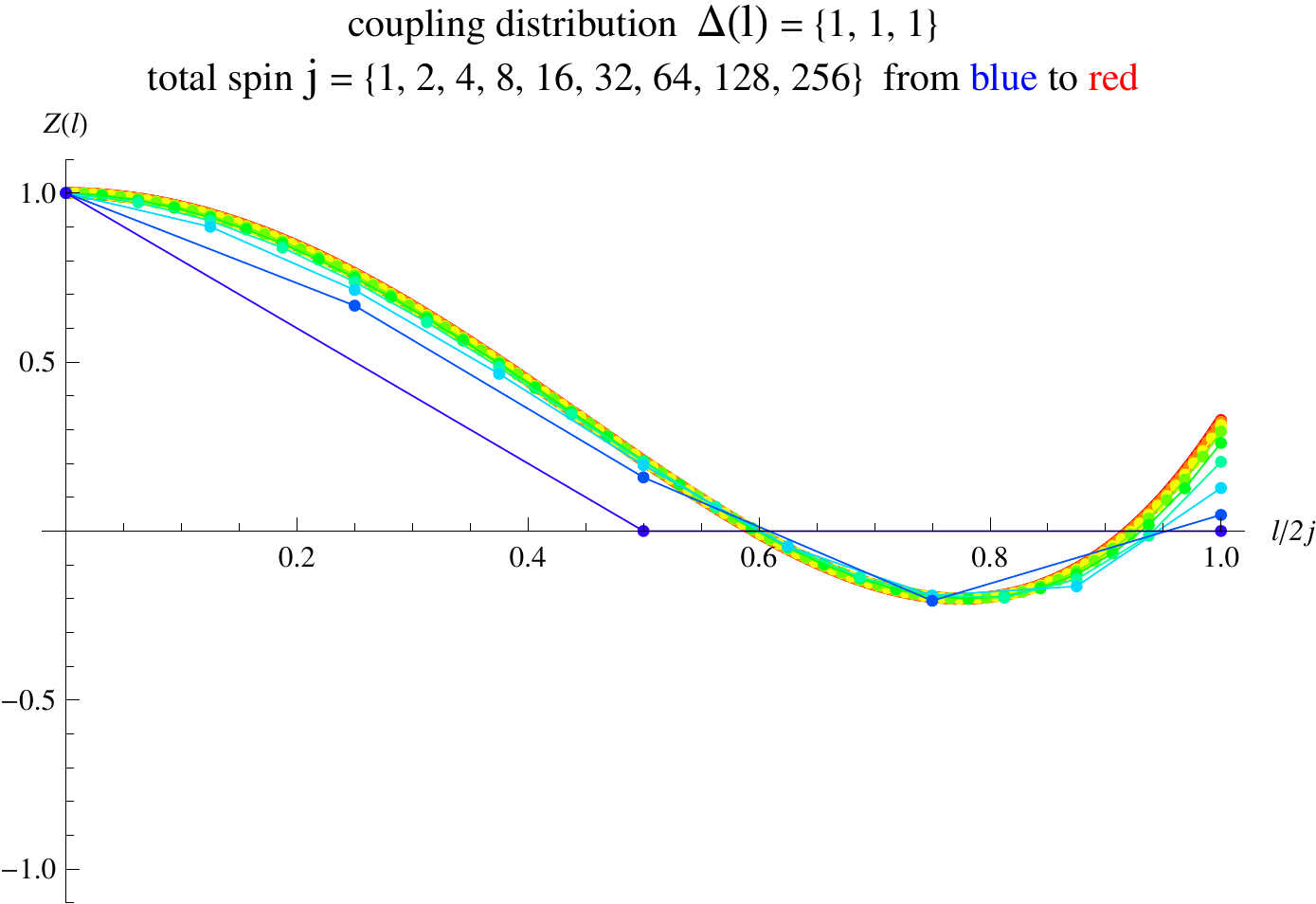}
\includegraphics[width=\Zplotwidth]{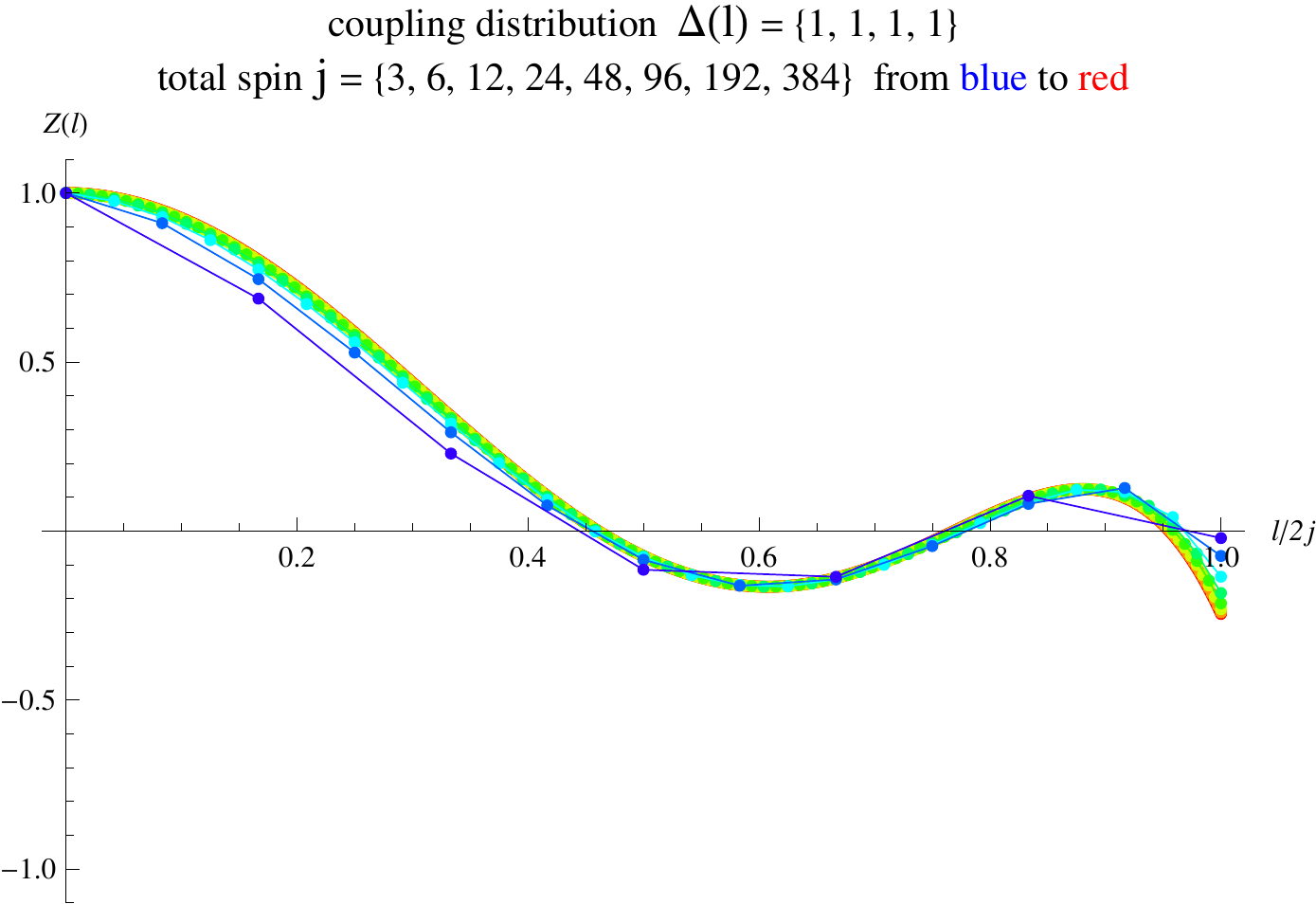} 
\includegraphics[width=\Zplotwidth]{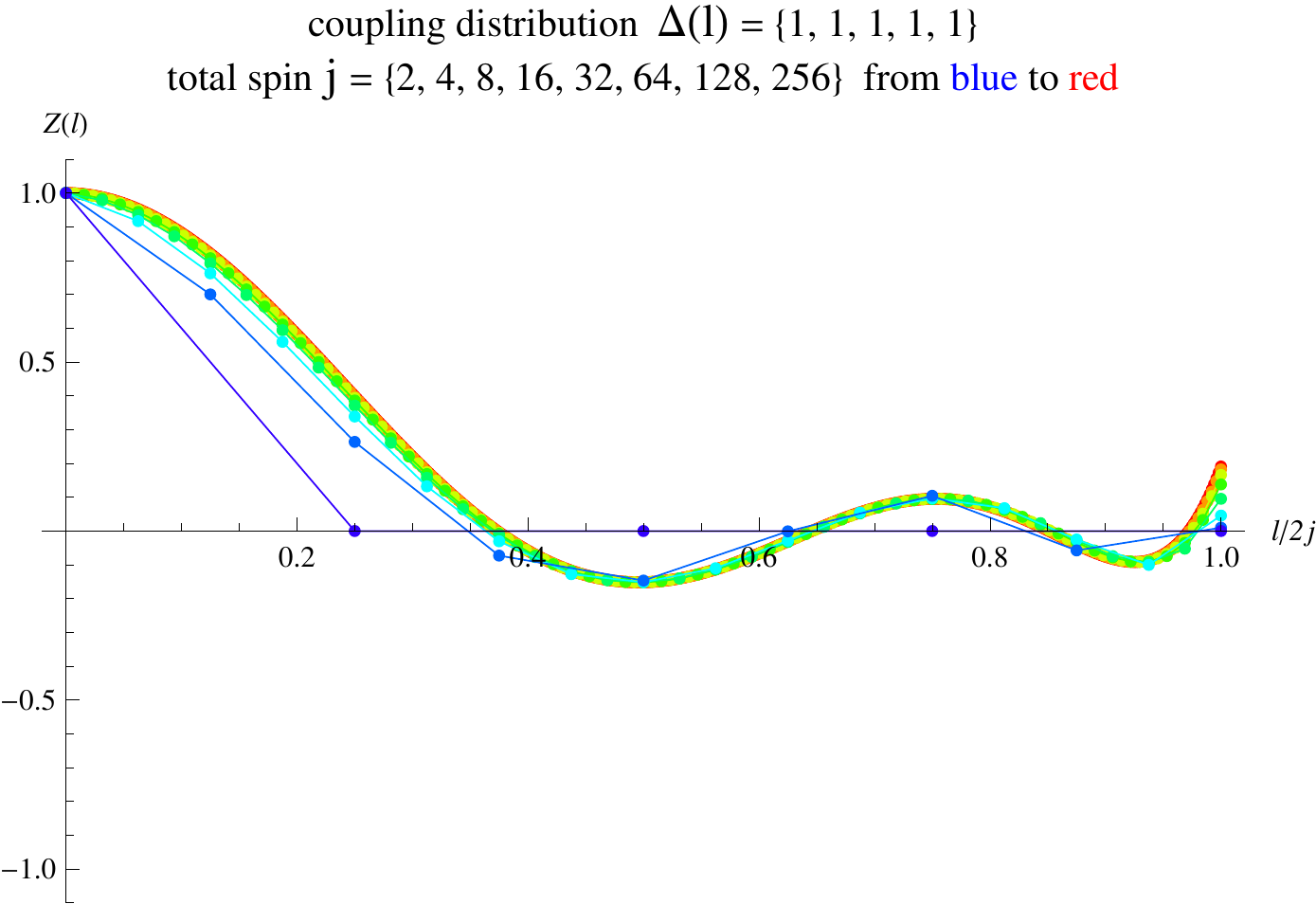} 
\includegraphics[width=\Zplotwidth]{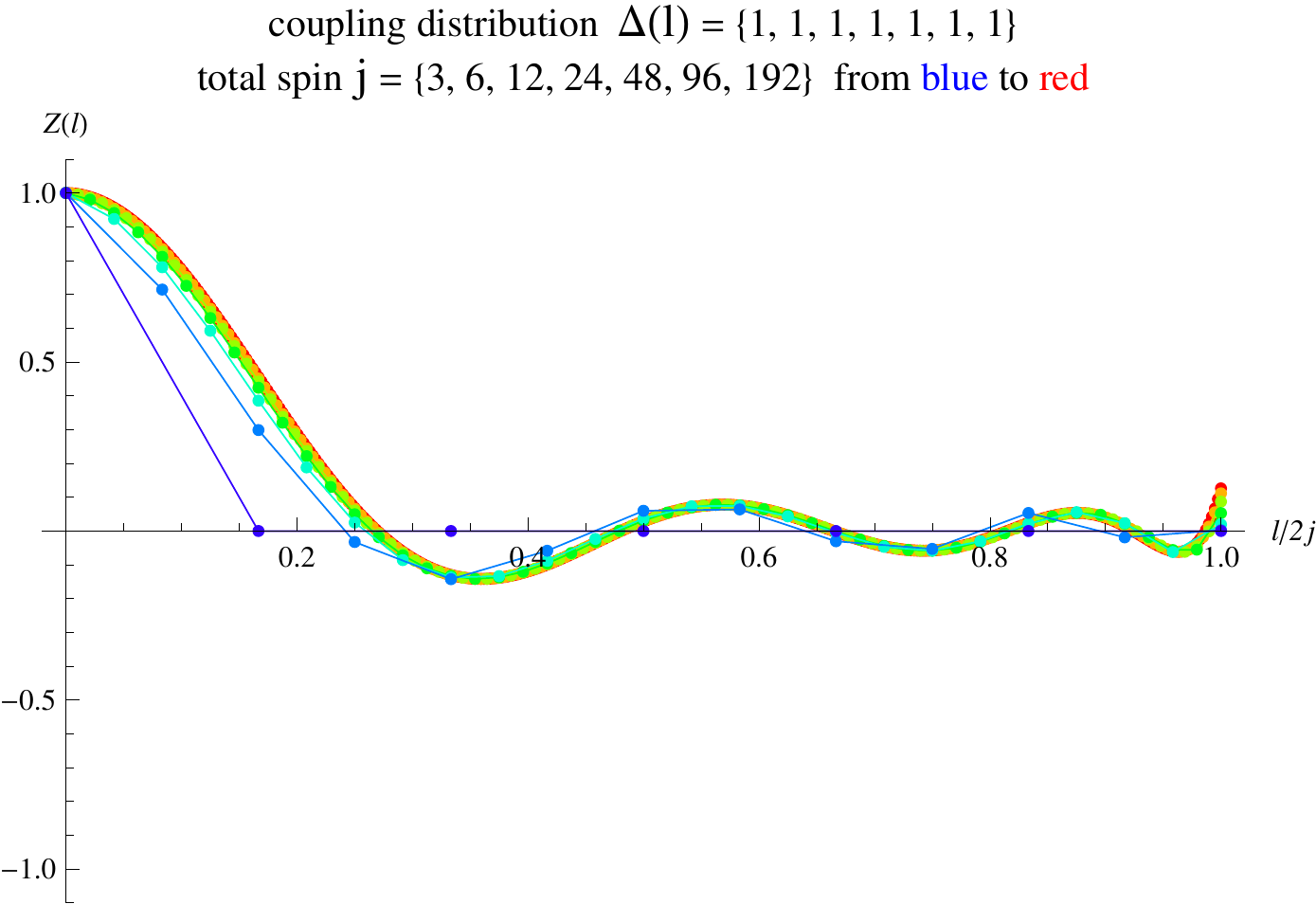} 
\includegraphics[width=\Zplotwidth]{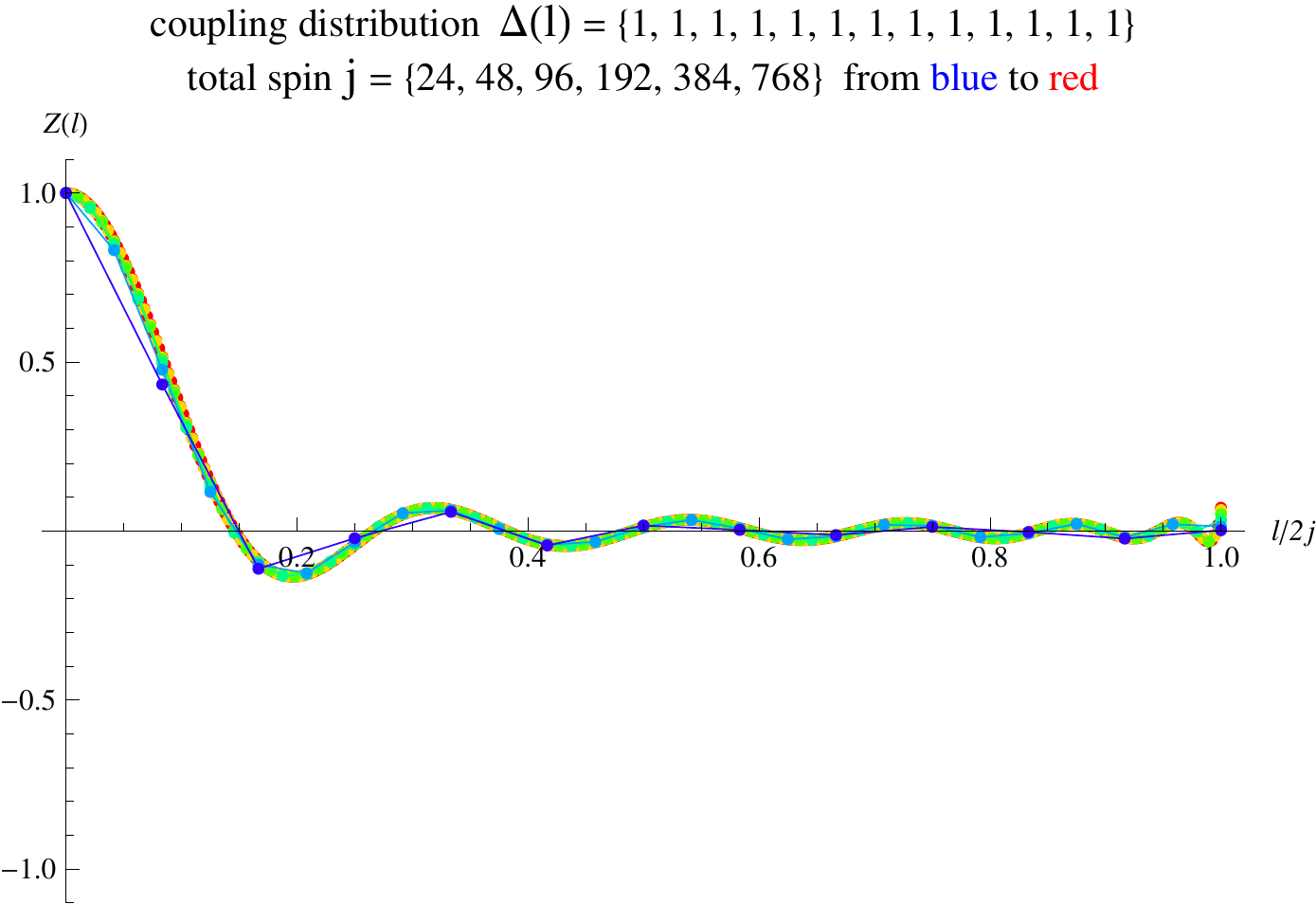} 
\caption{$Z(l)$ for various $l_0$ and $j$, all $\Delta(l)$'s are equal}
\label{fZDeEq}
\end{center}
\end{figure}

\newpage
\begin{figure}[p]
\begin{center}
\includegraphics[width=\Zplotwidth]{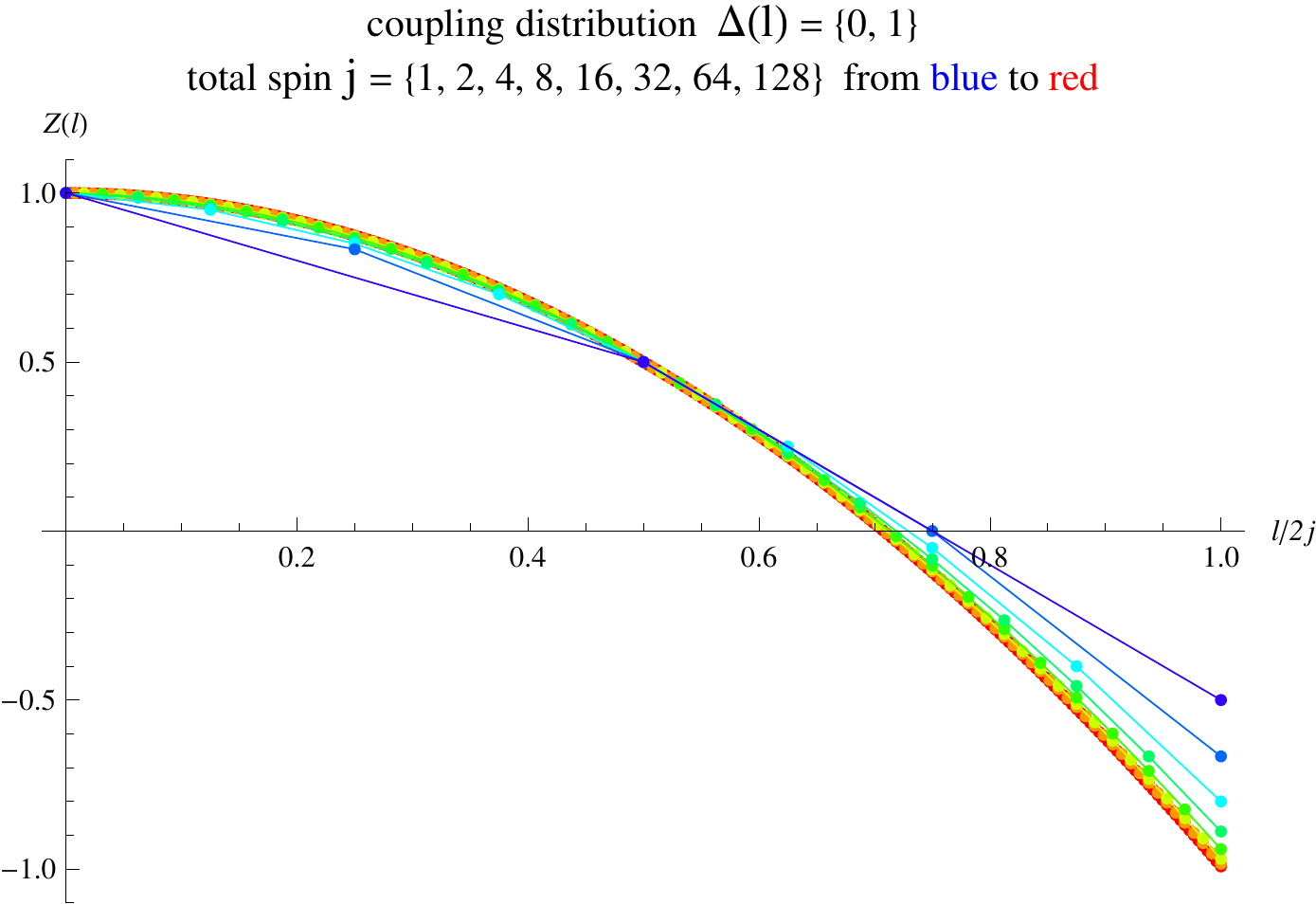} 
\includegraphics[width=\Zplotwidth]{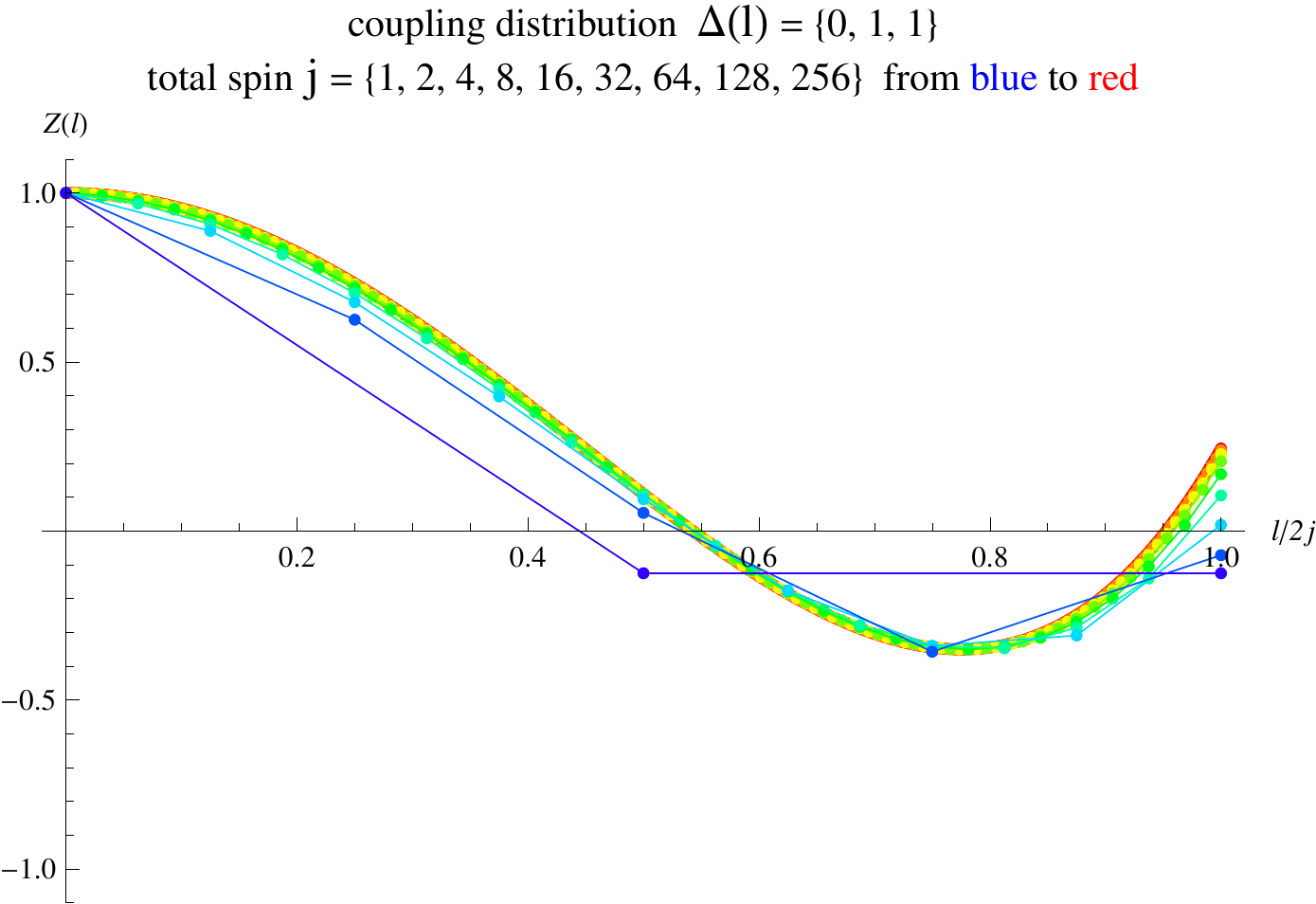}
\includegraphics[width=\Zplotwidth]{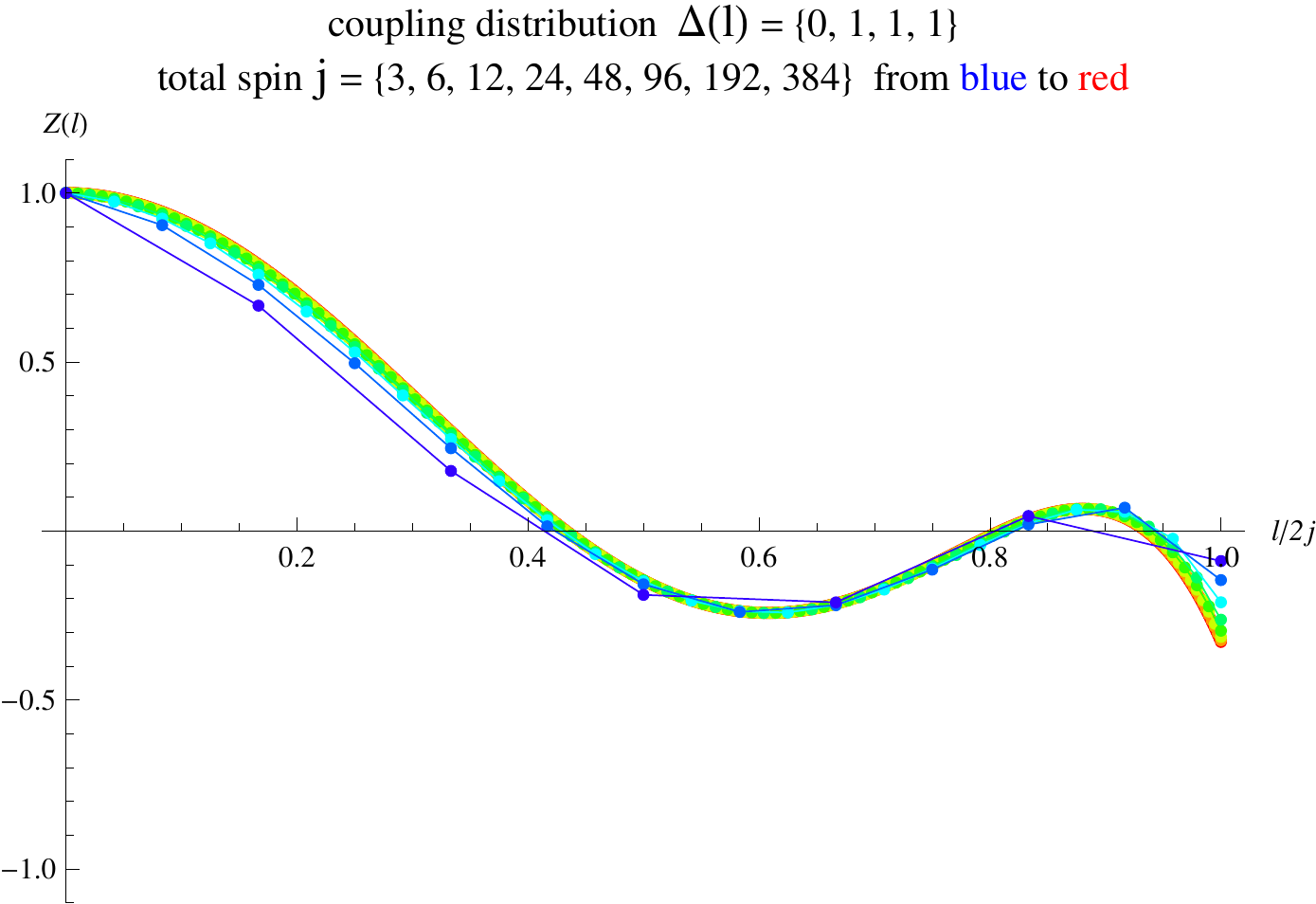}
\includegraphics[width=\Zplotwidth]{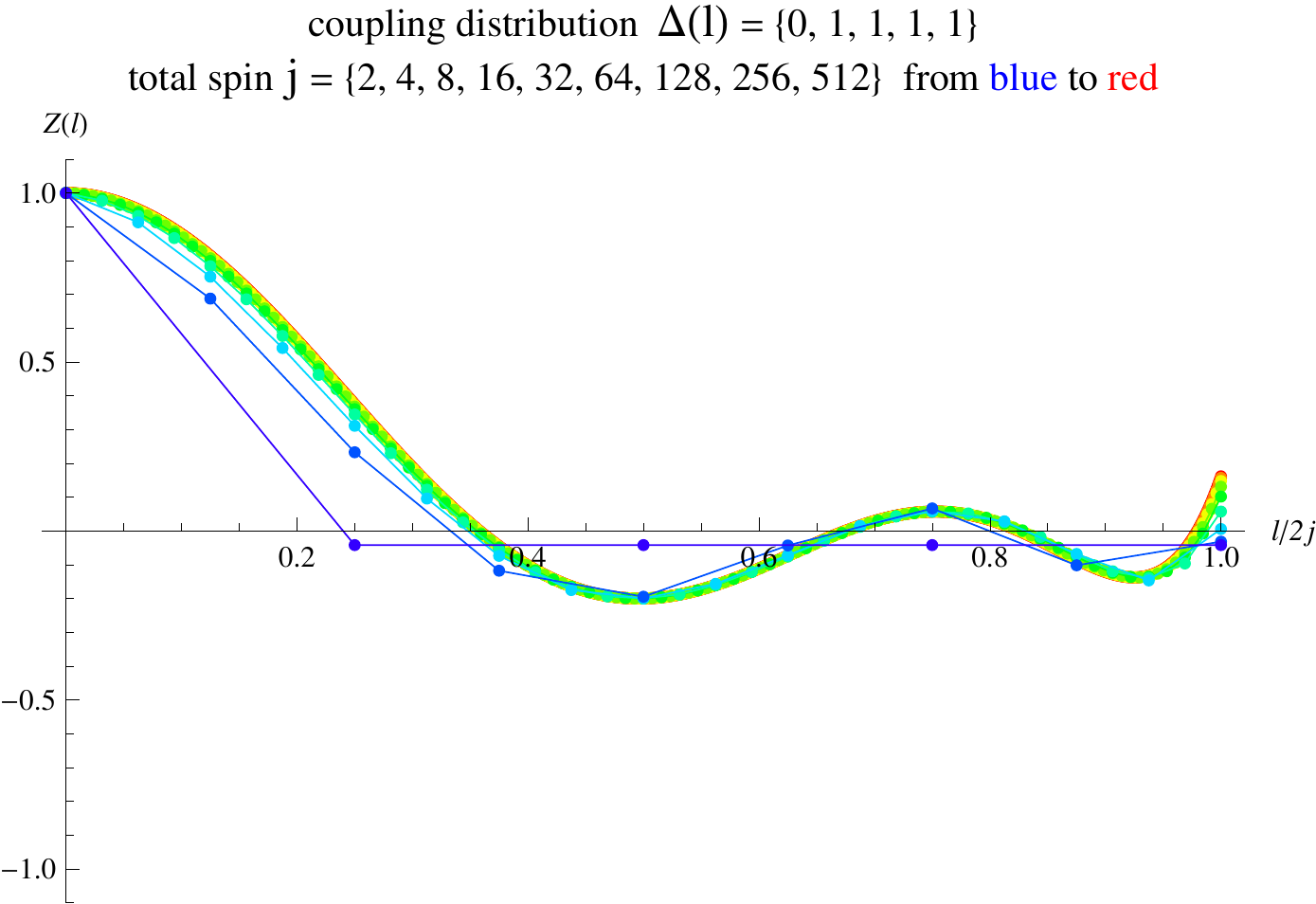}
\includegraphics[width=\Zplotwidth]{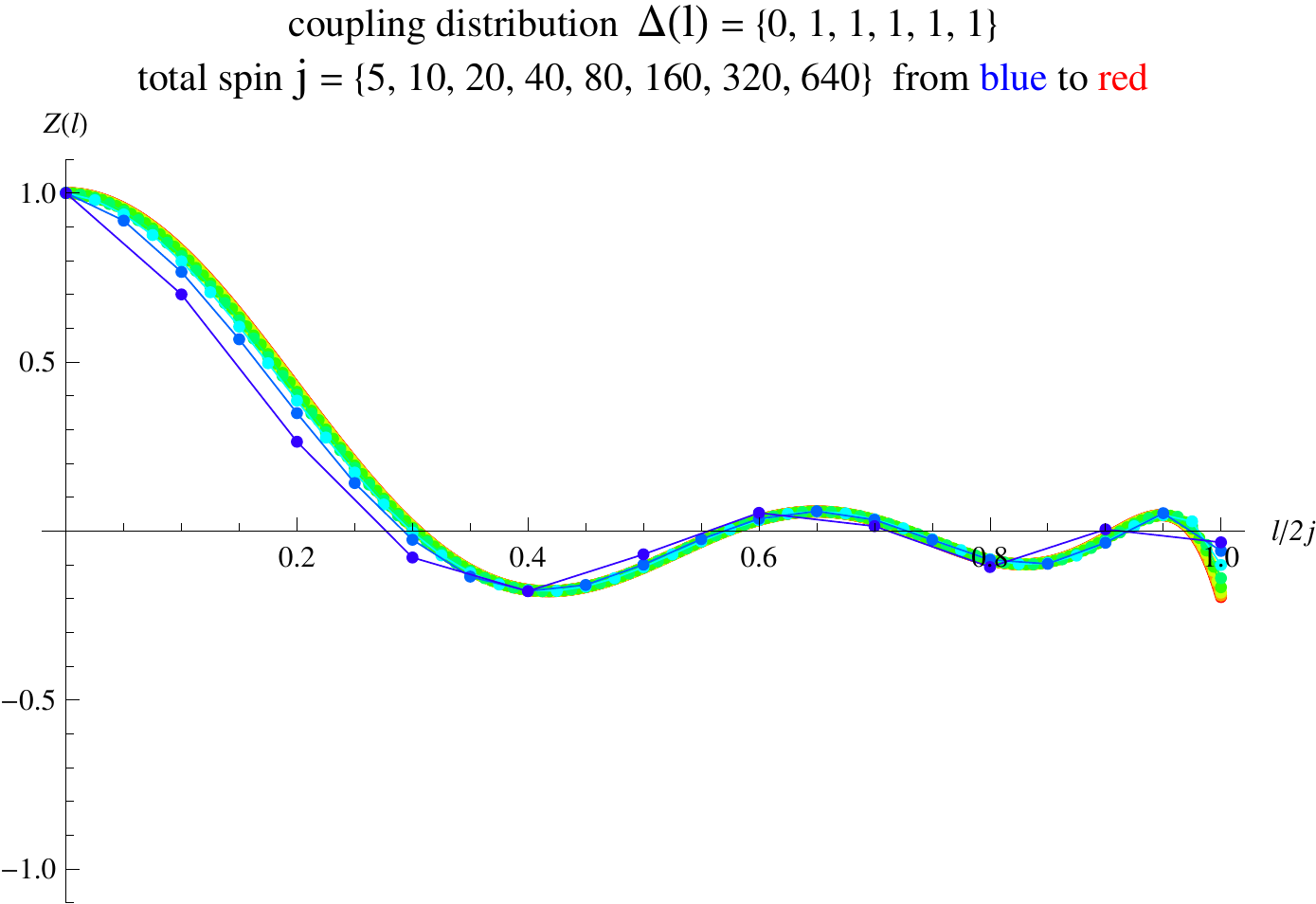} 
\includegraphics[width=\Zplotwidth]{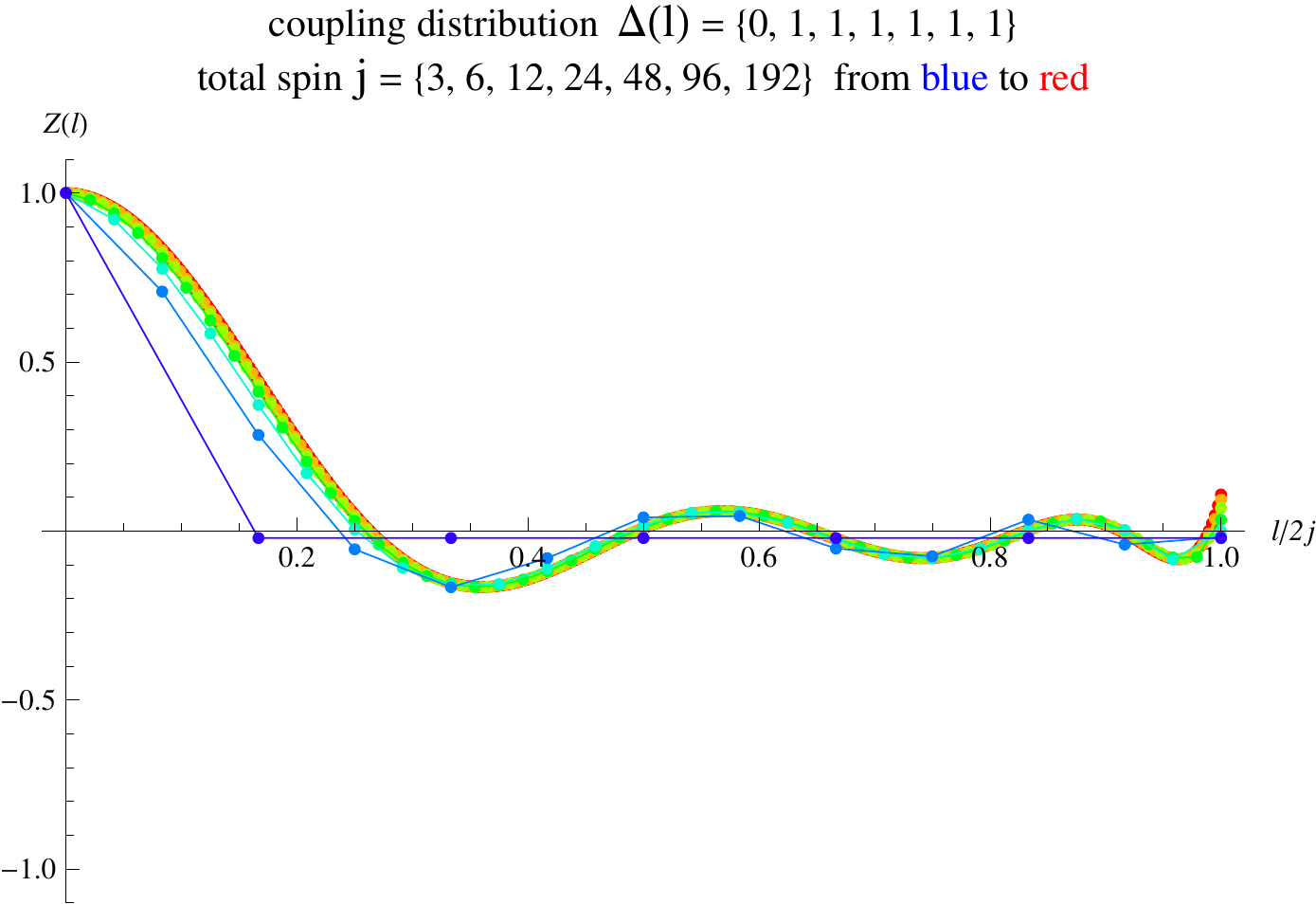} 
\caption{$Z(l)$ for various $l_0$ and $j$, all $\Delta(l)$'s are  equal but $\Delta(0)=0$}
\label{fZDelb0}
\end{center}
\end{figure}

\newpage

\begin{figure}[p]
\begin{center}
\includegraphics[width=\Zplotwidth]{plotZ20.pdf} 
\includegraphics[width=\Zplotwidth]{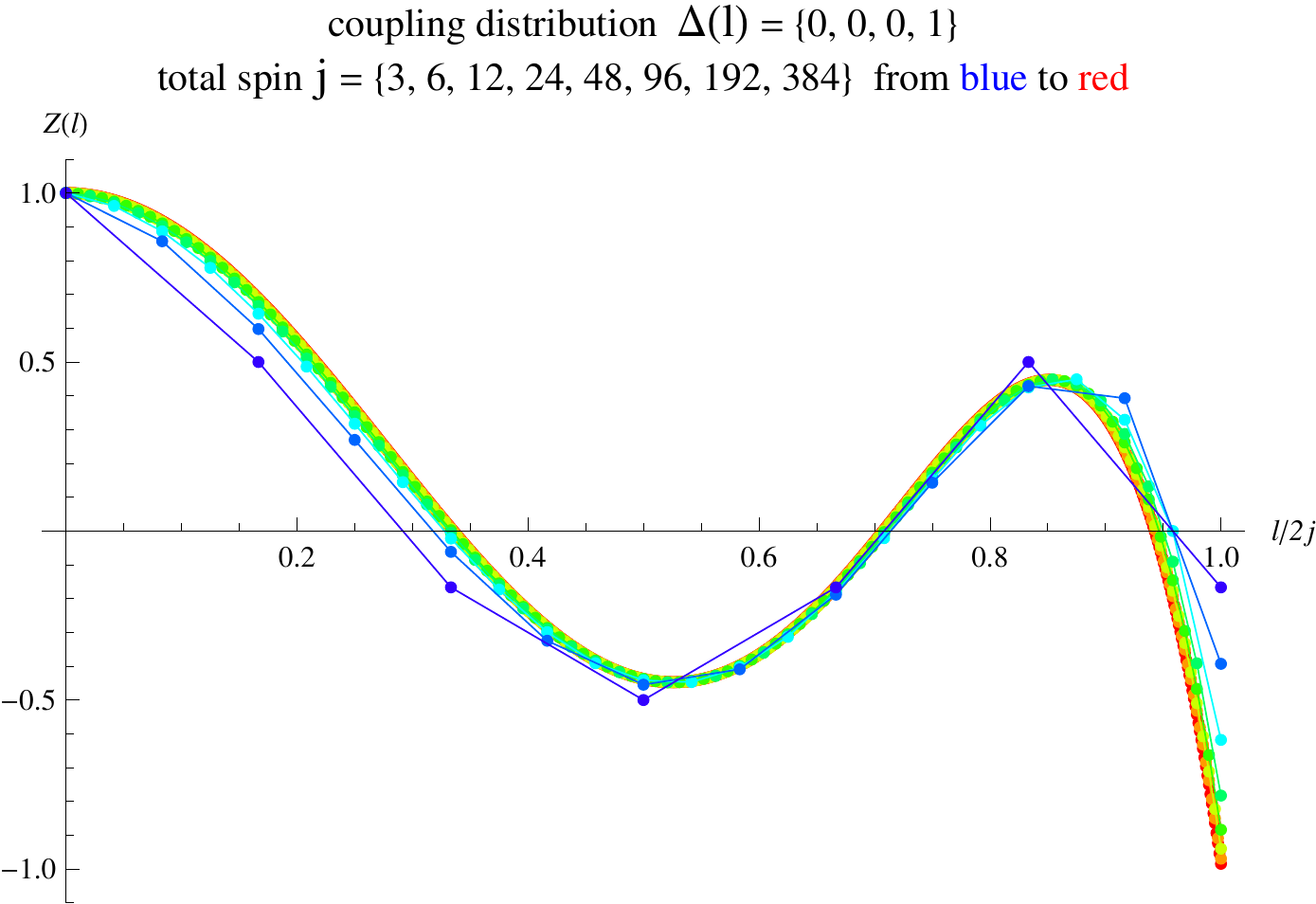} 
\includegraphics[width=\Zplotwidth]{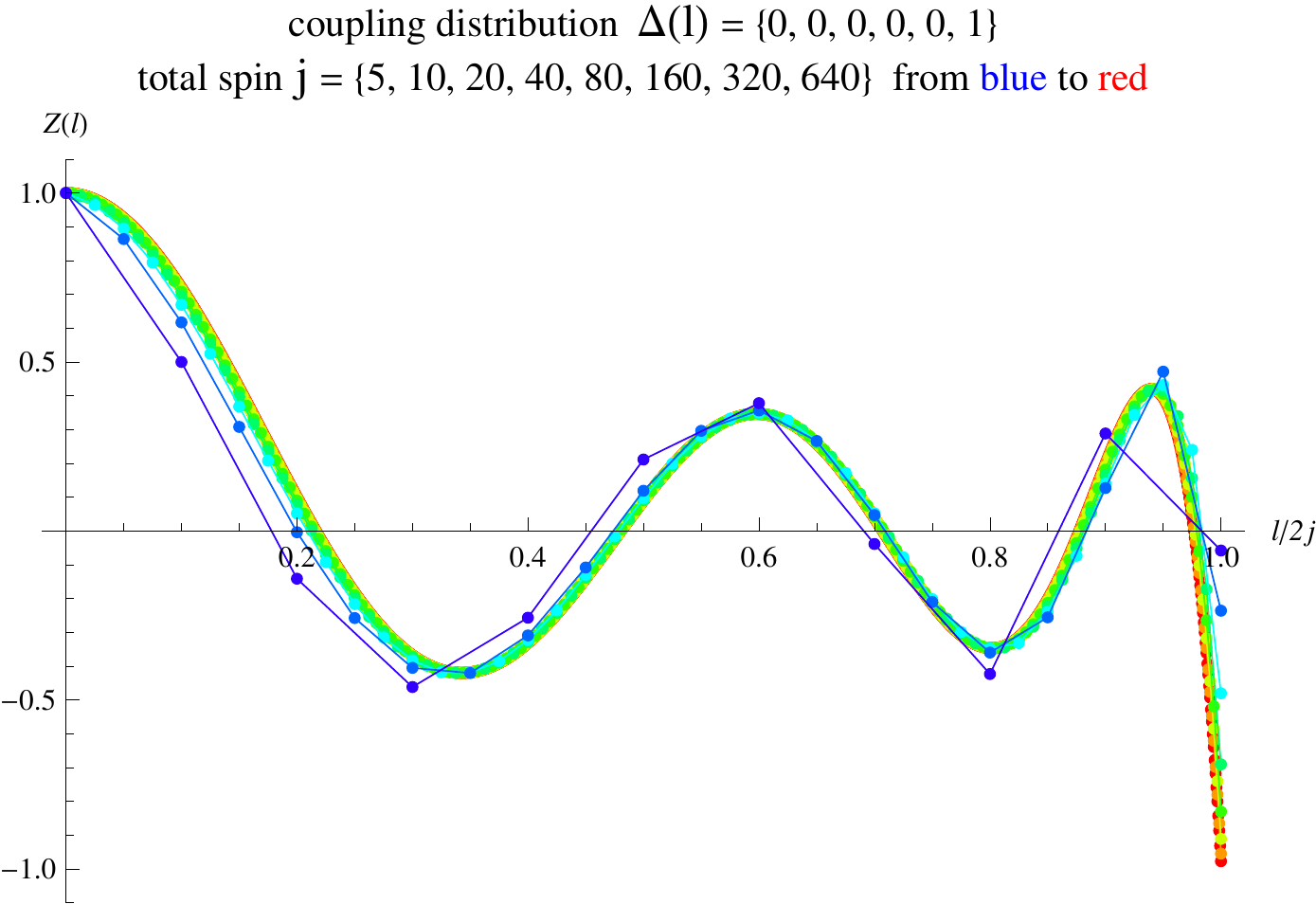} 
\includegraphics[width=\Zplotwidth]{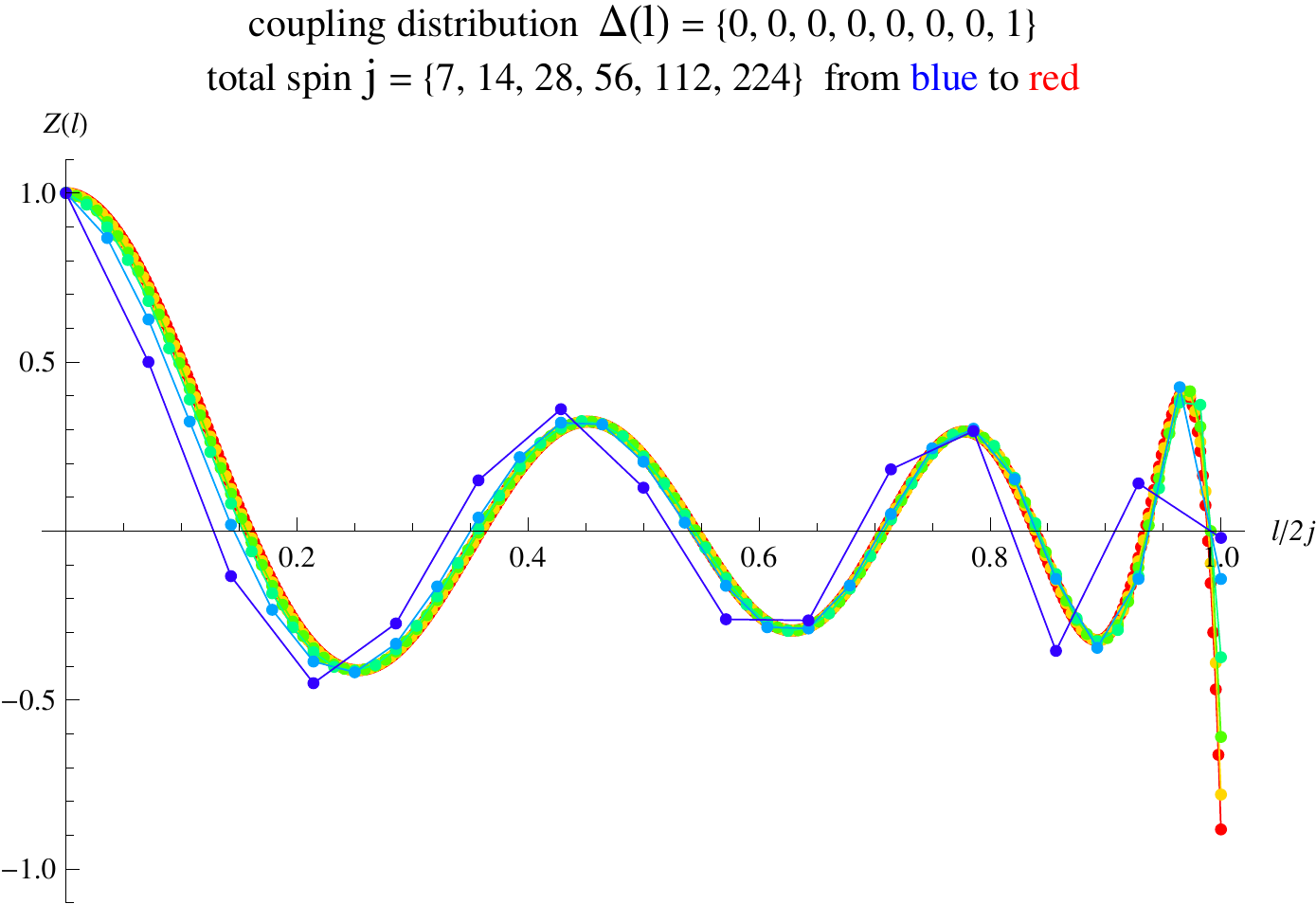} 
\caption{$Z(l)$ for various $l_0$ and $j$, one odd $l$ only}
\label{FZoneOdd}
\end{center}
\end{figure}

%\newpage

\begin{figure}[p]
\begin{center}
\includegraphics[width=\Zplotwidth]{plotZ20.pdf} 
\includegraphics[width=\Zplotwidth]{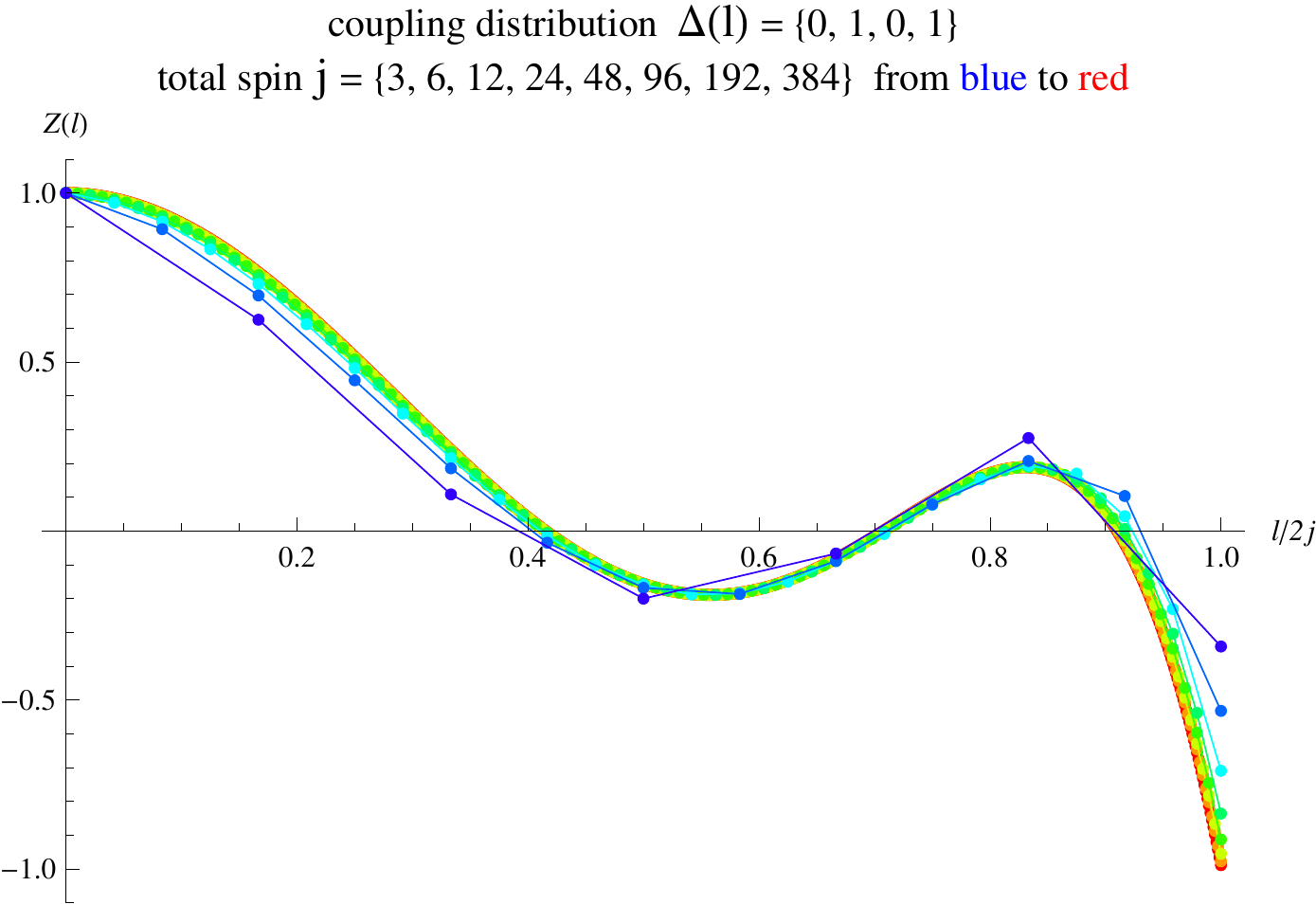} 
\includegraphics[width=\Zplotwidth]{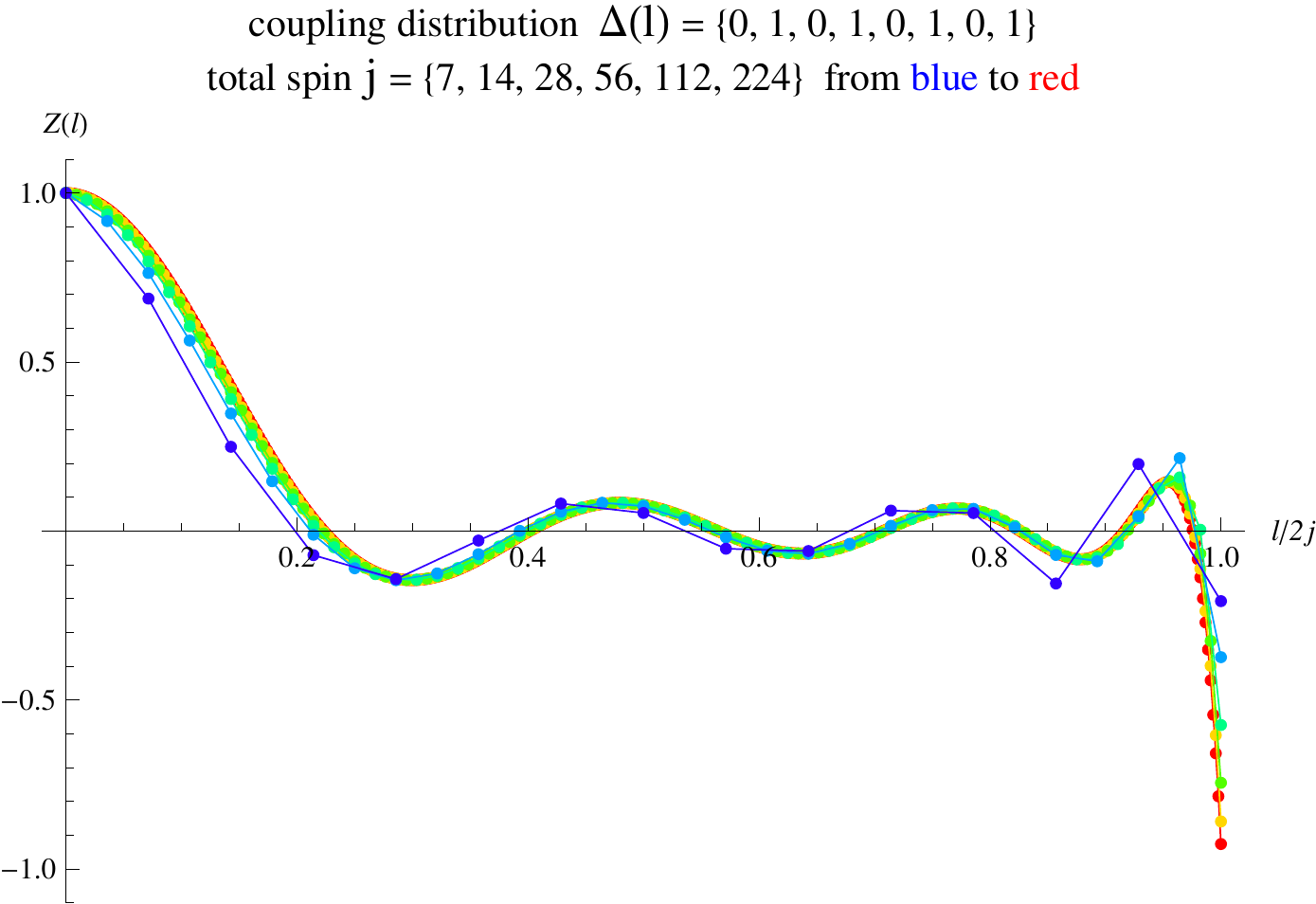} 
\includegraphics[width=\Zplotwidth]{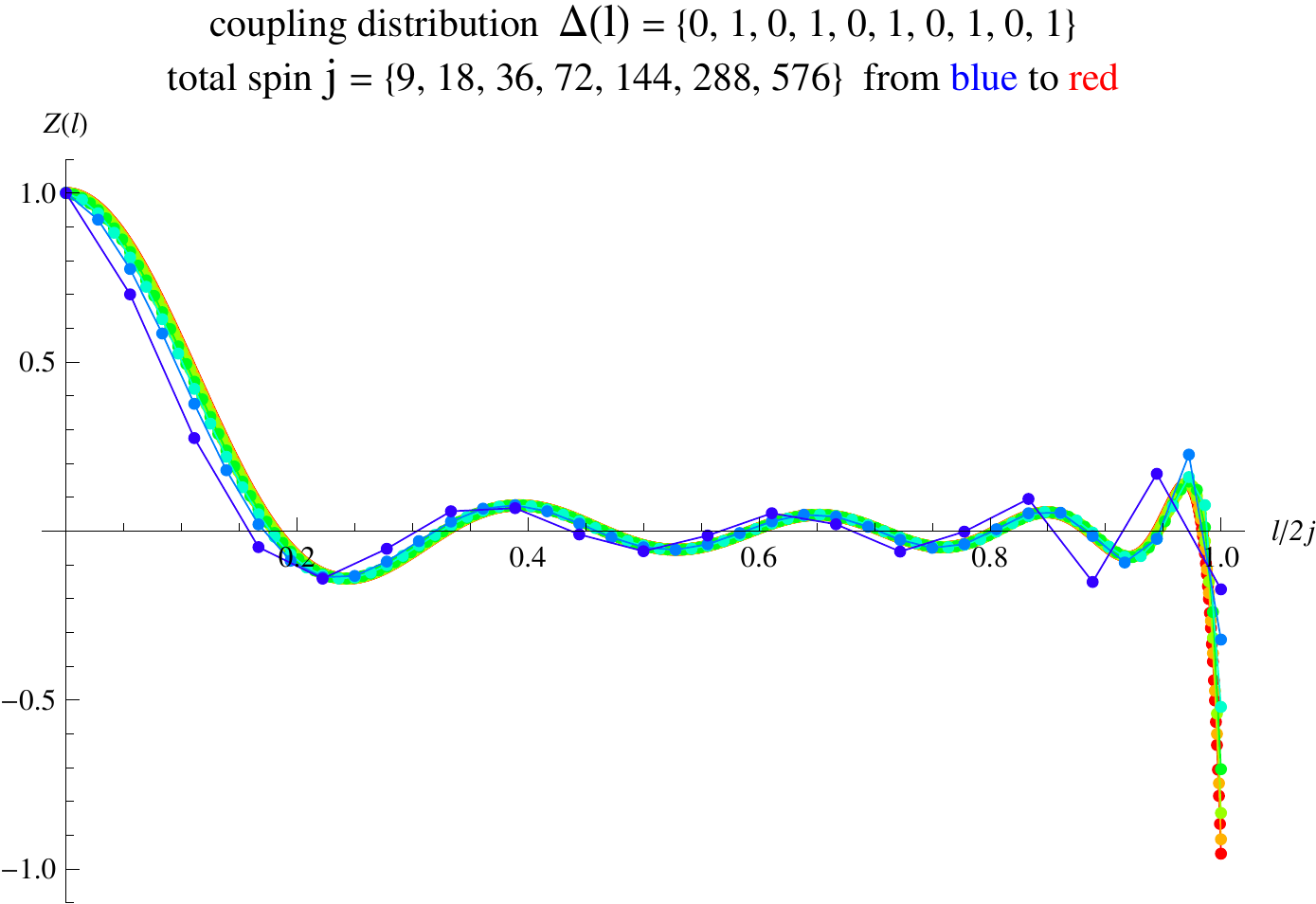} 
\caption{$Z(l)$ for various $l_0$ and $j$, only odd $l$'s , all equal}
\label{FZallOdd}
\end{center}
\end{figure}

\newpage

\begin{figure}[p]
\begin{center}
\includegraphics[width=\Zplotwidth]{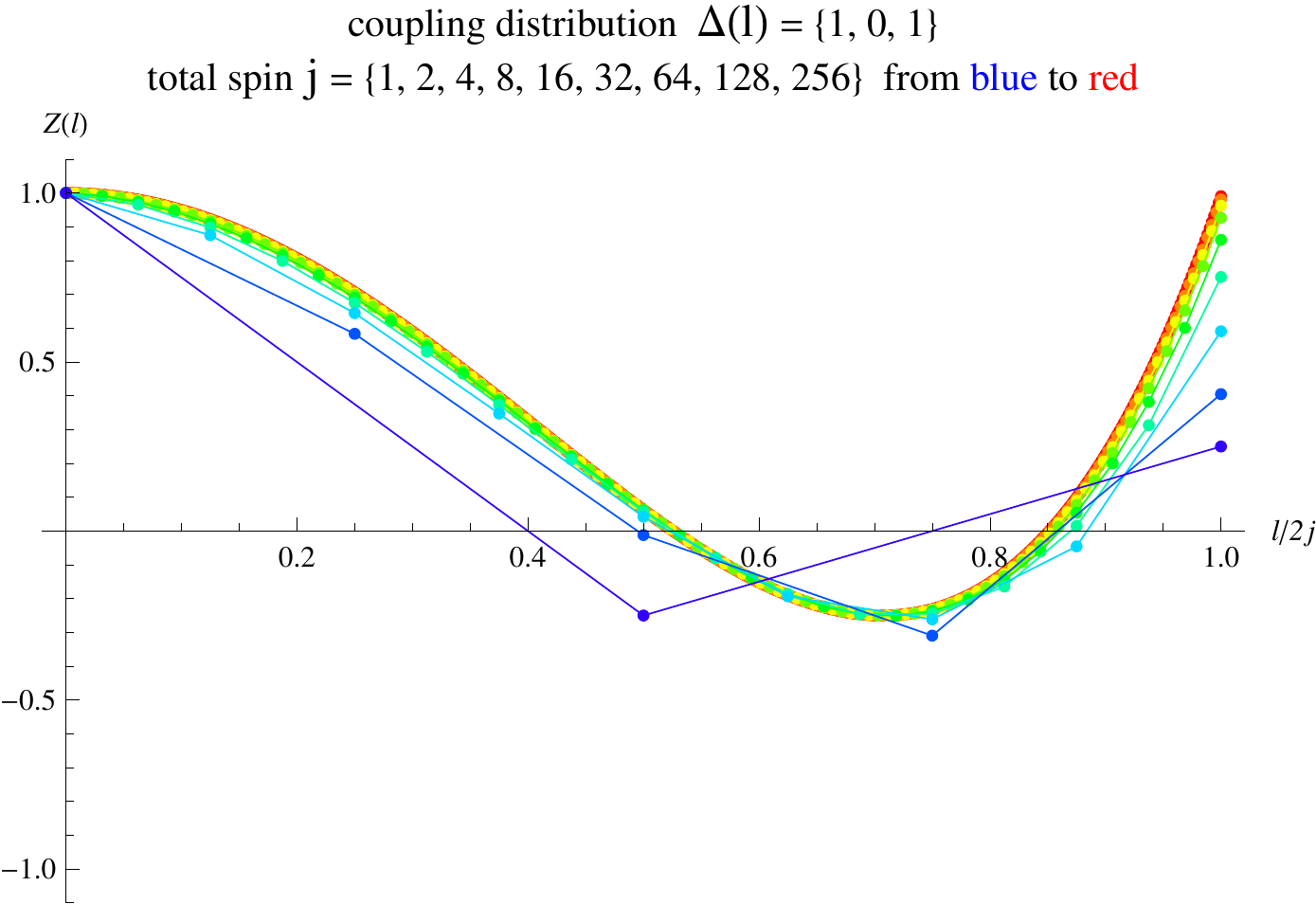} 
\includegraphics[width=\Zplotwidth]{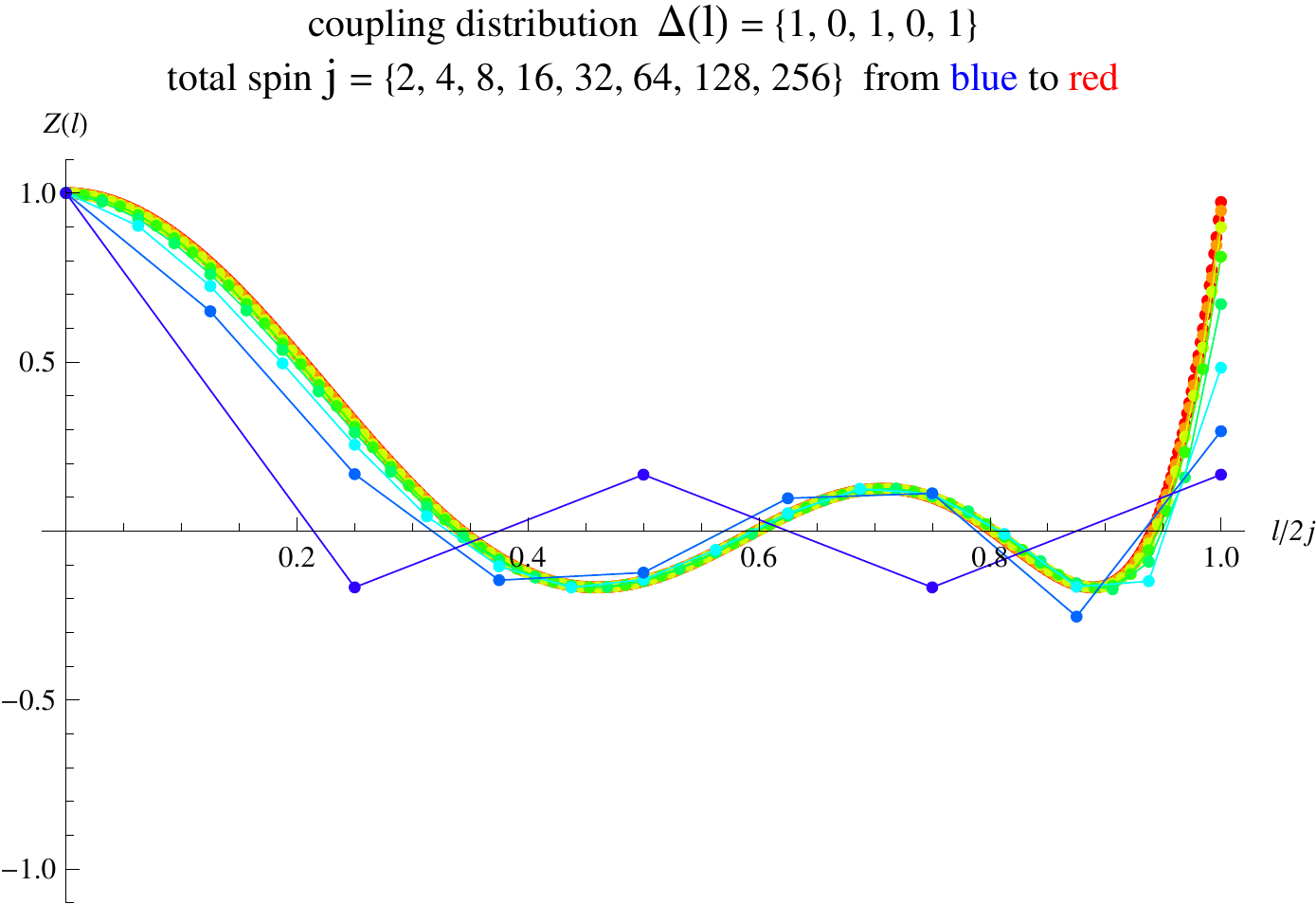} 
\includegraphics[width=\Zplotwidth]{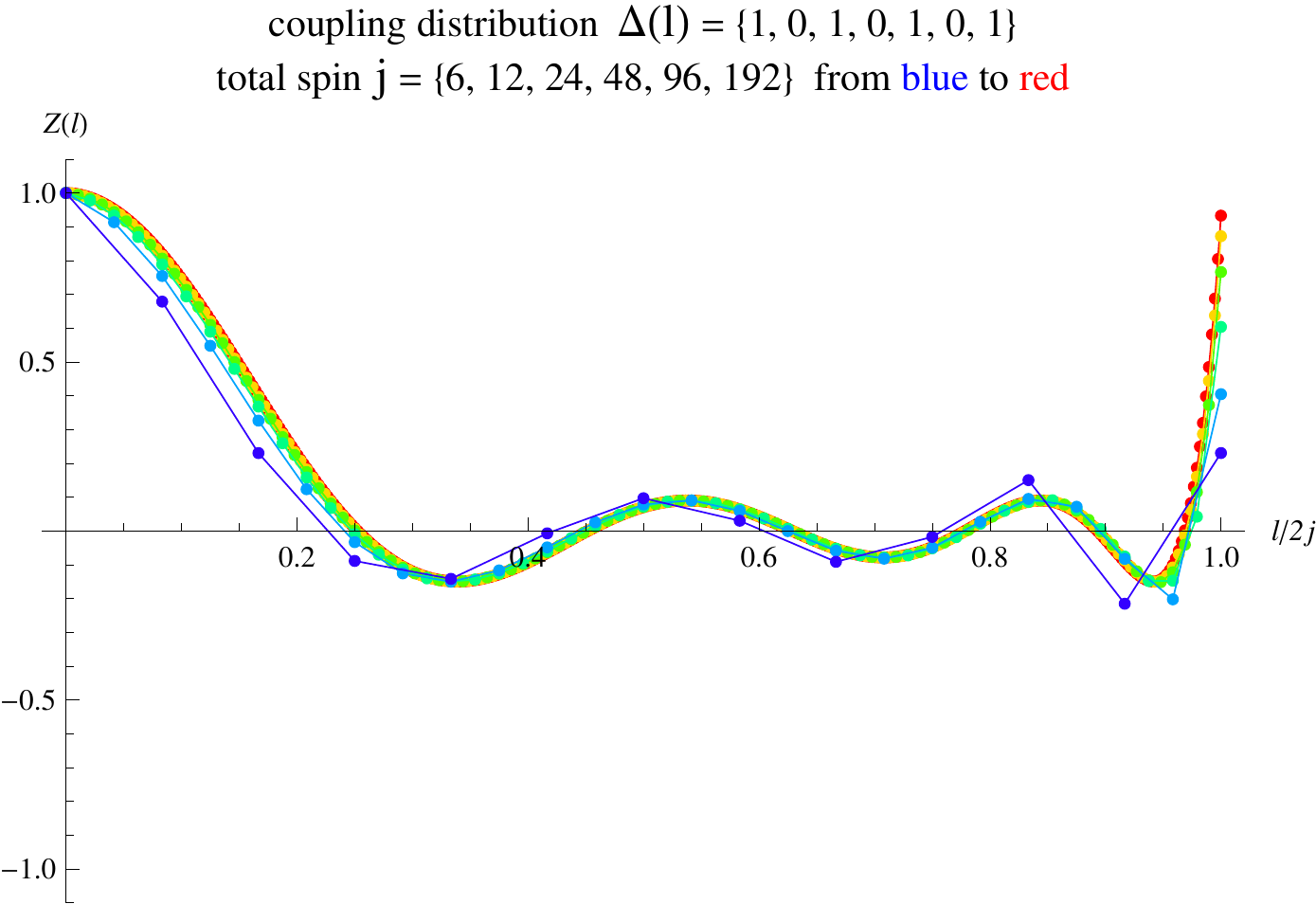} 
\includegraphics[width=\Zplotwidth]{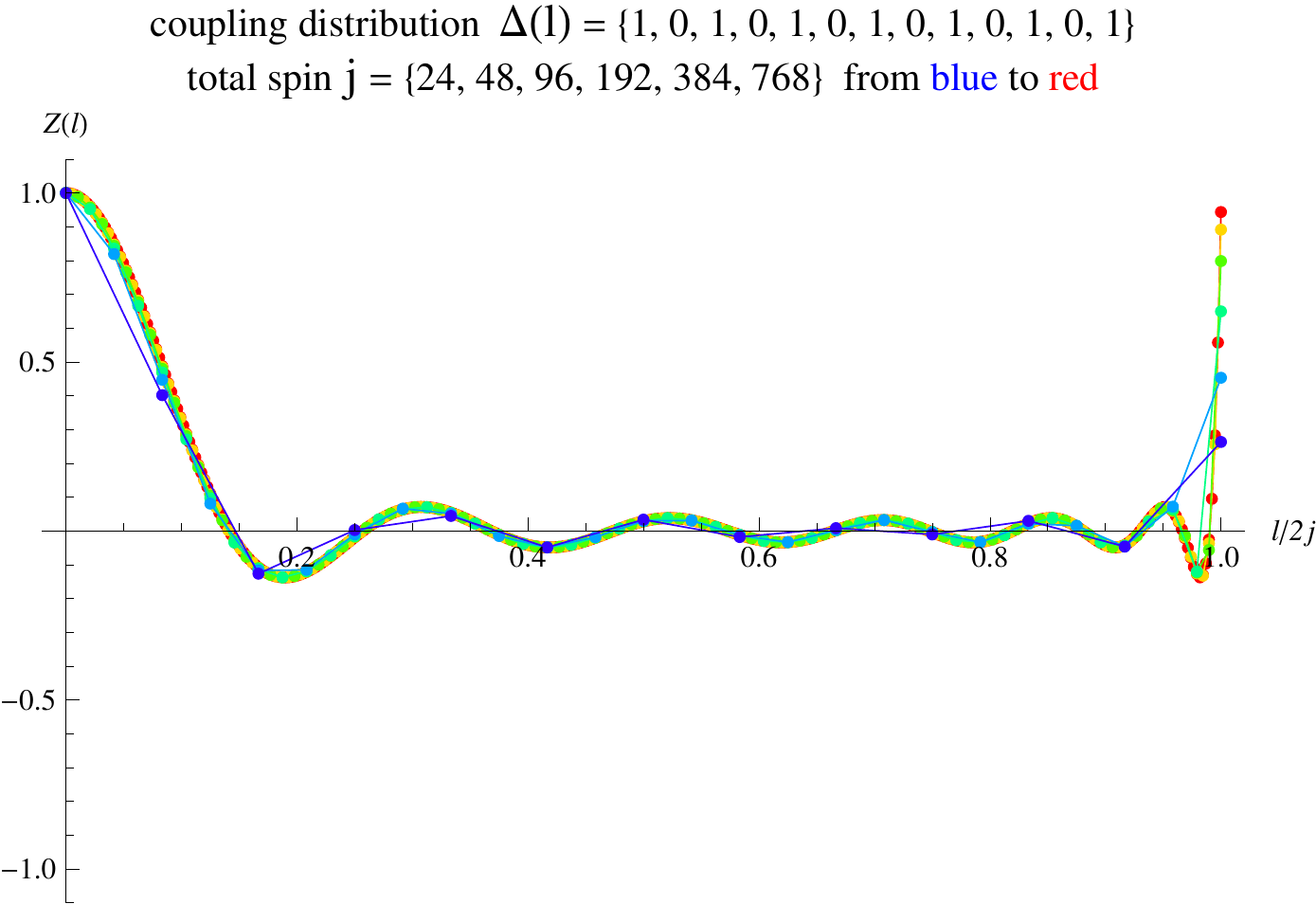} 
\caption{$Z(l)$ for various $l_0$ and $j$, only even $l$'s , all equal}
\label{FZallEven}
\end{center}
\end{figure}

\newpage

\begin{figure}[p]
\begin{center}
\includegraphics[width=\Zplotwidth]{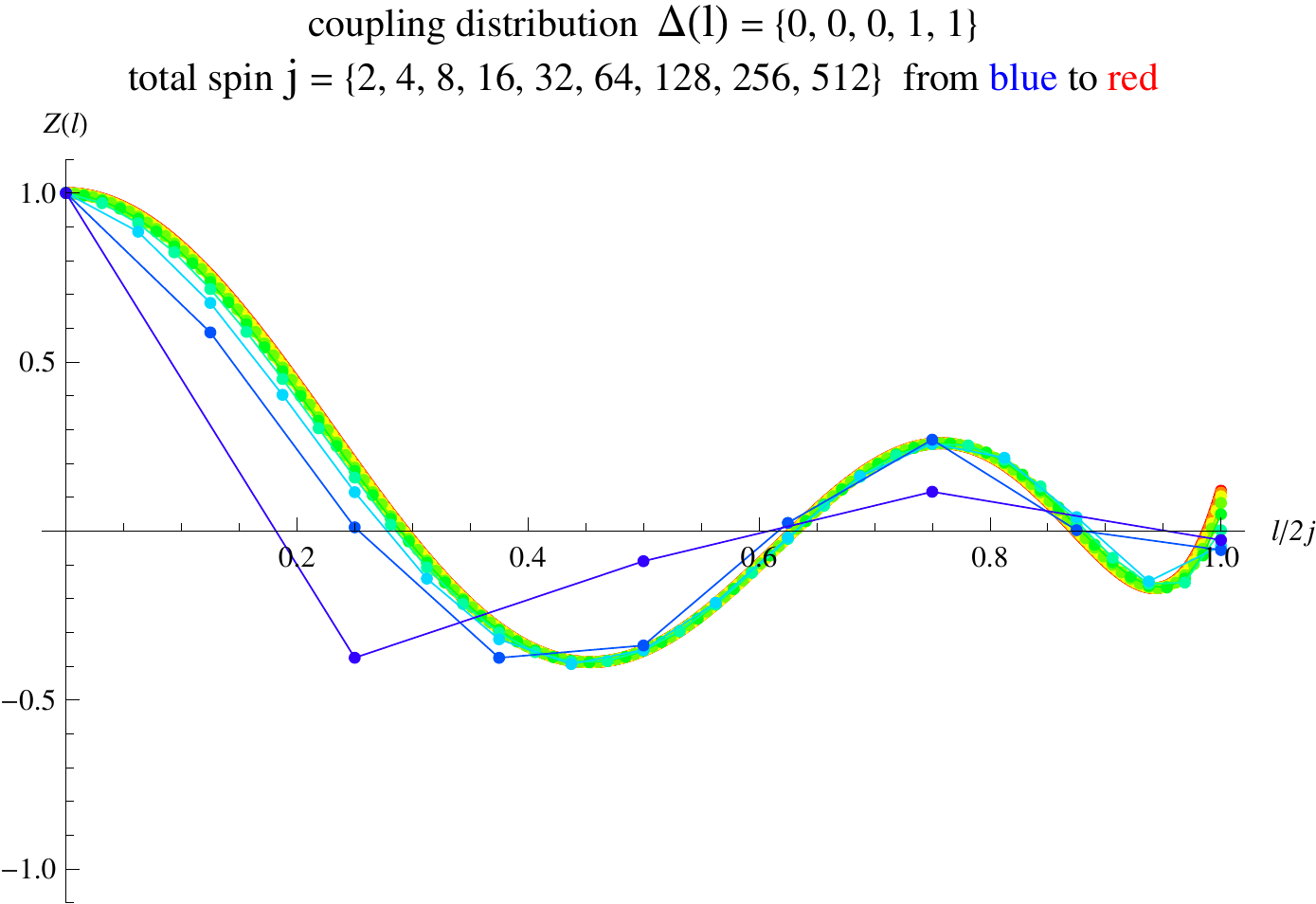} 
\includegraphics[width=\Zplotwidth]{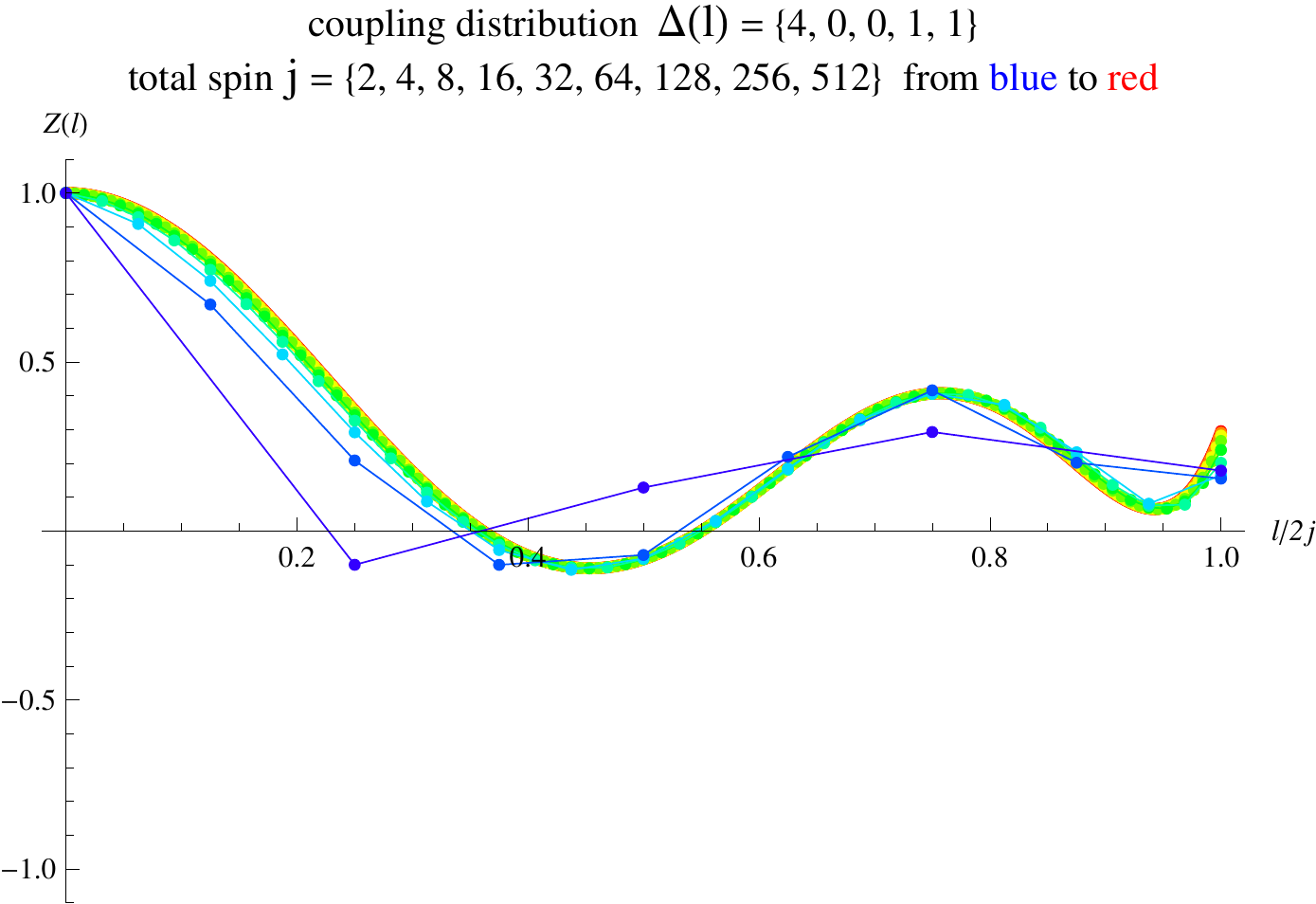} 
\includegraphics[width=\Zplotwidth]{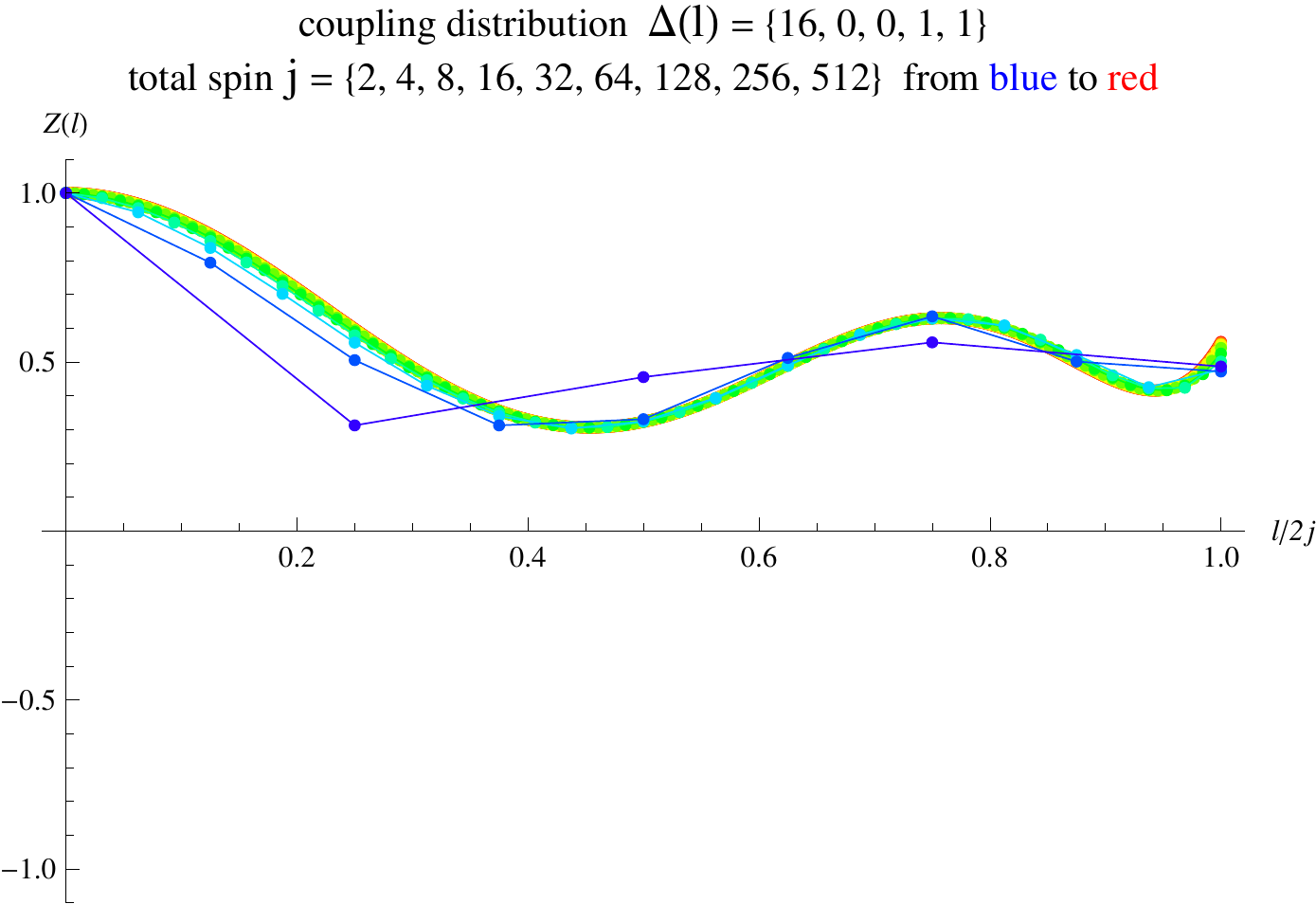} 
\includegraphics[width=\Zplotwidth]{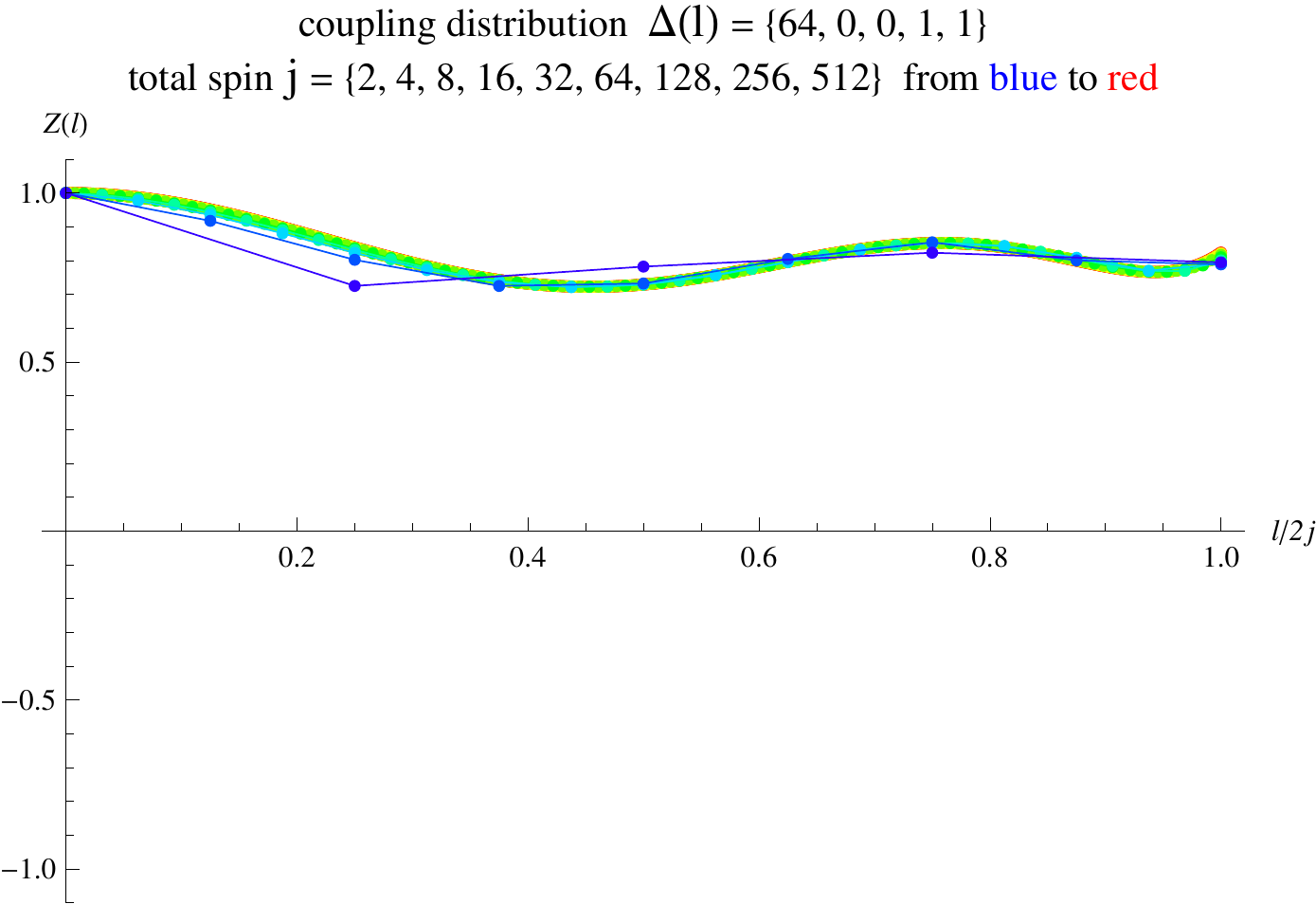} 
\caption{$\Delta(0)$ becomes large, the others are fixed}
\label{FZlargeZ0}
\end{center}
\end{figure}

%\newpage

\begin{figure}[p]
\begin{center}
\includegraphics[width=\Zplotwidth]{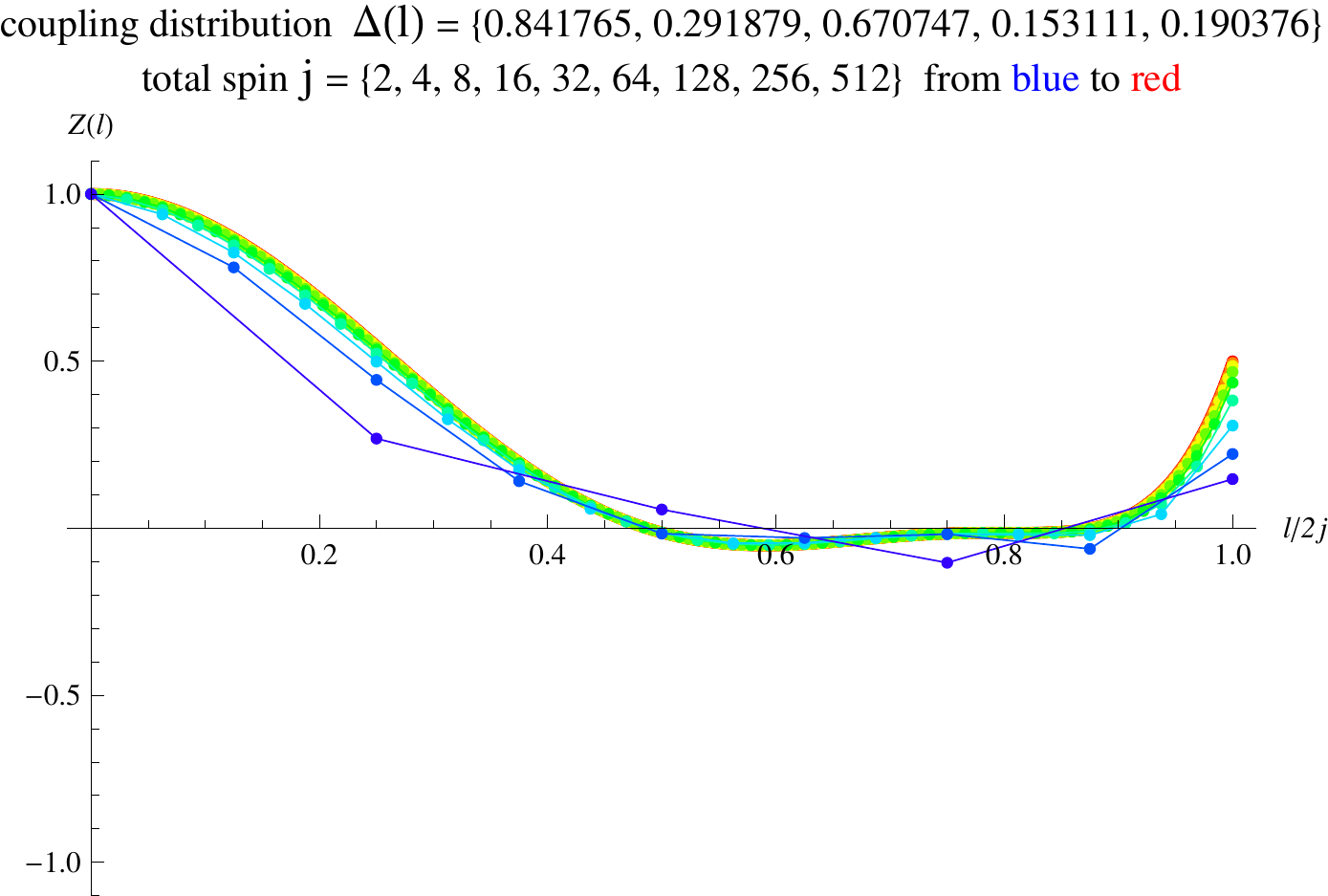} 
\includegraphics[width=\Zplotwidth]{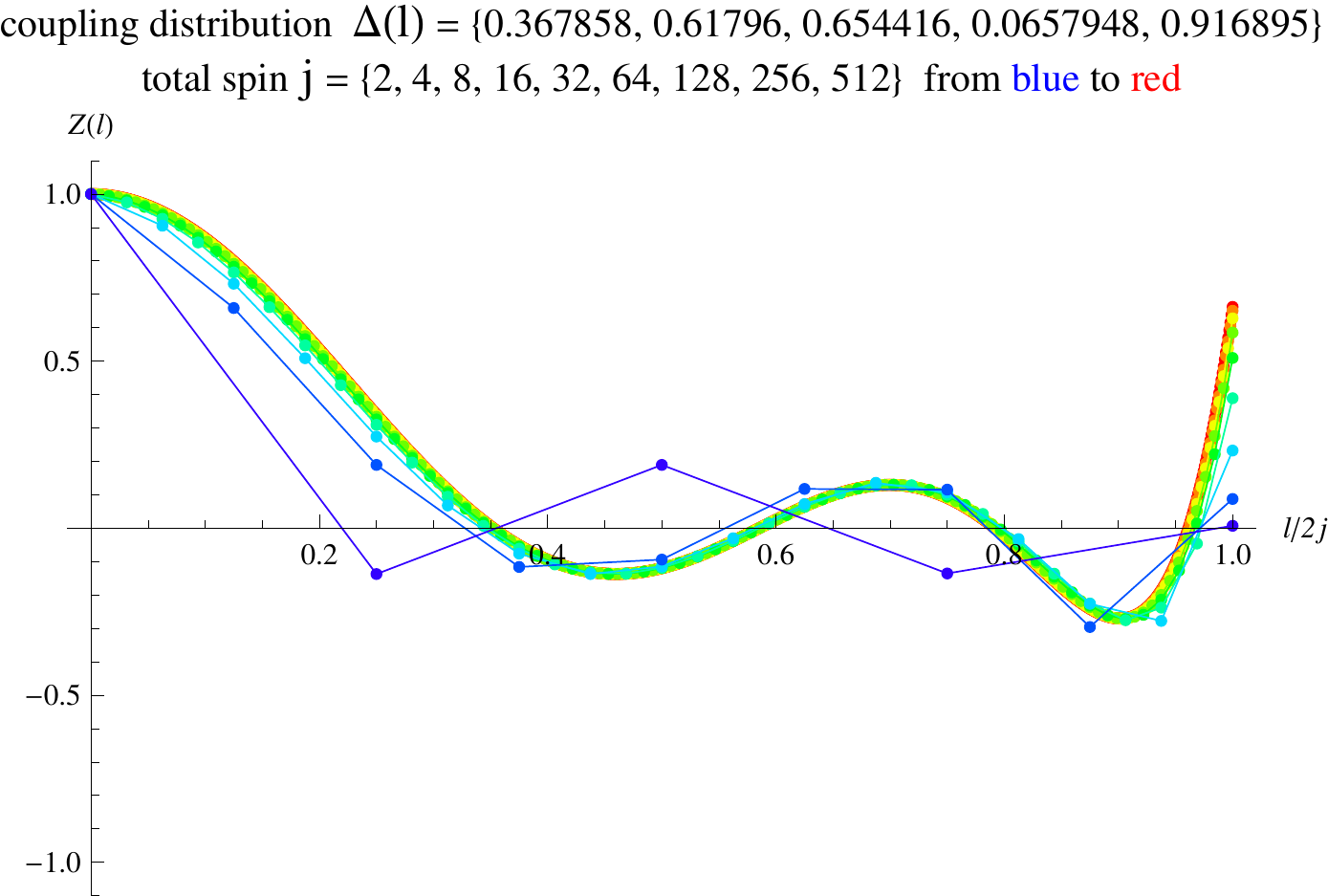} 
\caption{Random $\Delta(l)$'s}
\label{FZranDel}
\end{center}
\end{figure}

\end{appendix}

\end{document}